\documentclass[preprint,pre,floats,showkeys,aps,amsmath,amssymb,superscriptaddress,11pt,nofootinbib]{revtex41}
\pdfoutput=1
\usepackage{graphicx}
\usepackage{natbib}
\setcitestyle{square,comma,numbers,sort&compress}
\usepackage{amsmath,amsfonts,amssymb}
\usepackage{array}
\usepackage{enumerate}
\usepackage[shortlabels]{enumitem}
\usepackage{tabularx}
\usepackage{enumerate}
\usepackage[left = 1.0in, top=1.0in,right=1.0in,bottom=1.0in,a4paper]{geometry}
\usepackage[dvipsnames]{xcolor}
\usepackage{colortbl}
\usepackage{comment}
\usepackage{siunitx}
\usepackage[caption=false]{subfig}
\usepackage{tikz}
\usepackage[compat=1.1.0]{tikz-feynman}
\usepackage{cancel}
\usepackage{hyperref}

\newcommand{\be}{\begin{equation}}
\newcommand{\ee}{\end{equation}}
\newcommand{\ben}{\begin{equation*}}
\newcommand{\een}{\end{equation*}}

\def\bea{\begin{eqnarray}}
\def\eea{\end{eqnarray}}
\def\bean{\begin{eqnarray*}}
\def\eean{\end{eqnarray*}}

\def\l2{\log_2\,}
\newcommand{\barr}{\begin{array}}
\newcommand{\earr}{\end{array}}

\newcommand{\bed}{\begin{displaymath}}
\newcommand{\eed}{\end{displaymath}}
\newcommand{\bal}{\begin{array}{ll}}
\newcommand{\eal}{\end{array}}

\def\sc#1{\textstyle{\scriptstyle #1}}

\def\mc#1{\mathcal#1}

\definecolor{BrickRed}{rgb}{0.8, 0.25, 0.33}

\newcolumntype{Y}{>{\centering\arraybackslash}X}
 
\newcommand{\g}[1]{\mathbf{#1}}
\newcommand{\vev}[1]{\langle #1 \rangle_0}
\newcommand{\gb}[1]{\bar{\mathbf{#1}}}

\newcommand{\reply}[1]{\textcolor{black}{#1}}

\newcommand{\blu}[1]{\textcolor{black}{#1}}

\newcommand{\myboxed}[1]{%
  \rlap{\hspace*{\dimexpr\fboxrule+\fboxsep\relax}%
    \phantom{\m@th$\displaystyle#1$}}%
    \smash{\boxed{#1}}}

\renewcommand{\thesection}{\arabic{section}{\Large}}
\renewcommand{\thesubsection}{\thesection.\arabic{subsection}}
\renewcommand{\thesubsubsection}{\thesubsection.\arabic{subsubsection}}

\makeatletter
\renewcommand{\p@subsection}{}
\renewcommand{\p@subsubsection}{}
\makeatother

\begin{document}

\title{\Large Low-scale Resonant Leptogenesis in $SU(5)$ GUT \\with $\mathcal{T}_{13}$ Family Symmetry \\\vspace*{1cm}}

\author{Chee Sheng Fong} 
\email[Email: ]{sheng.fong@ufabc.edu.br}
\affiliation{\small{Centro de Ciências Naturais e Humanas,
Universidade Federal do ABC, 09.210-170, Santo André, SP, Brazil}}

\author{Moinul Hossain Rahat} 
\email[Email: ]{mrahat@ufl.edu}
\affiliation{\small{Institute for Fundamental Theory, Department of Physics,
University of Florida, Gainesville, FL 32611, USA }}

\author{Shaikh Saad} 
\email[Email: ]{shaikh.saad@unibas.ch}
\affiliation{\small{Department of Physics, University of Basel,\\Klingelbergstrasse\ 82, CH-4056 Basel, Switzerland} \vspace{1cm}}

\begin{abstract}
\vskip 0.5cm
We investigate low-scale resonant leptogenesis in an $SU(5) \times \mc T_{13}$ model where a single high energy phase in the complex Tribimaximal seesaw mixing produces the yet-to-be-observed low energy Dirac and Majorana ${CP}$ phases. A fourth right-handed neutrino, required to generate viable light neutrino masses within this scenario, also turns out to be necessary for successful resonant leptogenesis where $CP$ asymmetry is produced by the same high energy phase. We derive a lower bound on the right-handed neutrino mass spectrum in the $\rm GeV$ range, where part of the parameter space, although in tension with Big Bang Nucleosynthesis constrains, can be probed in planned high intensity experiments like DUNE. We also find the existence of a curious upper bound ($\text{TeV}$-scale) on the right-handed neutrino mass spectrum in majority of the parameter space due to significant washout of the resonant asymmetry by lighter right-handed neutrinos. While in most of the parameter space of our model classical Boltzmann equations are sufficient, when right-handed neutrino masses are below the electroweak sphalerons freeze-out temperature we resort to the more general density matrix formalism to take into account decays and oscillations of the right-handed neutrinos as well as their helicities which are relevant when they are relativistic.
\end{abstract}

\maketitle
\tableofcontents
\section{Introduction}

While leptonic mixing angles and neutrino mass squared differences have been measured rather precisely, the mechanism behind them is still an open question. A plausible explanation of the small neutrino mass scale $m_\nu$ is through the seesaw mechanism \cite{Minkowski:1977sc,Yanagida:1979as,GellMann1979,Glashow:1979nm,Mohapatra:1979ia,*Harvey:1980je} $m_\nu \sim c v^2/\Lambda$ where $v = 174$ $\text{GeV}$ is the Higgs vacuum expectation value (VEV), $c$ is a dimensionless number and $\Lambda$ is a new physics (seesaw) scale where lepton number is violated. This type of model also provides an opportunity to explain the observed baryon asymmetry through leptogenesis \cite{Fukugita:1986hr}\footnote{For reviews on leptogenesis, see for example Refs.  \cite{Buchmuller:2004nz,Nir:2007zq,Davidson:2008bu,Pilaftsis:2009pk,DiBari:2012fz,Fong:2013wr,Chun:2017spz,Dev:2017trv, Bodeker:2020ghk}.} at the cosmic temperature $T \sim \Lambda$. As long as a lepton asymmetry is generated above the temperature $T_{sph} \sim 131.7$ $\text{GeV}$ \cite{DOnofrio:2014rug} when the electroweak sphaleron interactions are in thermal equilibrium, a baryon asymmetry will also be induced.

The scale of the neutrino mass $m_\nu \sim 0.1$ $\rm eV$ imposes $\Lambda/c \sim 10^{14}$ $\text{GeV}$; depending on the details of the ultraviolet-complete model, leptogenesis can be viable for $\Lambda$ ranging from $\text{GeV}$ to $10^{14}$ $\text{GeV}$. In the simplest scenario, one introduces some heavy right-handed neutrinos with Majorana mass $M \sim \Lambda$ and they are singlets under the Standard Model (SM) gauge interactions. The $CP$ asymmetry from the decays of right-handed neutrinos can be estimated to be $\epsilon \sim {m_\nu M}/{(16\pi v^2)}$. A sufficiently large $CP$ asymmetry $\epsilon \gtrsim 10^{-7}$ to generate adequate baryon asymmetry imposes the so-called Davidson-Ibarra bound $M \gtrsim 10^9$ $\text{GeV}$ \cite{Davidson:2002qv}. Nevertheless, if a pair of right-handed neutrinos are quasi-degenerate in mass, the $CP$ asymmetry can be resonantly enhanced \cite{Covi:1996fm,Pilaftsis:1997jf} and the lower bound on $M$ comes only from the requirement that sufficient baryon asymmetry is induced at $T > T_{sph}$. This type of low-scale seesaw models have the virtue of being directly probed in experiments.

In the resonant leptogenesis scenario, leptogenesis can proceed through oscillations \cite{Akhmedov:1998qx} (see a recent review \cite{Drewes:2017zyw}) among the right-handed neutrinos or through their decays \cite{Pilaftsis:1997jf} (see a recent review \cite{Dev:2017wwc}). In the former scenario, the typical mass scale is $M \sim$ $0.1 - 1$ $\text{GeV}$ since oscillations take place at a higher scale $T \gg M$ (see some recent studies \cite{Abada:2018oly, Domcke:2020ety, Bondarenko:2021cpc}), while in the latter scenario, the requirement that sufficient amount of the right-handed neutrinos should decay above $T_{sph}$ imposes a lower bound on their mass $M \gtrsim T_{sph}$. The lower bound can in fact be much lower down to $\text{GeV}$ scale for the case of zero initial abundance of right-handed neutrinos since baryon asymmetry is generated at an earlier stage $T > M$ through inverse decays. This is confirmed in several recent studies \cite{Granelli:2020ysj,Klaric:2020lov} and is also consistent with our finding in this work. 
\reply{As shown in Ref.~\cite{Klaric:2020lov}, there is some overlap in the parameter space between resonant leptogenesis from oscillation and from decay, though in the latter case the required mass splitting of quasi-degenerate right-handed neutrinos is required to be much smaller. When $M \lesssim T_{sph}$, relativistic effects \cite{Biondini:2017rpb} and in particular, the effect from the helicities of $N_i$ \cite{Shaposhnikov:2008pf, Ghiglieri:2017gjz, Eijima:2017anv, Antusch:2017pkq, Bodeker:2019rvr,  Eijima:2018qke, Abada:2018oly, Klaric:2020lov, Klaric:2021cpi} become relevant. In this regime, we resort to using the density matrix equations \cite{Abada:2018oly} which take into account the oscillations among families of right-handed neutrinos of respective helicities.
}


\reply{In this work, we focus on the regime where resonant leptogenesis predominantly occurs through decays (though the effects of oscillation are also taken into account)} of the right-handed neutrinos within a specific model based on $SU(5)$ grand unified theory\footnote{Grand unified theories were initially proposed in Refs. \cite{Pati:1974yy, Georgi:1974sy, Georgi:1974yf, Georgi:1974my, Fritzsch:1974nn, Gursey:1975ki}.} (GUT) supplemented by a discrete family symmetry $\mathcal{T}_{13}$ \cite{Rahat:2018sgs, Perez:2019aqq, Perez:2020nqq, Rahat:2020mio}. With the family symmetry broken by the familon VEVs around the GUT scale, this model is able to explain the observed hierarchical mass spectra in the charged fermion sector from symmetry arguments \cite{Rahat:2018sgs,Perez:2019aqq,Perez:2020nqq}. Large mixing angles observed in the lepton sector, unlike the quark sector, may also be directly linked with the family symmetry\footnote{For recent reviews on flavor puzzle, see for example Refs. \cite{Babu:2009fd,Feruglio:2015jfa,Xing:2019vks, Feruglio:2019ktm}.}. The large atmospheric and solar angles in the PMNS matrix can be  naturally explained via the well-motivated Tribimaximal (TBM) mixing matrix \cite{harrison2002tri, tbm2, xing2002nearly, he2003some, wolfenstein1978oscillations, Luhn:2007sy, Chen:2014wiw}, and $\mathcal{T}_{13}$ is a suitable family symmetry group  to produce this TBM structure\footnote{For studies of the Standard Model supplemented by $\mathcal{T}_{13}$ group, see Refs. \cite{Kajiyama:2010sb,Parattu:2010cy,Ding:2011qt,Hartmann:2011pq}.}. In this framework, TBM neutrino mixing arises from alignment of the vacuum structure of a minimal number of familons that give rise to the Dirac Yukawa and the Majorana matrices in the seesaw formula. In order to explain the neutrino observables, a total of four right-handed neutrinos are needed in this $SU(5)$ set-up \cite{Perez:2020nqq,Rahat:2020mio}, which has non-trivial consequence on the resonant leptogenesis process we study in this work. 

\reply{
This model contains a very few free parameters, which set the scale of the Dirac Yukawa and Majorana matrices. In particular, the seesaw scale remains arbitrary since one has the freedom to adjust the scale of Dirac Yukawa couplings accordingly such that the light neutrino mass matrix remains unchanged~\cite{Perez:2020nqq}. 
Since the mass scale of the right-handed neutrinos is not fixed, it is interesting to explore the implication of both the high scale and the low scale masses on leptogenesis. In Ref.~\cite{Rahat:2020mio}, high scale masses of $\mc O(10^{11-12})$ GeV have been shown to generate successful flavored leptogenesis. The goal of this paper is to explore if the mass scale can be low, preferably of $\mc O(\text{GeV}-\text{TeV})$, and baryon asymmetry can still be explained via resonant leptogenesis. This opens up the possibility to test the model at low energy experiments to be detailed later in the text.}



As shown in Ref.~\cite{Rahat:2020mio}, the total $CP$ asymmetry is always vanishing in unflavored leptogenesis and leptogenesis can only proceed when the lepton flavor effects are relevant, i.e., $T \lesssim 10^{12}$ $\text{GeV}$. Ref. \cite{Rahat:2020mio} also shows that a symmetric familon VEV configuration to generate the Majorana mass entries of the right-handed neutrinos cannot lead to a successful leptogenesis. Hence, we focus on the next simplest scenario where one component of the familon VEV is lifted by a factor $f \neq 1$. Since the masses of all the right-handed neutrinos are related, this allows us to identify a unique quasi-degenerate mass pair \textcolor{black}{consisting of the heaviest right-handed neutrinos in the spectrum}, for which resonant leptogenesis is viable. It turns out that the fourth right-handed neutrino warranted to explain the neutrino observables is also crucial for viable resonant leptogenesis within this scenario.

Imposing the resonance condition that the mass difference between the resonant pairs is of the order of their average decay width, we derive lower bounds on the right-handed neutrino mass spectrum as a function of the factor $f$. Nontrivially, we also obtain upper bounds on right-handed neutrino mass spectrum for large $f$. This upper bound appears due to the existence of lighter right-handed neutrinos which can wash out the asymmetry generated by the heavier quasi-degenerate pair. 
Since all the parameters are related at resonance, we are able to plot the mixing elements of right-handed neutrinos with active neutrinos as function of their masses. Interestingly, some of the parameter space is constrained by the Big Bang Nucleosynthesis (BBN) observables while the rest would be interesting for experiments searching for heavy neutral leptons (such as SHiP \cite{Bonivento:2013jag, SHiP:2018xqw, Gorbunov:2020rjx}  and DUNE \cite{Ballett:2019bgd, Abi:2020evt}).

This paper is organized as follows.  In Sec. \ref{sec2}, we review the motivation for and the key features of the $SU(5)\times \mathcal{T}_{13}$ model. Conditions for resonant leptogenesis are discussed in Sec. \ref{sec3}, and further specifications of resonant leptogenesis in the context of the $SU(5)\times \mathcal{T}_{13}$ framework are detailed in Sec. \ref{sec:resonantasym}. Our results and experimental constraints are presented in Sec. \ref{sec:results}, and finally we conclude in Sec. \ref{sec6}. 
\reply{In App. \ref{app:VEV}, we discuss other possible choices of flavon VEV configurations of the model while in App. \ref{app:density_matrix}, we collect the density matrix equations that we use in the regime when $M \lesssim T_{sph}$.}

\section{The $SU(5) \times \mc T_{13}$ Model}\label{sec2}
The $SU(5) \times \mc T_{13}$ model constructs the ``asymmetric texture'' \cite{Rahat:2018sgs} aiming to generate the required ``Cabibbo haze'' \cite{Datta:2005ci, everett2006viewing, Everett:2006fq, Kile:2013gla, Kile:2014kya, Ramond:2019fsr} in order to supplement the TBM seesaw mixing  to reproduce the observed PMNS mixing angles. Its structure is inspired by the $SU(5)$ Georgi-Jarlskog texture \cite{Georgi:1979df} where the down-type quark and charged-lepton Yukawa matrices $Y^{(-\frac{1}{3})}$ and $Y^{(-1)}$, respectively, are generated by coupling of the $\gb{5}$ and $\overline{\g{45}}$ Higges to the fermions in $\gb{5}$ and $\g{10}$ representations of $SU(5)$, and are related by
\begin{align*}
Y^{(-\frac{1}{3})} = Y_{\gb{5}} + Y_{\overline{\g{45}}},\;\;\;\;\; Y^{(-1)} = Y_{\gb{5}}^T -3 Y_{\overline{\g{45}}}^T.
\end{align*}
Assuming a diagonal hierarchical up-type quark Yukawa matrix $Y^{(\frac{2}{3})}$, a bottom-up approach finds that symmetric Yukawa textures fall short in explaining the nonzero reactor angle \cite{Kile:2014kya, Rahat:2018sgs}. The minimal required asymmetry results in the following set of Yukawas \cite{Rahat:2018sgs}:
\begin{align}
\begin{aligned}
Y^{(\frac{2}{3})} \sim~ &\mathrm{diag}\ (\lambda^8, \lambda^4, 1), \\
Y^{(-{\frac{1}{3}})} \sim 
\begin{pmatrix}
 \frac{2}{3}\sqrt{\rho^2+\eta^2} \lambda ^4 & \frac{\lambda ^3}{3} & A\sqrt{\rho^2+\eta^2} \lambda ^3 \cr
 \frac{ \lambda ^3}{3} & \frac{ \lambda ^2}{3} & A \lambda ^2 \cr
 \frac{2 \lambda}{3A}  & A \lambda ^2 & 1 \end{pmatrix},
&\quad
Y^{(-1)} \sim 
\begin{pmatrix}
 \frac{2}{3}\sqrt{\rho^2+\eta^2} \lambda ^4 & \frac{\lambda ^3}{3} & \frac{2 \lambda}{3A} \cr
 \frac{ \lambda ^3}{3} & - \lambda ^2 & A \lambda ^2 \cr
 A\sqrt{\rho^2+\eta^2} \lambda ^3  & A \lambda ^2 & 1 \end{pmatrix}. \label{texture}
\end{aligned}
\end{align}

\noindent Here $A \simeq 0.81$, $\lambda \simeq 0.225 $, $\rho \simeq 0.135$, and $\eta \simeq 0.35$ are the Wolfenstein parameters \cite{Wolfenstein:1983yz}. The only $\overline{\g{45}}$ coupling is in the $(22)$ element of $Y^{(-\frac{1}{3})}$ and $Y^{(-1)}$. Unitary diagonalization of the Yukawa matrices give $Y^{(q)} =$ $ \mc U^{(q)} \mc D^{(q)} \mc V^{(q)^\dagger}$, where $\mc U^{(-\frac{1}{3})} = \mc U_{CKM}$ and 
\begin{equation}
\mc U^{(-1)} =\left(
\begin{array}{ccc}
 1-\left(\frac{2}{9 A^2}+\frac{1}{18}\right) \lambda ^2 & \frac{\lambda }{3} & \frac{2 \lambda }{3 A} \\[0.5em]
 -\frac{\lambda }{3} & 1-\frac{\lambda ^2}{18} & A \lambda ^2 \\[0.5em]
 -\frac{2 \lambda }{3 A} & \left(-A-\frac{2}{9 A}\right) \lambda ^2 & 1-\frac{2 \lambda ^2}{9 A^2} \\
\end{array}
\right)+\mathcal{O}(\lambda^3). \label{ulep}
\end{equation}

The matrices in Eq.~\eqref{texture} yield the GUT-scale mass ratios of quarks and charged leptons and CKM mixing angles of quarks. The lepton mixing PMNS matrix is an overlap between $\mc U^{(-1)}$ and the TBM seesaw mixing with a single phase $\delta$,
\begin{align}
    \mc U_\text{PMNS} &= \mc U^{(-1)^\dagger}\ \mc U_\text{TBM} (\delta) \nonumber,\\
\text{where}, \qquad
    \mc U_\text{TBM} (\delta) &= \text{diag}(1,1,e^{i\delta}) \begin{pmatrix}
        \sqrt{\frac{2}{3}} & \frac{1}{\sqrt{3}} & 0 \\ 
        -\frac{1}{\sqrt{6}} &  \frac{1}{\sqrt{3}} &  \frac{1}{\sqrt{2}} \\
        \frac{1}{\sqrt{6}} & - \frac{1}{\sqrt{3}} &  \frac{1}{\sqrt{2}} 
    \end{pmatrix}. \label{cTBM}
\end{align}
\textcolor{black}{The phase $\delta$ is crucial to reproduce the experimentally observed PMNS angles. We note that its placement in Eq.~\eqref{cTBM} is minimal (only one phase is required) and it is unique in the sense that if it were placed in any other entry of the diagonal phase matrix and/or if the phase matrix were placed to the right of the real TBM matrix, the values of the PMNS angles would no longer be consistent with PDG data \cite{pdglive}.} 

The TBM phase $\delta$ generates both the Dirac $\cancel{CP}$ phase $\delta_{CP}$ and the Majorana phases in the PMNS matrix to be discussed later in the text. For $66^\circ \leq |\delta| \leq 85^\circ$ \cite{Rahat:2018sgs,Rahat:2020mio}, all the PMNS mixing angles are generated within $3\sigma$ of corresponding PDG values \cite{pdglive} and Dirac $CP$ phase $\delta_{CP}$ is predicted to be $1.27 \pi \leq |\delta_{CP}| \leq 1.35 \pi$ \cite{Rahat:2018sgs,Rahat:2020mio}, consistent with the PDG fit $\delta_{CP}^{PDG} = 1.36 \pm 0.17 \pi$ \cite{pdglive}. The sign of $\delta$ remains unresolved at this stage and a negative sign of $\delta$ corresponds to a positive sign of $\delta_{CP}$ in the above range. 

Seeking to motivate the asymmetry in the texture from a discrete family symmetry, the order $39$ subgroup $\mc T_{13} \equiv \mc Z_{13} \rtimes \mc Z_{3}$ \cite{bovier1981finite, bovier1981representations, fairbairn1982some, ding2011tri, hartmann2011neutrino, Hartmann:2011dn, ishimori2012introduction, Ramond:2020dgm} of $SU(3)$ appears to be the best candidate \cite{Perez:2019aqq}. This group has two different complex triplet representations, required to generate an asymmetric term naturally. In Refs.~\cite{Perez:2019aqq} and \cite{Perez:2020nqq}, the structure and key features of the Yukawas of Eq.~\eqref{texture} and the complex TBM mixing of Eq.~\eqref{cTBM} are explained by constructing an $SU(5) \times \mc T_{13}$ model augmented by a $Z_{12}$ `shaping' symmetry.\footnote{Ref.~\cite{Perez:2020nqq} also discusses a slightly different model with a $Z_{14}$ `shaping' symmetry, which will remain out of the scope of this paper. Also see Ref.~\cite{CentellesChulia:2019ldn} for a different realization of the TBM phase from residual and generalized $CP$ symmetries.} Introducing four right-handed neutrinos, the fourth required to resolve the discrepancy with oscillation data \cite{pdglive}, this model predicts normal ordering of the light neutrino masses through the seesaw mechanism.

The seesaw sector of the model which is relevant to our discussion is described by the following Lagrangian \cite{Perez:2020nqq}:
\begin{align}
    \mc L_{ss} &\supset  y_{\mc A} F \Lambda \bar{H}_{\g{5}} + y'_{\mc A} \bar{N} \overline{\Lambda} \varphi_{\mc A} + y_{\mc B} \bar{N} \bar{N} \varphi_{\mc B}   + M_\Lambda \overline{\Lambda} \Lambda + y_v' \bar{N}_4 \overline{\Lambda} \varphi_v + M \bar{N}_4 \bar{N}_4. \label{seesawlag}
\end{align}
The charged-leptons contained in the field $F$ couple to the right-handed neutrinos $\bar{N}$ and $\bar{N}_4$ through a heavy vector-like messenger $\Lambda$ and familons $\varphi_{\mc A}$, $\varphi_{\mc B}$, and $\varphi_v$. $y_X$ are dimensionless Yukawa couplings;  $M_\Lambda$ and $M$ are masses of $\Lambda$ and $\bar{N}_4$. We assume that the messenger $\Lambda$ is heavier than the family symmetry breaking scale and can be integrated out. The Lagrangian in Eq.~\eqref{seesawlag} then becomes
\begin{align}
    \mc L_{ss} \supset \frac{1}{M_\Lambda} y_{\mc A} y_{\mc A}' F \bar{N} \bar{H}_{\g{5}} \varphi_{\mc A} + \frac{1}{M_\Lambda} y_{\mc A} y_{v}' F \bar{N}_{4} \bar{H}_{\g{5}} \varphi_{v} + y_{\mc B} \bar{N} \bar{N} \varphi_{\mc B} + M \bar{N}_4 \bar{N}_4. \label{seesawlag2}
\end{align}
The transformation properties of the fields under $SU(5)$, $\mc T_{13}$ and $\mc Z_{12}$ symmetries  are given in Table \ref{table:model}.
\begin{table}[ht]\centering
\renewcommand\arraystretch{1.1}
\begin{tabularx}{\textwidth}{@{}l | Y Y Y Y Y Y Y Y @{}}
\hline \hline
     & $F$ & $\bar{N}$ & $\bar{N}_4$ & $\bar{H}_{\g{5}}$ & $\Lambda$  & $\varphi_{\mc A}$ & $\varphi_{\mc B}$ & $\varphi_v$ \\ 
\hline
$SU(5)$    & $\overline{ \g{5}}$ & $\g{1}$ & ${ \g{1}}$ & ${\g{5}}$ & $\g{1}$ & ${ \g{1}}$ & $\g{1}$ & $\g{1}$ \\ 
$\mc T_{13}$ &  $\g{3}_1$ & $\g{3}_2$ & $\g{1}$ & $\g{1}$ & $\gb{3}_1$  & $\gb{3}_2$ & $\g{3}_2$ & $\gb{3}_1$  \\ 
$\mathcal{Z}_{12}$    & $\g{\omega}$ & $\g{\omega^3}$ & $\g{1}$ & $\g{\omega^9}$ & $\g{\omega^2}$  & $\g{\omega^{11}}$ & $\g{\omega^6}$ & $\g{\omega^2}$   \\ 
\hline \hline
\end{tabularx} 
\caption{Transformation properties of matter, Higgs, messenger and familon fields in the seesaw sector. Here $\omega^{12} = 1$. The $\mc Z_{12}$ `shaping' symmetry prevents unwanted tree-level operators.}
\label{table:model}
\end{table}

The $\mc T_{13} \times \mc Z_{12}$ symmetry is spontaneously broken by the following chosen VEVs of the familons (for details see Ref. \cite{Perez:2020nqq}):
\begin{align}
\begin{aligned}
y_{\mc A} y_{\mc A}' \vev{\varphi_{\mc A}} &= \frac{M_{\Lambda}}{v} \sqrt{m_\nu b_1 b_2 b_3}\ (-b_2^{-1} e^{i\delta}, b_1^{-1}, b_3^{-1}),\\ 
y_{\mc B} \vev{\varphi_{\mc B}} &= (b_1, b_2, b_3), \\
y_{\mc A} y_{v}' \vev{\varphi_v} &= \frac{M_{\Lambda}}{v} \sqrt{Mm_v'}\ (2,-1,e^{i\delta}),
\end{aligned}    \label{VEV}
\end{align}
where $b_1, b_2, b_3, M \neq 0$ and $v=174\ \text{GeV}$ is the VEV of the Standard Model Higgs. Notice that in Eq. \eqref{VEV},  the VEV of the familon $\varphi_{\mc A}$ is related to the VEV of $\varphi_{\mc B}$. Alignment of these VEVs are assumed \cite{Perez:2020nqq} to ensure that the seesaw matrix is diagonalized by the complex TBM matrix, and the TBM phase $\delta$ which eventually generates both the Dirac and Majorana phases in the PMNS matrix arises from the vacuum structure of the familons. The last two terms of Eq.~\eqref{seesawlag2} give the following Majorana matrix:
\begin{align}
    \mathcal{M} &\equiv \left(
\begin{array}{cccc}
 0 & b_2 & b_3 & 0 \\
 b_2 & 0 & b_1 & 0 \\
 b_3 & b_1 & 0 & 0 \\
 0 & 0 & 0 & M \\
\end{array}
\right), \label{Mmat}
\end{align}
whereas the first two terms generate the operators $F \bar{N} \bar{H}_{\g{5}}$ and $F \bar{N}_4 \bar{H}_{\g{5}}$ that yield the following Yukawa matrix:
\begin{align}
    Y^{(0)} &\equiv \frac{\sqrt{b_1 b_2 b_3 m_\nu}}{v} \left(
\begin{array}{cccc}
 0 &  b_3^{-1} & 0 & 2 \sqrt{\frac{Mm_v'}{b_1 b_2 b_3 m_\nu}}  \\[0.5em]
  b_1^{-1} & 0 & 0 & - \sqrt{\frac{Mm_v'}{b_1 b_2 b_3 m_\nu}} \\[0.5em]
 0 & 0 & - e^{i \delta}b_2^{-1} &  e^{i \delta} \sqrt{\frac{Mm_v'}{b_1 b_2 b_3 m_\nu}} \\
\end{array}
\right). \label{Y0mat}
\end{align}

\noindent Correspondingly, the seesaw matrix is defined in terms of the Yukawa and Majorana matrices and its diagonalization with the complex TBM matrix of Eq.~\eqref{cTBM} yields the light neutrino masses: 
\begin{align}
    \mc S = Y^{(0)}\ \mc M^{-1}\ Y^{(0)^T} =\ \mc U_\text{TBM}(\delta)\ \text{diag}\ (m_{\nu_1}, m_{\nu_2}, m_{\nu_3})\ \mc U_\text{TBM}^T(\delta),
\end{align}
where \cite{Perez:2020nqq}
\begin{align}
    m_{\nu_1} = -m_\nu + 6m_v', \quad m_{\nu_2} = \frac{1}{2} m_{\nu}, \quad m_{\nu_3} = -m_{\nu}.\label{ambi}
\end{align}
We note that with only the three right-handed neutrinos $\bar{N}$, one gets $m_v'=0$ and $m_{\nu_1}$ is degenerate with $m_{\nu_3}$, in contradiction with the oscillation data \cite{pdglive}. Adding the fourth right-handed neutrino breaks this degeneracy. Using $|m_{\nu_2}| = \frac{1}{2} |m_{\nu_3}|$ together with oscillation data, we determine \cite{Perez:2020nqq}: 
\begin{align}
    |m_\nu| = 57.8\ \text{meV},  \quad |m_v'| = 5.03\ \text{or}\ 14.2\ \text{meV}, \label{mnu}
\end{align}
so that the light neutrino masses are given by \cite{Perez:2020nqq}: 
\begin{align}
    |m_{\nu_1}| = 27.6\ \text{meV},\quad |m_{\nu_2}| = 28.9\ \text{meV}, \quad |m_{\nu_3}| = 57.8\ \text{meV}, \label{m_v} 
\end{align}
nearly saturating the upper limit on their sum $\sum_i |m_{\nu_i}| < 120\ \text{meV}$ set by the Planck collaboration \cite{aghanim2018planck} (see also \cite{Vagnozzi:2017ovm}).

\textcolor{black}{Notice that this model can only accommodate normal mass ordering spectrum $|m_{\nu_1}| < |m_{\nu_2}| < |m_{\nu_3}|$ which implies both $m_\nu$ and $m_v'$ must have the same sign in Eq.~\eqref{mnu}. The ambiguity in the magnitude of $m_v'$ corresponds to two possible relative signs between $m_{\nu_1}$ and, $m_{\nu_2}$ and $m_{\nu_3}$ ($m_{\nu_2}$ and $m_{\nu_3}$ always have the opposite signs).}
Expressing the PMNS matrix in terms of the mixing angles $\theta_{ij}$, Dirac phase $\delta_{CP}$ and Majorana phases $\alpha_{21}$ and $\alpha_{31}$ in the PDG convention \cite{tanabashi2018review}:
\begin{equation*}\label{eq:mnsppdg}
 \mc U_\text{PMNS}\!=\!\!\left(\!\!\!
\begin{array}{ccc}
 c_{12} c_{13} & c_{13} s_{12} & e^{-i \delta_{CP} } s_{13} \\
 -c_{23} s_{12}-c_{12} s_{13} s_{23}e^{i \delta_{CP} }  & c_{12} c_{23}- s_{12} s_{13} s_{23}e^{i \delta_{CP} } & c_{13} s_{23} \\
 s_{12} s_{23}-c_{12} c_{23}  s_{13}e^{i \delta_{CP} } & -c_{12} s_{23}-c_{23}  s_{12} s_{13}e^{i \delta_{CP} } & c_{13} c_{23} \\
\end{array}
\!\!\!\right) \!\!\!
\left(\!\!
\begin{array}{ccc}
    1 & &  \\
     & e^{i\frac{\alpha_{21}}{2}} & \\
     & & e^{i\frac{\alpha_{31}}{2}} \\
\end{array}
\!\!\!\right)
\end{equation*}
where $s_{ij} \equiv \sin{\theta_{ij}}$ and $c_{ij} \equiv \cos{\theta_{ij}}$, the Jarlskog and Majorana invariants are given by 
\begin{equation}\label{eq:jarlskogpert}
\begin{aligned}
&c_{12} c_{13}^2 c_{23} s_{12} s_{13} s_{23} \sin {\delta_{CP} } = \frac{\lambda \sin {\delta }}{9 A}-\frac{\lambda ^2 \sin {\delta }}{27 A} + \mc O(\lambda^3),\\
&c_{12}^2 c_{13}^4 s_{12}^2 \sin {\alpha_{21}} = \frac{4 \lambda  \sin {\delta }}{9 A}-\frac{2 \lambda ^2 \sin {\delta }\ (A-2 \cos {\delta})}{27 A^2} + \mc O(\lambda^3), \\
&c_{12}^2 c_{13}^2 s_{13}^2 \sin {(\alpha_{31}-2 \delta_{CP}) } = \frac{4 \lambda ^2 \sin {\delta }\ (A+2 \cos {\delta })}{27 A^2} + \mc O(\lambda^3),
\end{aligned}
\end{equation}
and $\delta = \mp 78^\circ$ yields \cite{Perez:2020nqq}
\begin{equation}\label{eq:sinphases}
\begin{aligned}
\sin \delta_{CP} = \pm 0.854, \quad
\sin \alpha_{21} = \pm 0.515,\quad
\sin(\alpha_{31}-2\delta_{CP})= \pm 0.809.
\end{aligned}
\end{equation}
Moreover, the effective Majorana mass parameter in neutrinoless double-beta decay \cite{bilenky2015neutrinoless}:
\begin{align}
    | m_{\beta \beta}  | &= \left| c_{13}^2 c^2_{12} m_{\nu_1} + c_{13}^2 s^2_{12} e^{i\alpha_{21}} m_{\nu_2} + s_{13}^2 m_{\nu_3} e^{i(\alpha_{31}-2\delta_{CP})} \right|, \label{mbbac}
\end{align}
\reply{assuming the contribution of the right-handed neutrinos is negligible,}\footnote{\reply{In the end of Sec.~\ref{sec:results}, we will show that the contributions from the right-handed neutrinos with their masses determined by resonant leptogenesis can have considerable impact on this.}} is given by 
\begin{align} \label{mbbactive}
| m_{\beta \beta}  | &= 10.70\ \mathrm{ meV}\quad \text{or}\quad 27.12\ \text{meV},
\end{align}
depending on the two different values of $m_v'$ as mentioned above. Both of these values are below the upper limit $61$-$165\ \text{meV}$ set by the KamLAND-Zen experiment \cite{Dolinski:2019nrj,gando2016search}. Note that the sign ambiguity in $\delta$, and therefore in Eq.~\eqref{eq:sinphases}, has no implication on $|m_{\beta \beta}|$. The set of equations given in Eq. \eqref{eq:jarlskogpert} explicitly show how the Dirac phase as well as two Majorana phases are related to the only phase $\delta$ of the theory.

For concreteness, in this paper we will adopt $\delta = -78^\circ$ (which yields all PMNS angles close to their central PDG values \cite{pdglive}), $m_\nu = 57.8\ \text{meV}$ and $m_v' = 5.03\ \text{meV}$. This leaves four undetermined mass parameters $b_1$, $b_2$, $b_3$ and $M$. The first three are related to the scale of family symmetry breaking. Although $M$ is treated as an independent bare mass parameter, it could originate from the VEV of a singlet familon and thus be linked to the family symmetry breaking scale.
Note that the light neutrino masses and the lepton mixing angles are derived irrespective of the values of $b_1, b_2, b_3$ and $M$ \cite{Perez:2020nqq}. An important objective of the present work is to relate these unresolved parameters in the context of leptogenesis.




We now express the matrices relevant for leptogenesis in the so-called \emph{weak} basis, where the charged-lepton Yukawa matrix and the right-handed Majorana matrix are diagonal with real, positive entries \cite{Zhang:2015qia}. After spontaneously breaking the GUT and family symmetry to the Standard Model gauge group, the lepton sector of the Lagrangian contains the terms
\begin{align}
    \mc L &\supset \ell^\dagger Y^{(-1)} \bar{e}H + \ell^\dagger Y^{(0)}\bar{N}H^* + \bar{N}^T \mc M \bar{N} \nonumber \\
    &= \ell^\dagger \mc U^{(-1)} \mc D^{(-1)} \mc V^{(-1)\dagger} \bar{e}H + \ell^\dagger Y^{(0)}\bar{N}H^* + \bar{N}^T \mc U_m \mc D_m \mc U_m^T \bar{N}. \label{lagrangian}
\end{align}
In Eq.~\eqref{lagrangian} we have expressed the charged-lepton Yukawa matrix and the Majorana mass matrix in terms of unitary and diagonal matrices:
\begin{align}
    Y^{(-1)} = \mc U^{(-1)} \mc D^{(-1)} \mc V^{(-1)\dagger}, \qquad \mc M = \mc U_m \mc D_m \mc U_m^T. 
\end{align}
Applying the transformations $\ell \rightarrow \mc U^{(-1)}\ell$, $\bar{e} \rightarrow \mc V^{(-1)}\bar{e}$, and $\bar{N} \rightarrow \mc U_m^* \bar{N}$, they become
\begin{align}
    \mc L \supset \ell^\dagger \mc D^{(-1)} \bar{e}H + \ell^\dagger \mc U^{(-1)\dagger} Y^{(0)} \mc U_m^* \bar{N} H^* + \bar{N}^T \mc D_m \bar{N}.
\end{align}
The first and third terms contain the real and positive diagonal mass matrices of the charged leptons and the right-handed neutrinos, respectively. From the second term, we identify the light neutrino Yukawa matrix:
\begin{align}
    Y_\nu = \mc U^{(-1)\dagger} Y^{(0)} \mc U_m^*. \label{first}
\end{align}
%

\blu{Here we briefly revisit the result of Ref.~\cite{Rahat:2020mio} that the total $CP$ violation vanishes for unflavored leptogenesis when $T \gg 10^{12}\ \text{GeV}$ and leptogenesis has to proceed when lepton flavor effects are relevant, i.e., $T \lesssim 10^{12}\ \text{GeV}$. If $\mc M$ is real (all the familon VEVs are real), ${\mc U}_m$ is real and orthogonal up to a possible right diagonal matrix with some entries of $i$ (the eigenvalues must be real, but some can be negative and they can be made positive by multiplying ${\mc U}_m$ with right diagonal matrix with corresponding entries of $i$). Next, notice that $Y^{(0)\dagger} Y^{(0)}$ is real. Hence $Y_\nu^\dagger Y_\nu = {\mc U}_m^T Y^{(0)^\dagger} Y^{(0)} {\mc U}_m^*$ can only have off-diagonal terms which are purely real or imaginary. Since the total $CP$ violation in unflavored leptogenesis is proportional to ${\rm Im}[(Y_\nu^\dagger Y_\nu)_{ij}^2]$ \cite{Buchmuller:2004nz}, it is identically zero and unflavored leptogenesis fails. We conclude that, in this model, leptogenesis must proceed taking into account of the lepton flavor effects.} In the next section we briefly review the formalism of flavored leptogenesis in the resonant regime.


\section{Boltzmann Equations for Resonant Leptogenesis}\label{sec3}

\reply{Since we are focusing on the resonant leptogenesis scenario which occurs at relatively low temperature $T < 10^4$ $\text{GeV}$ where the interactions mediated by all the charged lepton Yukawa couplings are in thermal equilibrium, we will consider the Boltzmann equations in the three-flavor regime} \cite{Barbieri:1999ma, Abada:2006fw, Blanchet:2006be, Vives:2005ra, Nardi:2006fx, Abada:2006ea, Dev:2017trv, Samanta:2020tcl, Samanta:2019yeg}\footnote{\reply{Here order of one spectator effects \cite{Buchmuller:2001sr} have been neglected.}}
\begin{align}
    \frac{dN_{N_i}}{dz} &= - D_i (N_{N_i} - N_{N_i}^{\rm eq}), \quad i = 1, 2, 3, 4,\label{eq:Nni1}\\
    \frac{dN_{\Delta\alpha}}{dz} &= -\sum_{i} \varepsilon_{i\alpha} D_i (N_{N_i} - N_{N_i}^{\rm eq}) - N_{\Delta \alpha} \sum_i P_{i\alpha}  W_i,\quad \alpha = e, \mu, \tau, \label{eq:NBLflav}
\end{align}
where $z = M_{min}/T$ and $M_{min} = \min{(M_i)}$. $N_{N_i}$ is the number density of the Majorana neutrino $N_i$ and $N_{\Delta \alpha}$ is the $B/3-L_\alpha$ asymmetry, both normalized by the photon density. Introducing the notation $x_i \equiv M_i^2 / M_{min}^2$ and $z_i \equiv z\sqrt{x_i}$,
the \emph{equilibrium number density} can be expressed in terms of the modified Bessel functions of the second kind: 
\begin{align}
    N_{N_i}^{\rm eq} (z_i) &= \frac{3}{8} z_i^2 \mc K_2(z_i),
\end{align}
The \emph{decay factor} $D_i$ and the \emph{washout term} $W_i$ are respectively given by
\begin{align}
    D_i \equiv K_i x_i z \frac{\mc K_1(z_i)}{\mc K_2(z_i)} \label{eq:Di1},\;\;\; \textrm{and}\;\;\; W_i \equiv \frac{1}{4} K_i \sqrt{x_i} \mc K_1(z_i) z_i^3, 
\end{align}
where we have defined the \emph{decay parameter} $K_i \equiv \frac{\Gamma_i }{H(z_i = 1)} = \frac{(Y^\dagger_\nu Y_\nu)_{ii} v^2}{M_i m_*}$ with $\Gamma_i$, the \emph{total decay width} of $N_i$, given by
\begin{equation}
\Gamma_i = \frac{(Y_\nu^\dagger Y_\nu)_{ii} M_i}{8\pi}, \label{Gammai}
\end{equation}
and the \emph{equilibrium neutrino mass} $m_* \simeq 1.08\ \text{meV}$.
The \emph{branching ratio} for $N_i$ decaying into $\ell_\alpha$ is given by
\begin{align} \label{bratio}
    P_{i\alpha} = \frac{|(Y_{\nu})_{\alpha i}|^2}{\sum_\gamma |(Y_{\nu})_{\gamma i}|^2}.
\end{align}

For resonant leptogenesis, we consider the $CP$ asymmetry parameter from mixing and oscillation among the right-handed neutrinos \cite{Dev:2014laa,Bambhaniya:2016rbb,Brivio:2019hrj}
\begin{align} \label{epsires}
    \varepsilon_{i\alpha} &= \sum_{j\neq i}  \frac{\text{Im} [ (Y_\nu^*)_{\alpha i} (Y_\nu)_{\alpha j} (Y_\nu^\dagger Y_\nu)_{ij} ] + \frac{M_i}{M_j} \text{Im} [ (Y_\nu^*)_{\alpha i} (Y_\nu)_{\alpha j} (Y_\nu^\dagger Y_\nu)_{ji} ]}{(Y_\nu^\dagger Y_\nu)_{ii} (Y_\nu^\dagger Y_\nu)_{jj}} (f_{ij}^{\text{mix}} + f_{ij}^{\text{osc}}),
\end{align}
where \textcolor{black}{the self-energy regulators are given by} 
\begin{align}
    f_{ij}^{\text{mix}} &= \frac{(M_i^2 - M_j^2) M_i \Gamma_j}{(M_i^2 - M_j^2)^2 + M_i^2 \Gamma_j^2},  \label{fmix} \\
    f_{ij}^{\text{osc}} &= \frac{(M_i^2 - M_j^2) M_i \Gamma_j}{(M_i^2 - M_j^2)^2 + (M_i \Gamma_i +  M_j \Gamma_j)^2 \frac{\det{[\text{Re}([Y_\nu^\dagger Y_\nu]_{ij})]}}{(Y_\nu^\dagger Y_\nu)_{ii} (Y_\nu^\dagger Y_\nu)_{jj}}}, \label{fosc}
\end{align}
and $[Y_\nu^\dagger Y_\nu]_{ij}$ in the denominator of Eq.~\eqref{fosc} is the $2\times 2$ submatrix 
\begin{align}
[Y_\nu^\dagger Y_\nu]_{ij} \equiv \left(
\begin{array}{cc}
 (Y_\nu^\dagger Y_\nu)_{ii}  & (Y_\nu^\dagger Y_\nu)_{ij}  \\
 (Y_\nu^\dagger Y_\nu)_{ji}  & (Y_\nu^\dagger Y_\nu)_{jj}  \\
\end{array}
\right).
\end{align}
The $CP$ asymmetry can be resonantly enhanced when at least two of the right-handed neutrino masses are nearly degenerate and their mass difference is of the order of their average decay width \cite{Pilaftsis:1997dr}. \blu{In this paper we will employ the condition} 
\begin{align}
    |M_i - M_j| = \frac{1}{2} \left(\frac{\Gamma_i + \Gamma_j}{2}\right), \label{rescondgen}
\end{align}
for resonant leptogenesis. Approximating $M_i + M_j \simeq 2M_{i,j}$, the \textcolor{black}{regulators} 
can be expressed as
\begin{align}
    f_{ij}^{\rm mix} &\approx \frac{2 \Gamma_j (\Gamma_i + \Gamma_j)}{(\Gamma_i+\Gamma_j)^2 + 4\Gamma_j^2}, \label{fmixres}\\
    f_{ij}^{\rm osc} &\approx \frac{2\Gamma_j}{(\Gamma_i + \Gamma_j)\left(1+4\frac{\text{det}[\text{Re}([Y^\dagger_\nu Y_\nu]_{ij})]}{(Y^\dagger_\nu Y_\nu)_{ii}(Y^\dagger_\nu Y_\nu)_{jj}}\right)} \label{foscres}.
\end{align}
For $\Gamma_i \simeq \Gamma_j$, Eq.~\eqref{fmixres} attains the maximum value $f_{ij}^{\rm mix} = 1/2$. 

\reply{It is important to note that we have included the contribution from the oscillation among the heavy neutrinos in Eq.~\eqref{foscres} which is relevant when the mass splitting is small $|M_i - M_j| \ll M_i$ as the oscillations cannot be averaged over \cite{Garny:2011hg, Garbrecht:2011aw, Garbrecht:2018mrp}. The approximate expression in Eq.~\eqref{foscres} is derived in Ref.~\cite{Dev:2014laa} starting from the density matrix formalism and is shown to be accurate for $|M_i - M_j| \ll M_i$ and when the decay rates of $N_i$ are faster than the Hubble rate (see the discussion in the next section about Fig.~\ref{fig:Ki}).
When $M \lesssim T_{sph}$, since the relevant asymmetry is produced at $T > M \gtrsim T_{sph}$, relativistic effects \cite{Biondini:2017rpb,Shaposhnikov:2008pf, Ghiglieri:2017gjz, Eijima:2017anv, Antusch:2017pkq, Bodeker:2019rvr,  Eijima:2018qke, Abada:2018oly, Klaric:2020lov, Klaric:2021cpi} should be taken into account and we use the density matrix equations as collected in the App. \ref{app:density_matrix}.
}

Solving the system of equations, \eqref{eq:Nni1} with the initial condition (i) ``zero initial abundance'': $N_{N_i}(z=0) = 0$ or (ii) ``thermal initial abundance'': $N_{N_i} (z=0) = N_{N_i}^{\rm eq}(z=0)$,\footnote{\reply{Heavy neutrinos are gauge singlets in this model and so the most appropriate initial condition is the zero initial abundance where they will be populated only from the neutrino Yukawa interactions. We also consider the case for thermal initial abundance to take into account the possibility that they could be in thermal equilibrium due to additional interactions beyond the current model (new gauge interactions, interactions with inflaton etc.).}} and \eqref{eq:NBLflav} with the initial condition $N_{\Delta \alpha} (z=0) = 0$, the final value of the $B-L$ asymmetry is evaluated at $T_f = T_{sph} \simeq 131.7$ $\text{GeV}$ \cite{DOnofrio:2014rug} 
when electroweak sphaleron processes freeze out: 
\begin{align}
N_{B-L}^f = \sum_\alpha N_{\Delta \alpha} (z_f = M_{min}/T_f), \label{NBLf}
\end{align}
and is related to the baryon asymmetry by
\begin{align}
    \eta_B = \frac{a_{sph}}{f_d} N_{B-L}^f \simeq 1.28 \times 10^{-2} N_{B-L}^f, \label{formulaB}
\end{align}
where the sphaleron conversion coefficient is $a_{sph} = 28/79$ \cite{Kuzmin:1985mm, Khlebnikov:1988sr, Harvey:1990qw} and the dilution factor is $f_d \equiv N^{rec}_\gamma/N_{\gamma}^* = 2387/86$, calculated assuming photon production from the beginning of leptogenesis to recombination \cite{Buchmuller:2004nz}. Successful leptogenesis requires $\eta_B$ to match the measured value from Cosmic Microwave Background (CMB) data \cite{Akrami:2018vks}:
\begin{align}
    \eta_B^{CMB} &= (6.12 \pm 0.04) \times 10^{-10}. \label{measuredB}
\end{align}

\section{Resonant Leptogenesis in the $SU(5) \times \mc T_{13}$ Model} \label{sec:resonantasym}

In this section we discuss resonant leptogenesis in the context of the $SU(5) \times \mc T_{13}$ model. We  analyze the mass spectrum of the right-handed neutrinos for a simple choice of VEV of the seesaw familons and identify a unique case of quasi-degeneracy relevant for resonant leptogenesis. 


In the $SU(5) \times \mc T_{13}$ model there are four undetermined parameters from the VEV of the familon $\vev{\varphi_{\mc B}} \equiv (b_1, b_2, b_3)$ and the bare mass of the fourth right-handed neutrino $M$. For simplicity as well as minimality (this choice is minimal with respect to the number of undetermined parameters introduced in the theory), we choose the following VEV structure\footnote{The simplest choice $(b_1, b_2, b_3) \equiv b(1,1,1)$ does not generate resonant enhancement to $CP$ asymmetry, as we will discuss later in this section. In App.~\ref{app:VEV} we discuss two other variants of the VEV structure: $(b_1, b_2, b_3) \equiv b\ (f,1,1)$ and $(b_1, b_2, b_3) \equiv b\ (1,1,f)$, and argue that they yield qualitatively similar phenomenology.}
\begin{align}
    (b_1, b_2, b_3) \equiv b\ (1,f,1)
\end{align}
for the remainder of the paper. Here $f$ is a dimensionless unknown parameter. We further define
\begin{align}
    a \equiv \frac{M}{b},
\end{align}
so that the undetermined parameters of the model are $a, b$, and $f$. 
This results in the following Yukawa and Majorana matrices:
\begin{align}
    Y^{(0)} &= \frac{\sqrt{b f m_\nu}}{v} \left(
\begin{array}{cccc}
 0 & 1 & 0 & 2\beta   \\[-0.1em]
 1 & 0 & 0 & -\beta  \\[-0.1em]
 0 & 0 & -f^{-1} e^{i \delta} & \beta e^{i \delta}  \\
\end{array}
\right), \label{Y0}
\end{align}
where $\beta \equiv \sqrt{\frac{a m_v'}{f m_\nu }}$, and 
\begin{align}
    \mc M &=  b\left(
\begin{array}{cccc}
 0 & f & 1 & 0 \\[-0.1em]
 f & 0 & 1 & 0 \\[-0.1em]
 1 & 1 & 0 & 0 \\[-0.1em]
 0 & 0 & 0 & a \\
\end{array}
\right). \label{Majo}
\end{align}
Since $b$ is given by the Yukawa coupling $y_{\mc B}$ and the scale of the family symmetry breaking $\vev{\varphi_{\mc B}}$, cf. Eq.~\eqref{VEV}, a small value of $b$ can be attributed to a small value of $y_{\mc B}$ (a single Yukawa coupling) rather than a small family symmetry breaking scale. The choice of small $b$, as required to realize low-scale leptogenesis, involves somewhat fine-tuning, which we accept. Note however that fine-tuning is inherently present in GUT models to successfully implement doublet-triplet mass splitting to build a realistic model. $M$ on the other hand is a bare mass parameter, which a priori,  cannot be determined.

$\mc M$ is a symmetric matrix and its Takagi factorization \cite{horn2012matrix} 
$
    \mc M = \mc U_m\ \mc D_m\ \mc U_m^T,
$
where $\mc D_m \equiv \text{diag}(M_1, M_2, M_3, M_4)$, yields
\begin{align}
\begin{split}
    M_1 &= bf, \quad M_2 = \frac{b}{2} \left(\sqrt{f^2+8}-f\right),\quad
    M_3 = \frac{b}{2} \left(\sqrt{f^2+8}+f\right),\ \quad M_4 = a b, \label{Mmass}
\end{split}
\end{align}
\begin{align}
    \text{and}\qquad \mc U_m = \left(
\begin{array}{cccc}
 -\frac{i}{\sqrt{2}} & \frac{-i}{2} \sqrt{1-\frac{f}{\sqrt{f^2+8}}} & \frac{1}{2} \sqrt{1+\frac{f}{\sqrt{f^2+8}}} & 0 \\ [1.0em]
 \frac{i}{\sqrt{2}} & \frac{-i}{2}  \sqrt{1-\frac{f}{\sqrt{f^2+8}}} & \frac{1}{2} \sqrt{1+\frac{f}{\sqrt{f^2+8}}} & 0 \\[1.0em]
 0 & \frac{i}{\sqrt{2}} \sqrt{1+\frac{f}{\sqrt{f^2+8}}} & \frac{1}{\sqrt{2}} \sqrt{1-\frac{f}{\sqrt{f^2+8}}} & 0 \\[1em]
 0 & 0 & 0 & 1 \\
\end{array}
\right). \label{Unitm}
\end{align}

The $CP$ asymmetry parameter is determined by the imaginary parts of $ (Y_\nu^*)_{\alpha i} (Y_\nu)_{\alpha j} (Y_\nu^\dagger Y_\nu)_{ij} $ and $ (Y_\nu^*)_{\alpha i} (Y_\nu)_{\alpha j} (Y_\nu^\dagger Y_\nu)_{ji} $, cf.~Eq.~\eqref{epsires}. Explicitly calculating, we get the Hermitian matrix
\begin{align} \label{YY}
    &Y_\nu^\dagger Y_\nu = \frac{bfm_\nu}{v^2} \nonumber \\ 
    &\times \left(\!\!
\begin{array}{cccc}
 1 & 0 & 0 & \frac{3 i \beta }{\sqrt{2}} \\
 * & \frac{1}{2}\! \left(1\!-\!\frac{f^3-f-\sqrt{f^2+8}}{f^2 \sqrt{f^2+8}}\!\right) & -\frac{i \sqrt{2} \left(f^2-1\right)}{f^2 \sqrt{f^2+8}} & \!\!-\frac{i \beta}{2 f}  \left(f \sqrt{1-\frac{f}{\sqrt{f^2+8}}} \!+\! \sqrt{2\!+\!\frac{2 f}{\sqrt{f^2+8}}}\!\right) \\
 * & * & \frac{1}{2} \left(1\!+\!\frac{f^3-f+\sqrt{f^2+8}}{f^2 \sqrt{f^2+8}}\!\right) & \frac{\beta}{2 f}\!  \left(\!f \sqrt{1\!+\!\frac{f}{\sqrt{f^2+8}}}-\sqrt{2\!-\!\frac{2 f}{\sqrt{f^2+8}}}\!\right) \\
 * & * & * & 6 \beta ^2 \\
\end{array}
\!\!\!\!\right),
\end{align}
where $*$ denotes the complex conjugate of the corresponding transposed element. 

\begin{figure}[!ht] 
    \centering
       \includegraphics[width=0.48\textwidth]{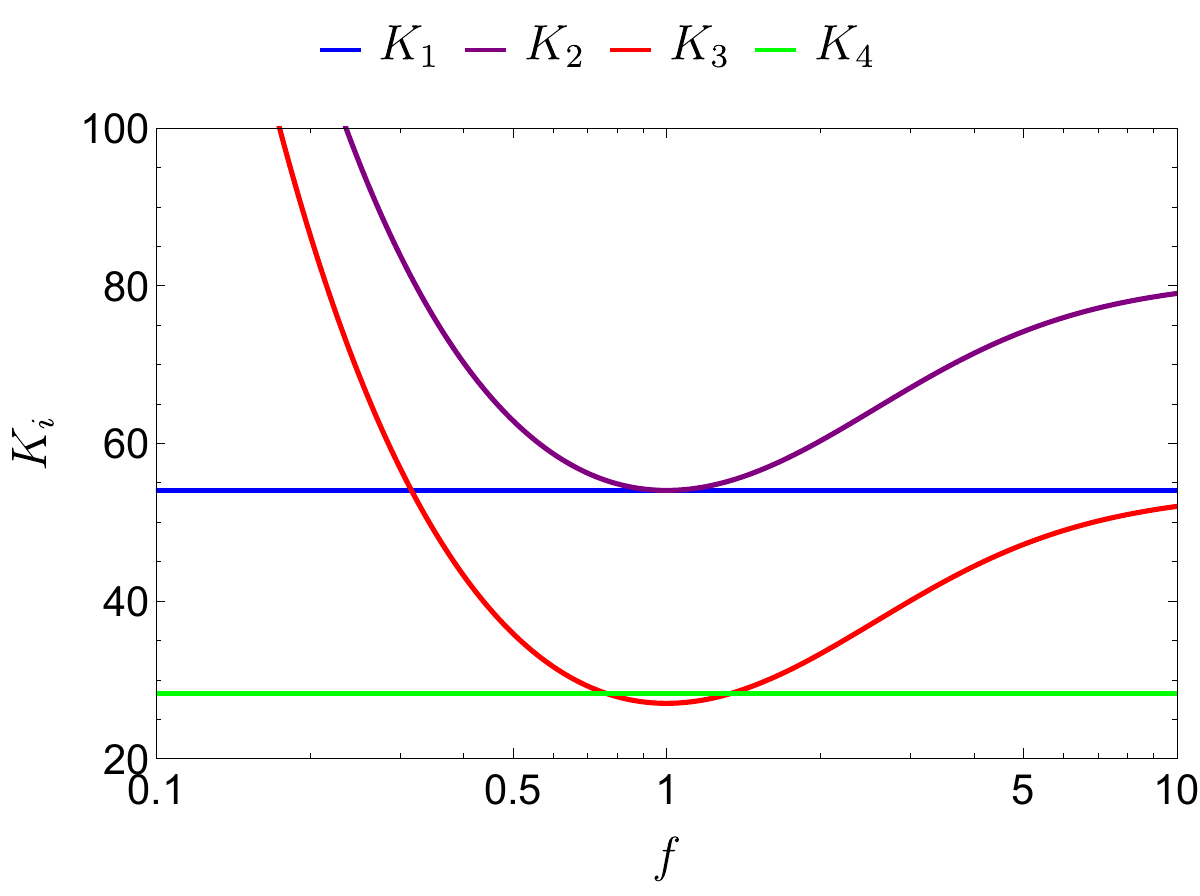}
    \caption{\reply{Decay parameters are large implying $N_i$ always achieve thermal equilibrium before decaying (see text for detailed discussion).}}
    \label{fig:Ki}
\end{figure}

\reply{In this model, the decay parameters $K_i$ are given by
\begin{align}
\begin{split}
    K_1 &= \frac{m_\nu}{ m_*},\quad\ \ K_2 = \frac{m_\nu}{m_*} \frac{f \left(1-f^2\right)+\left(f^2+1\right) \sqrt{f^2+8}}{f \sqrt{f^2+8} \left(\sqrt{f^2+8}-f\right)}, \\
    K_4 &= \frac{6m_v'}{m_*}, \quad K_3 = \frac{m_\nu}{m_*} \frac{f \left(f^2-1\right)+\left(f^2+1\right) \sqrt{f^2+8}}{f \sqrt{f^2+8} \left(\sqrt{f^2+8}+f\right)}.
\end{split}
\end{align}
%
In Fig.~\ref{fig:Ki}, we see that in all relevant parameter space, $K_i \gg 1$ indicating that in this model, $N_i$ always achieve thermal equilibrium before decaying at $T < M_i$. For the case $M \gg T_{sph}$, the final asymmetry is determined at the moment $T \ll M_i$ when the inverse decay is sufficiently Boltzmann suppressed. Hence in this regime, the asymmetry generated is not sensitive to the physics at $T > M_i$ and can be described accurately with the classical Boltzmann equations Eqs. \eqref{eq:Nni1} and \eqref{eq:NBLflav}.\footnote{Ref.~\cite{Salvio:2011sf} showed that altogether thermal corrections and scatterings give rise to a less than 10\% effect in the strong washout regime.}
For the case $M \lesssim T_{sph}$, we use the density matrix equations in App. \ref{app:density_matrix} taking into account the relativistic effects \cite{Biondini:2017rpb,Shaposhnikov:2008pf, Ghiglieri:2017gjz, Eijima:2017anv, Antusch:2017pkq, Bodeker:2019rvr,  Eijima:2018qke, Abada:2018oly, Klaric:2020lov, Klaric:2021cpi}. 
}

Let us first focus on the case with the symmetric VEV $\vev{\varphi_{\mc B}} \equiv b(1,1,1)$ setting $f=1$. In this  case the only nonzero off-diagonal entries are $(Y_\nu^\dagger Y_\nu)_{14}$ and $(Y_\nu^\dagger Y_\nu)_{24}$. Hence $CP$ violation can only arise from the interference between $N_1$ and $N_4$, and $N_2$ and $N_4$, respectively. For all other cases, the $CP$ asymmetry vanishes identically, for any flavor. Since both $(Y_\nu^\dagger Y_\nu)_{14}$ and $(Y_\nu^\dagger Y_\nu)_{24}$ are purely imaginary, the $CP$ violation from $N_4$ decay is proportional to the real part of $(Y_\nu^*)_{\alpha 4}(Y_\nu^*)_{\alpha 1}$ and $(Y_\nu^*)_{\alpha 4}(Y_\nu^*)_{\alpha 2}$ which are equal in magnitude and opposite in sign (see App. B of Ref.~\cite{Rahat:2020mio} for details) and as a result, $CP$ violation for each flavor vanishes identically. Due to the same reason, for the decays of the degenerate pairs $N_1$ and $N_2$, their $CP$ asymmetry parameters are also equal in magnitude and opposite in sign and cancel exactly when one considers both their contributions.

Since the simplest case with $f=1$ fails to yield successful leptogenesis, we now move to the more general scenario $f \neq 1$. The mass spectrum of the right-handed neutrinos in Eq.~\eqref{Mmass} show that there can be six cases in general for different values of $f$ and $a$, when at least two of the masses are quasi-degenerate:
\begin{enumerate}[(i)]
    \item $M_1 \simeq M_2$: for $f \simeq 1$ and any value of $a$,
    \item $M_1 \simeq M_3$: \textcolor{black}{for $f \gg 2\sqrt{2}$ and any value of $a$ },
    \item $M_2 \simeq M_3$: for $f \simeq 0$ and any value of $a$,
    \item $M_1 \simeq M_4$: for $f \simeq a$,
    \item $M_2 \simeq M_4$: for $a \simeq \frac{1}{2} (\sqrt{f^2 + 8}-f)$,
    \item $M_3 \simeq M_4$: for $a \simeq \frac{1}{2} (\sqrt{f^2 + 8}+f)$.
\end{enumerate}
\textcolor{black}{The first three cases specify $f$ only while $a$ remains unconstrained, whereas the last three cases relate $f$ with $a$. We will argue below that $CP$ asymmetry is not necessarily enhanced for all of the above cases and it depends on the structure of the neutrino Yukawa matrix $Y_\nu$ dictated by the $\mc T_{13}$ family symmetry.
}

A qualitative understanding of the above six cases can be achieved from analyzing the structure of the matrix in Eq.~\eqref{YY} in the context of the $CP$ asymmetry parameter.
Introducing the notation $(Y_\nu^*)_{\alpha i} (Y_\nu)_{\alpha j} \equiv p + iq$ and $(Y_\nu^\dagger Y_\nu)_{ij} \equiv r+is$, which implies $(Y_\nu^\dagger Y_\nu)_{ji} \equiv r-is$, 
the numerator of Eq.~\eqref{epsires} can be written as
\begin{align}
    \text{Im} [ (Y_\nu^*)_{\alpha i} (Y_\nu)_{\alpha j} (Y_\nu^\dagger Y_\nu)_{ij} ] + \text{Im} [ (Y_\nu^*)_{\alpha i} (Y_\nu)_{\alpha j} (Y_\nu^\dagger Y_\nu)_{ji} ]
    = qr\left( 1+\frac{M_i}{M_j}  \right) + ps \left(1-\frac{M_i}{M_j}\right).
\end{align}
In Eq.~\eqref{YY}, the off-diagonal elements are either real ($s = 0$) or imaginary ($r = 0$). In the latter case, the numerator of the $CP$ asymmetry is proportional to
\begin{align}
    ps\left( 1-\frac{M_i}{M_j} \right) &= ps \frac{(Y^\dagger_\nu Y_\nu)_{ii}+(Y^\dagger_\nu Y_\nu)_{jj}}{32 \pi M_j}
\end{align}
after applying the resonance condition in Eq.~\eqref{rescondgen}. Even for $p, s \lesssim \mc O(1)$, the other factor is suppressed by $\mc O \left( m_\nu/v^2 \right) \sim \mc O(10^{-17})$; hence the $CP$ asymmetry cannot account for the observed baryon asymmetry. From Eq.~\eqref{YY}, this situation arises for \textcolor{black}{(i) $M_1 \simeq M_2$, (ii) $M_1 \simeq M_3$,} (iv) $M_1 \simeq M_4$ and (v) $M_2 \simeq M_4$, and these cases can be ruled out. 

\blu{The case for (iii) $M_2 \simeq M_3$, occurring for $f \simeq 0$, is more subtle. In this case $r=0$, hence the $CP$ asymmetry is suppressed by $\mc O(m_\nu / v^2)$. However, if the $CP$ asymmetry were dependent on $1/f^n$ for $n> 0$, this suppression could be overcome by choosing an appropriately small $f$. }

\blu{We discuss the case $i=2$, $j=3$ in Eq.~\eqref{epsires} for $f\rightarrow 0$ ($i=3$, $j=2$ would yield similar conclusion). In this limit, the terms in the denominator of the $CP$ asymmetry, i.e., $(Y_\nu^\dagger Y_\nu)_{22}$ and $(Y_\nu^\dagger Y_\nu)_{33}$, both depend on $1/f$, cf. Eq.~\eqref{YY}:}
\begin{align}
    (Y^\dagger_\nu Y_\nu)_{22} \simeq (Y^\dagger_\nu Y_\nu)_{33} \rightarrow \frac{b m_\nu}{2 f v^2},
\end{align}
thus the denominator has an $1/f^2$ dependence. Since $r=0$ for $(Y_\nu^\dagger Y_\nu)_{23}$, the numerator of the $CP$ asymmetry is given by 
\begin{align*}
\text{Im} [ (Y_\nu^*)_{\alpha 2} (Y_\nu)_{\alpha 3} (Y_\nu^\dagger Y_\nu)_{23} ] + \text{Im} [ (Y_\nu^*)_{\alpha 2} (Y_\nu)_{\alpha 3} (Y_\nu^\dagger Y_\nu)_{32} ] = ps \frac{\left( M_2 - M_3 \right)}{ M_2},
\end{align*}
where 
\begin{align}
    s \equiv \text{Im} \left[(Y^\dagger_\nu Y_\nu)_{23}\right] &\rightarrow \frac{b m_\nu}{2 f v^2}, \qquad
    \text{and }\qquad \frac{|M_2 - M_3|}{M_2} \rightarrow\frac{f}{\sqrt{2}}. \label{delM}
\end{align}
To see how $p \equiv \text{Re}\left[ (Y_\nu)^*_{\alpha 2} (Y_\nu)_{\alpha 3} \right]$ depends on $f$, following Eq.~\eqref{first} we  write
\begin{align}
    (Y_\nu)^*_{\alpha 2} (Y_\nu)_{\alpha 3} &= \sum_{\beta, \gamma, k, l} {\mc U^{(-1)}}^T_{\alpha \beta}  {\mc U^{(-1)}}^\dagger_{\alpha \gamma} {Y^{(0)}}^*_{\beta k} {Y^{(0)}}_{\gamma l} (\mc U_m)_{k2} (\mc U_m^*)_{l3}.
\end{align}
\blu{From Eq.~\eqref{Unitm}, $(\mc U_m)_{k2} (\mc U_m^*)_{l3}$ is zero when either $k$ or $l$ is $4$, and is imaginary otherwise. For $f \rightarrow 0$, the nonzero elements are independent of $f$. Since $\mc U^{(-1)}$ is real, cf. Eq.~\eqref{ulep}, we extract $p$ from the imaginary part of ${Y^{(0)}}^*_{\beta k} {Y^{(0)}}_{\gamma l}$. Since $k$ and $l$ cannot be $4$, ${Y^{(0)}}^*_{\beta k} {Y^{(0)}}_{\gamma l}$ can have a nonzero imaginary part only when the $(33)$ element of $Y^{(0)}$ (or its complex conjugate) is multiplied with the $(12)$ or $(21)$ element, cf. Eq.~\eqref{Y0}. In either case, the $f$ dependence gets canceled:}
\begin{align*}
    \text{Im}\left[ {Y^{(0)}}^*_{33} {Y^{(0)}}_{12} \right] = \text{Im}\left[ {Y^{(0)}}^*_{33} {Y^{(0)}}_{21} \right] = \frac{b m_\nu \sin{\delta}}{v^2}.
\end{align*}
Hence $p$ is independent of $f$ for $f \rightarrow 0$. Combining this with Eq.~\eqref{delM}, the numerator is independent of $f$. 

The $CP$ asymmetry is proportional to $f^2$, due to the $1/f^2$ dependence of the denominator, and is suppressed as $f \rightarrow 0$. Therefore the case (iii) $M_2 \simeq M_3$ also fails to yield successful leptogenesis.

\begin{figure}[!ht] 
    \centering
       \includegraphics[width=0.48\textwidth]{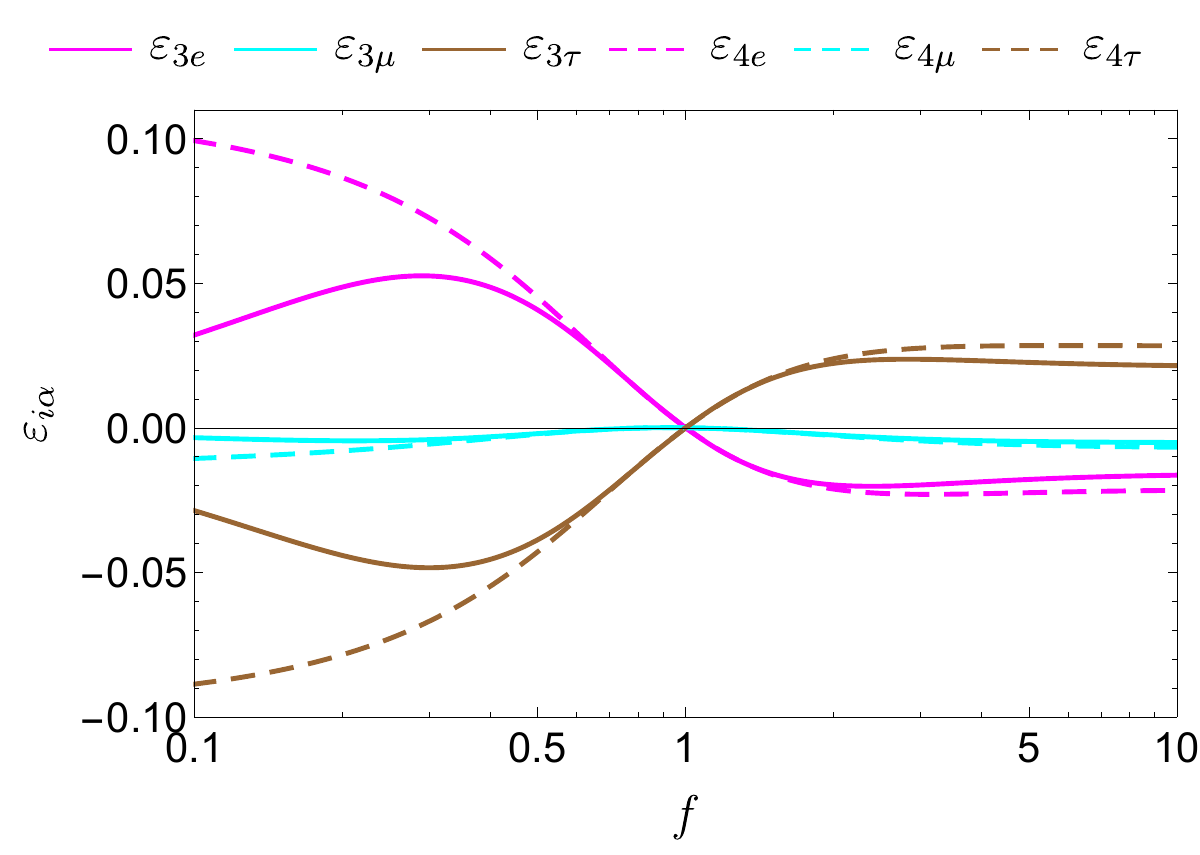}
    \caption{$CP$ asymmetry parameters at the resonance $M_3 \simeq M_4$. The sum of the flavored $CP$ asymmetries is zero, $\sum_\alpha \varepsilon_{i\alpha} = 0$, and hence unflavored leptogenesis is not successful in this model \cite{Rahat:2020mio}. An extreme example of this occurs at $f=1$, where the individual flavor components vanish, since $(Y_\nu^\dagger Y_\nu)_{34} = 0$.
    }
    \label{fig:CPasymmetry34}
\end{figure}

\blu{For the remaining case (vi) $M_3 \simeq M_4$, $(Y^\dagger_\nu Y_\nu)_{34}$ is real and the numerator of the $CP$ asymmetry is given by $qr(1+M_3/M_4)$. In this case the $CP$ asymmetry can be quite large, as shown in Fig.~\ref{fig:CPasymmetry34}. }
%

\blu{We will concentrate on the only viable $N_3$ - $N_4$ resonant leptogenesis scenario in the rest of paper. The resonance condition in Eq.~\eqref{rescondgen} translates into $|M_3 - M_4| = (\Gamma_3 + \Gamma_4)/4$, from which we can express the parameter $a$ in terms of $b$ and $f$. This reduces the number of undetermined parameters to two.} 
\blu{We will treat $b$ and $f$ as input parameters, assuming that $a$ can be determined from them applying the resonance condition.} \reply{It would be interesting to enhance the symmetry of the model that distinguishes the $N_3$ - $N_4$ quasi-degeneracy, i.e. the resonance condition. However, in this paper, we are interested in the phenomenological study of the viable parameter space of the model consistent with resonant leptogenesis and remain agnostic about symmetry reasons that can potentially explain the required degeneracy.}

\blu{The mass spectrum of the right-handed neutrinos up to the overall factor $b$ at resonance is shown in Fig.~\ref{fig:Mmassres} as a function of $f$.
\begin{figure}[!ht] 
    \centering
       \includegraphics[width=0.48\textwidth]{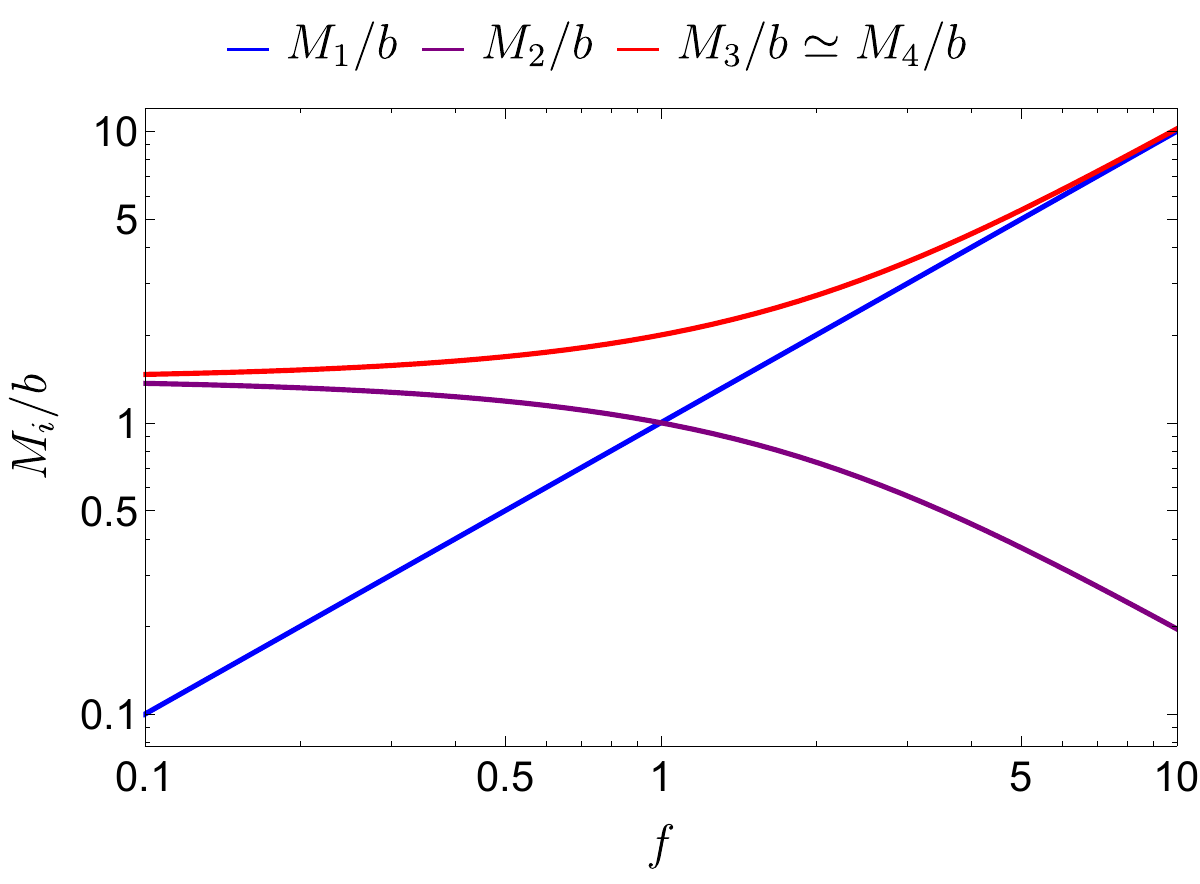}
    \caption{Mass spectrum of the right-handed neutrinos up to the overall factor $b$ at resonance $M_3 \simeq M_4$. }
    \label{fig:Mmassres}
\end{figure}
}
\blu{Other than the $M_3 \simeq M_4$ degeneracy, two other degeneracies are approached for $f \ll 1$ and $f \gg 1$. At $f \ll 1$, the mass spectrum can be approximated as
\begin{align}
     M_1 = bf, \qquad M_2 \simeq \sqrt{2}b, \qquad M_3 \simeq M_4 \simeq \sqrt{2}b, \label{masssmall}
\end{align}
In this regime $M_2 \simeq M_3 \simeq M_4$. On the other hand, for $f \gg 1$, we can express the masses as
\begin{align}
    M_1 = bf, \qquad M_2 \simeq \frac{2b}{f}, \qquad M_3 \simeq M_4 \simeq bf. \label{masslarge} 
\end{align}
and observe that $M_1 \simeq M_3 \simeq M_4$.}

\blu{Although the masses are directly proportional to $b$, the $CP$ asymmetry parameters at resonance do not have an explicit dependence on $b$. For the dominant terms $\varepsilon_{3\alpha}$ and $\varepsilon_{4\alpha}$, cf. Eq.~\eqref{epsires}, the prefactors of $\sqrt{b}$ in $Y_\nu$ in the numerator and denominator cancel out. $b$ dependence also drops out from the \textcolor{black}{regulators}
$f_{34}^{\rm mix}$ and $f_{34}^{\rm osc}$ at resonance, as can be seen from Eqs.~\eqref{fmixres} and \eqref{foscres}. The decay width $\Gamma_i$ are proportional to $b^2$, but this dependence vanishes between the denominator and numerator in these expressions.}  


In the next section, we determine the range of the right-handed neutrino masses required for successful leptogenesis at the resonance $M_3 \simeq M_4$, and discuss the mixing of the right-handed neutrinos with active neutrinos in connection with various experimental and cosmological bounds.

\section{Results} \label{sec:results}
\blu{In this section we discuss the numerical results of the resonant leptogenesis for $M_3 \simeq M_4$. The degeneracy between $M_3$ and $M_4$ can be approached either from the $M_3 \gtrsim M_4$ side ($a \lesssim \frac{1}{2} \left(\sqrt{f^2+8}+f\right)$) or from the $M_3 \lesssim M_4$ side ($a \gtrsim \frac{1}{2} \left(\sqrt{f^2+8}+f\right)$). The $CP$ asymmetry in Eq.~\eqref{epsires} flips sign as we move from one side to the other, since Eqs.~\eqref{fmix} and \eqref{fosc} contain the term $M_i^2 - M_j^2$ in the numerator. It should be noted that the $CP$ asymmetry is a function of $\sin \delta$, as showed in Ref.~\cite{Rahat:2020mio}; hence its sign can also be overturned by inverting the sign of $\delta$. However, in the following discussion we will adopt $\delta = -78^\circ$ and choose the appropriate side of the $M_3\simeq M_4$ degeneracy so that the generated baryon asymmetry is always positive to match the observed value. }

\reply{At this stage the model has three free parameters $a, b$ and $f$. Our strategy for exploring the parameter space of the model is as follows. We express $a$ in terms of $b$ and $f$ using the resonance condition in Eq.~\eqref{rescondgen}.  We vary $0.1<f<10$ and will argue later in this section that this captures the essential physics of the model. For every value of $f$, we determine $b$ by demanding that the baryon asymmetry calculated using \eqref{formulaB} and \eqref{NBLf} from the solution of the Boltzmann equations \eqref{eq:Nni1} and \eqref{eq:NBLflav} matches the measured CMB value given in Eq.~\eqref{measuredB}. }



\subsection{Lower bound on the right-handed neutrinos} \label{lowermass}
\begin{figure}[!ht]
    \centering
    \subfloat[Zero initial abundance \label{fig:minbzero}]{
        \includegraphics[width=0.48\textwidth]{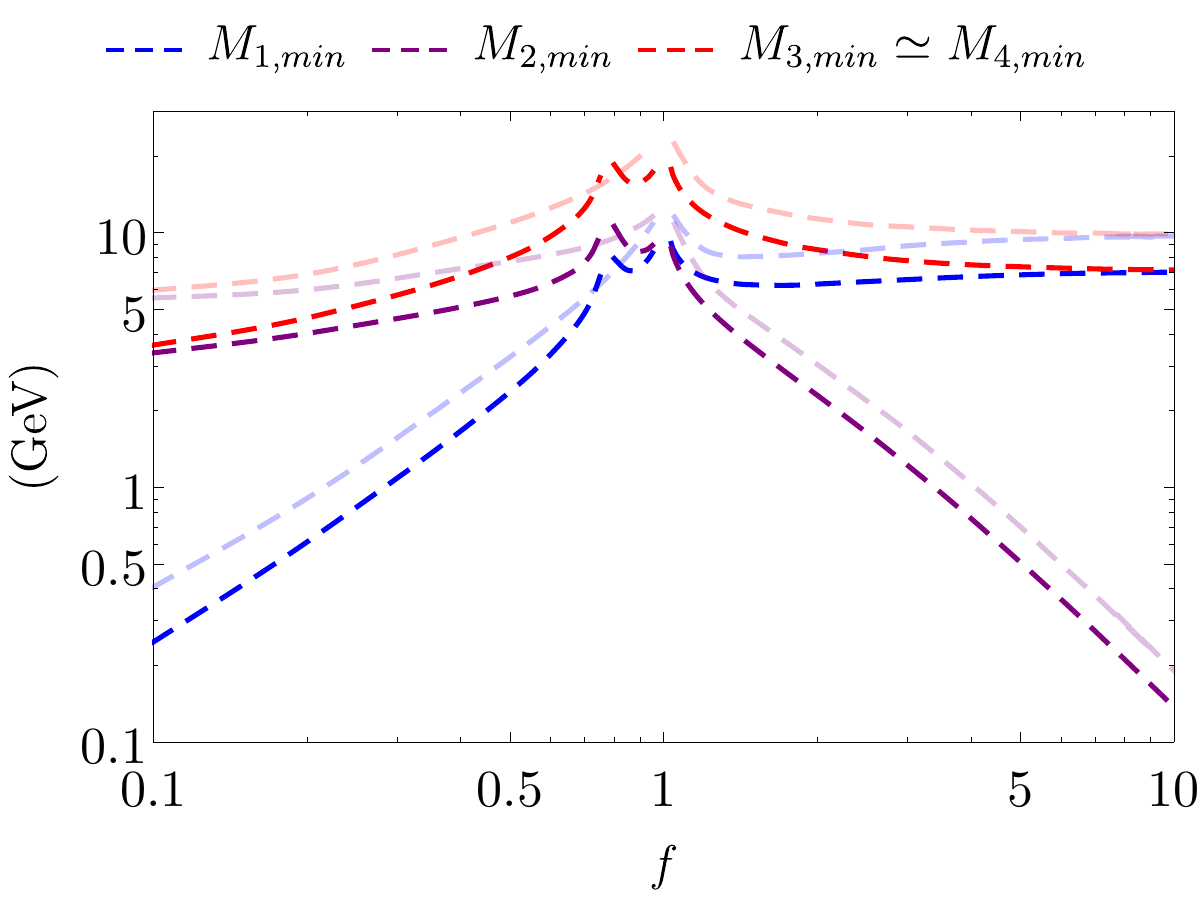}
    }
    \subfloat[Thermal initial abundance \label{fig:minbthermal}]{
        \includegraphics[width=0.48\textwidth]{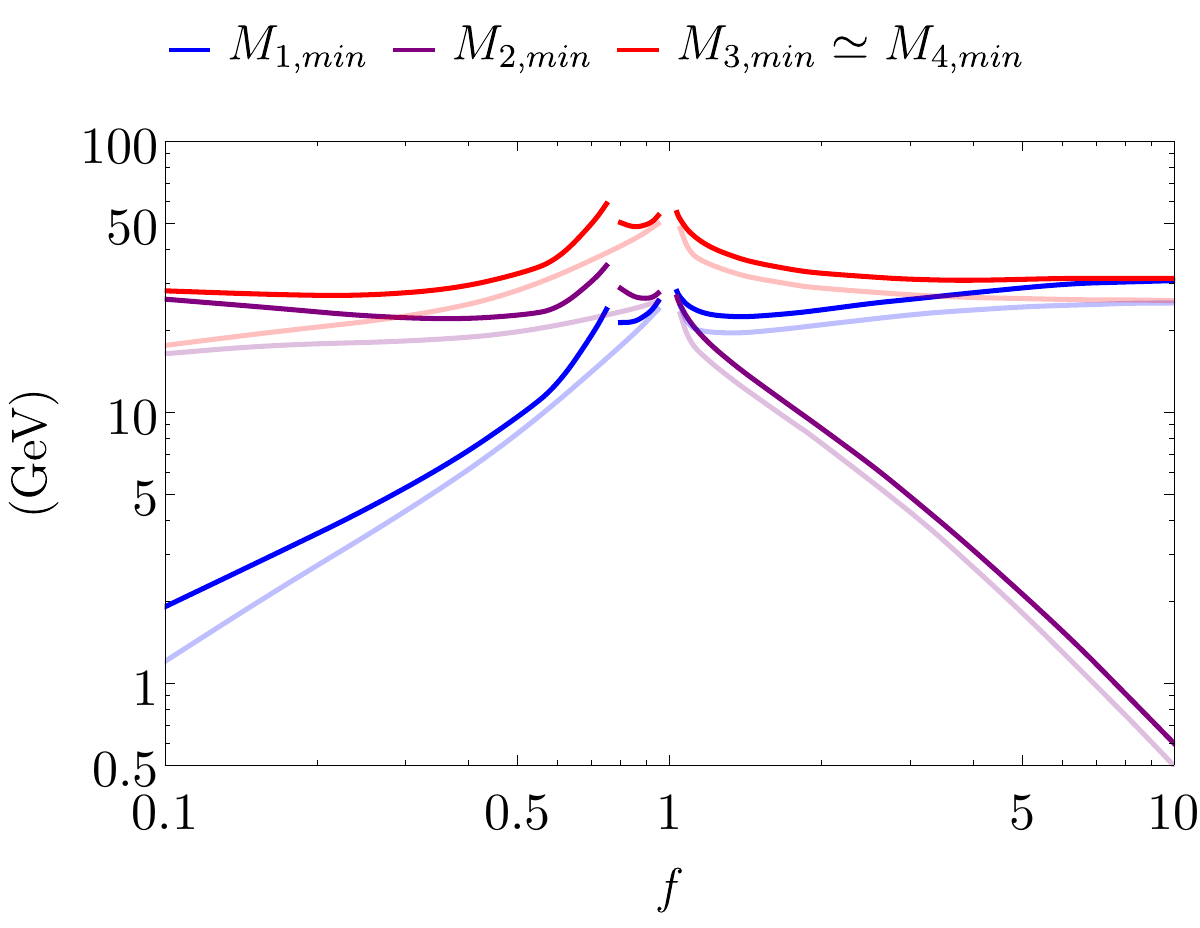} 
    }
    \caption{\reply{Minimum value of the right-handed neutrino masses for (a) zero ($N_{N_i} (z=0) = 0$) and (b) thermal initial abundance ($N_{N_i} (z=0) = N_{N_i}^{\rm eq}$) at the resonance $M_3 \simeq M_4$. Darker lines represent the results obtained from solving the density matrix equations. For comparison we also show the results obtained by solving the classical Boltzmann equations denoted by lighter lines. The results are of the same order of magnitude in the region of interest. At $f=1$, the $CP$ asymmetry vanishes identically and leptogenesis fails. As one approaches $f=1$ from either direction, the minimum mass scale increases to compensate for the suppression in the $CP$ asymmetry. At $f \simeq 0.74$, the total baryon asymmetry flips sign, an effect not seen when using the classical Boltzmann equations.}}
    \label{fig:minb}
\end{figure}

For a particular choice of $f$, given that the resonance condition is satisfied, there is a minimum value of $b$ for which the generated baryon asymmetry matches the CMB value.
\blu{The reason is as follows. All the masses of $N_i$ are proportional to $b$ as in Eq.~\eqref{Mmass}. As $b$ decreases, so do $M_3$ and $M_4$, and this results in longer lifetime of $N_3$ and $N_4$. Since they decay late close to the electroweak sphaleron freeze-out temperature $T_{sph}$, the amount of $B-L$ asymmetry which is being converted to baryon asymmetry will be limited by the decays which occur above $T_{sph}$. Hence, the smaller the $b$, the fewer the decays above $T_{sph}$ and the smaller the resulting baryon asymmetry.}
This puts a lower bound on all the right-handed neutrino masses. In Fig.~\ref{fig:minb}, we show the minimum masses required for successful leptogenesis at the resonance $M_3 \simeq M_4$. The lowest degenerate mass is of $\mc O(1)\ \text{GeV}$.
\reply{In this regime, $M \ll T_{sph}$ and we use the density matrix equations in App. \ref{app:density_matrix}. For comparison, we also show the results obtained from using the classical Boltzmann equations in Eqs. \eqref{eq:Nni1} and \eqref{eq:NBLflav}. The results agree within $\mc O(1)$ factors, except there is a sign flip of the produced asymmetry near $f \simeq 0.74$ when using the density matrix equations. The sign flip occurs because in the vicinity of $f=0.74$, electron and tau flavor contributions to the asymmetry are close in magnitude but opposite in sign, and their combined contribution has the opposite sign of the mu flavor contribution. For $f\lesssim 0.74$, the combined contribution of the electron and tau flavors win over the mu contribution. But for $f \gtrsim 0.74$, the mu flavor contribution becomes dominant and flips the overall sign. No such sign flip is observed in the solutions of the classical Boltzmann equations, where similar signs of the three flavors is observed, but the contribution of the mu flavor dominates for $f<1$.
}

From Fig.~\ref{fig:minb}, the lower bound on the masses is higher in the case of thermal initial abundance compared to the case of zero initial abundance of $N_i$. 
\reply{For the latter case, there is an asymmetry generation during the population of $N_i$ from the ``inverse decay'' $\ell H \to N_i$ and thermal scatterings (involving top quarks and gauge bosons) at high temperature $T\gtrsim M_i$. As for the case of thermal initial abundance, the asymmetry is only generated from $N_i$ decays which predominantly occur much later at $T \lesssim M_i$. Since substantial $B-L$ asymmetry is built up for the case of zero initial abundance at $T \gtrsim M_i$, $N_i$ can decay much later, resulting in more relaxed lower bound on their masses.}



We now look at the limiting behavior of the lower bound for large and small $f$. 
\reply{Though we use the density matrix equations where the $CP$ asymmetry parameter is implicit, as we will discuss below, the qualitative behavior of the results can be easily understood using the $CP$ asymmetry parameter defined in Eq. \eqref{epsires} (except the additional sign flip in baryon asymmetry around $f \simeq 0.74$).}
For $f \gg 1$, the right-handed neutrino masses are given by Eq.~\eqref{masslarge}.
In this limit $M_1 \simeq M_3 \simeq M_4$. 
At resonance, the $CP$ asymmetry parameters are independent of the mass scale $b$ and mildly dependent on $f$ as shown in Fig. \ref{fig:CPasymmetry34}. Hence the lower bound on $M_{3,4}$ is determined mainly from the amount of $B-L$ asymmetry that is generated above $T_{sph}$. As a result, the lower bound on $M_{3,4}$ will be approximately constant where the mild dependence on $f$ comes only from the mild dependence of $CP$ asymmetry parameters on $f$ for $f \gg 1$. On the other hand, $M_2$ being inversely proportional to $f$, continue to decrease for increasing $f$.


For $f \ll 1$, the right-handed neutrino masses are given by Eq.~\eqref{masssmall}.
In this limit $M_2 \simeq M_3 \simeq M_4$. 
\blu{For the same reason as in the case of $f\gg 1$, the lower bound on $M_{3,4}$ which is fixed by $T_{sph}$ will be approximately constant and the mild dependence on $f$ comes only from the dependence of the $CP$ asymmetry parameters on $f$ for $f\ll 1$ as shown in Fig. \ref{fig:CPasymmetry34}. Now $M_1$ being proportional to $f$ will decrease with decreasing $f$.}

At $f=1$, the resonant $CP$ asymmetries $\varepsilon_{3\alpha}$ and $\varepsilon_{4\alpha}$ nearly vanish as $(Y^\dagger_\nu Y_\nu)_{34} = 0$, cf. Eq.~\eqref{YY}. As one approaches $f=1$ from both directions, the $CP$ asymmetry is getting more suppressed and to compensate for this, higher mass scale is required.


\subsection{Upper bound on the right-handed neutrinos}

Next, we will discuss a rather unexpected result, namely, the existence of upper bound on the right-handed neutrino masses. This is due to the specific mass spectrum of the right-handed neutrinos as given in Eq.~\eqref{Mmass}, which is unique to the model under consideration. In general, there exists lighter $N_{1,2}$ than the resonant pairs $N_{3,4}$ that can result in substantial washout of asymmetry and hence limit the amount on the final asymmetry. At resonance, the $CP$ asymmetry parameters are independent of the mass scale $b$. As $b$ increases, all $N_{3,4}$ can decay much before $T_{sph}$ and hence the asymmetry generated from the resonant pair will be independent of $b$. Now, it is possible to have additional washout of asymmetry from lighter $N_{1,2}$. If this washout is significant, this will give an upper bound on how heavy $N_{1,2}$ can be. This in turns will translate to an upper bound on $b$ and hence an upper bound on the masses of all the right-handed neutrinos.

\begin{figure}[!ht]
    \centering
    \includegraphics[width=0.48\textwidth]{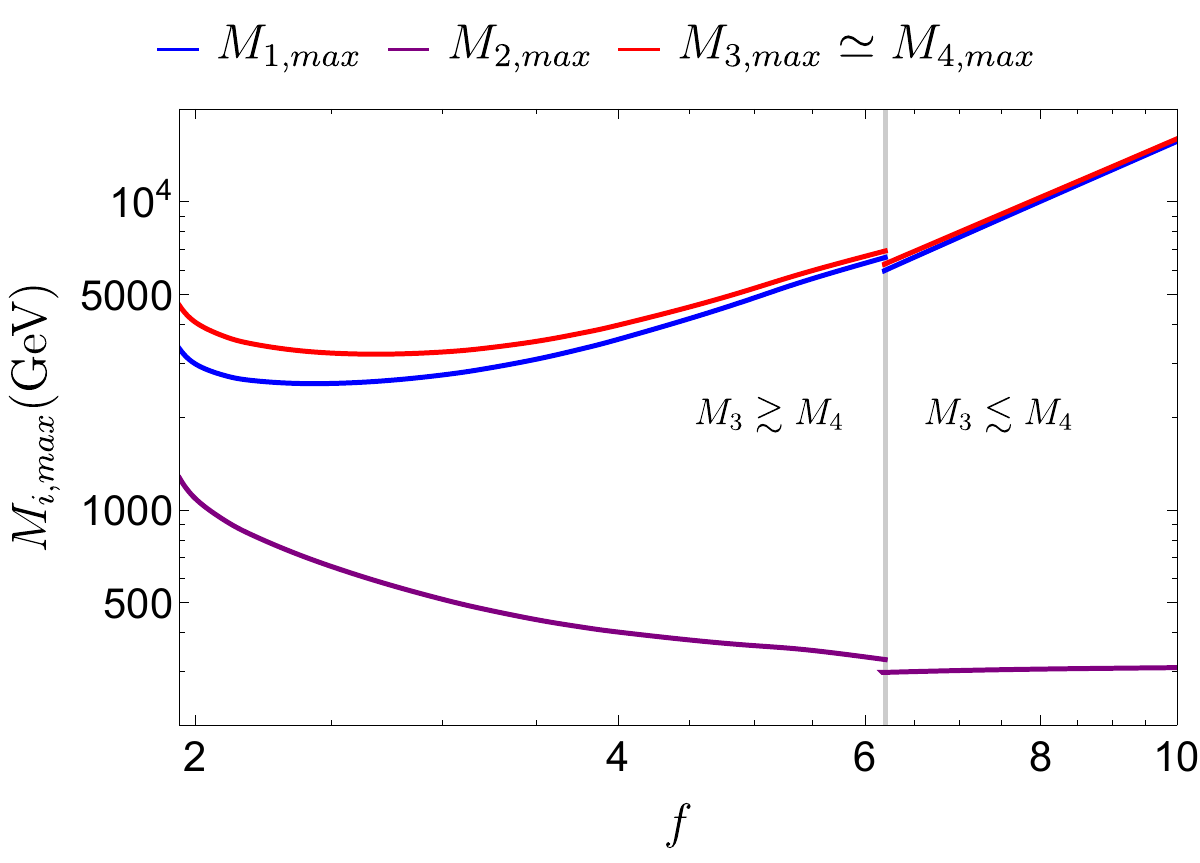}
\caption{Maximum value of the right-handed neutrino masses for both thermal and zero initial abundance near the resonance $M_3 \simeq M_4$. There is no upper bound on the masses for $f \lesssim 2$. Positive baryon asymmetry is generated for $M_3 \gtrsim M_4$ when $2 \lesssim f < 6.20$ and for $M_3 \lesssim M_4$ when $f > 6.20$. There is a sudden change of the mass values at $f=6.20$. } 
\label{fig:maxb}
\end{figure}

\blu{
Due to the flavor structure of $Y_\nu$ and mass spectrum of $N_i$, it turns out we only have an upper bound for $f \gtrsim 2$ as shown in Fig.~\ref{fig:maxb}.}
\begin{figure}[!ht]
    \centering
    \subfloat[\label{fig_PK_a}]{
        \includegraphics[width=0.48\textwidth]{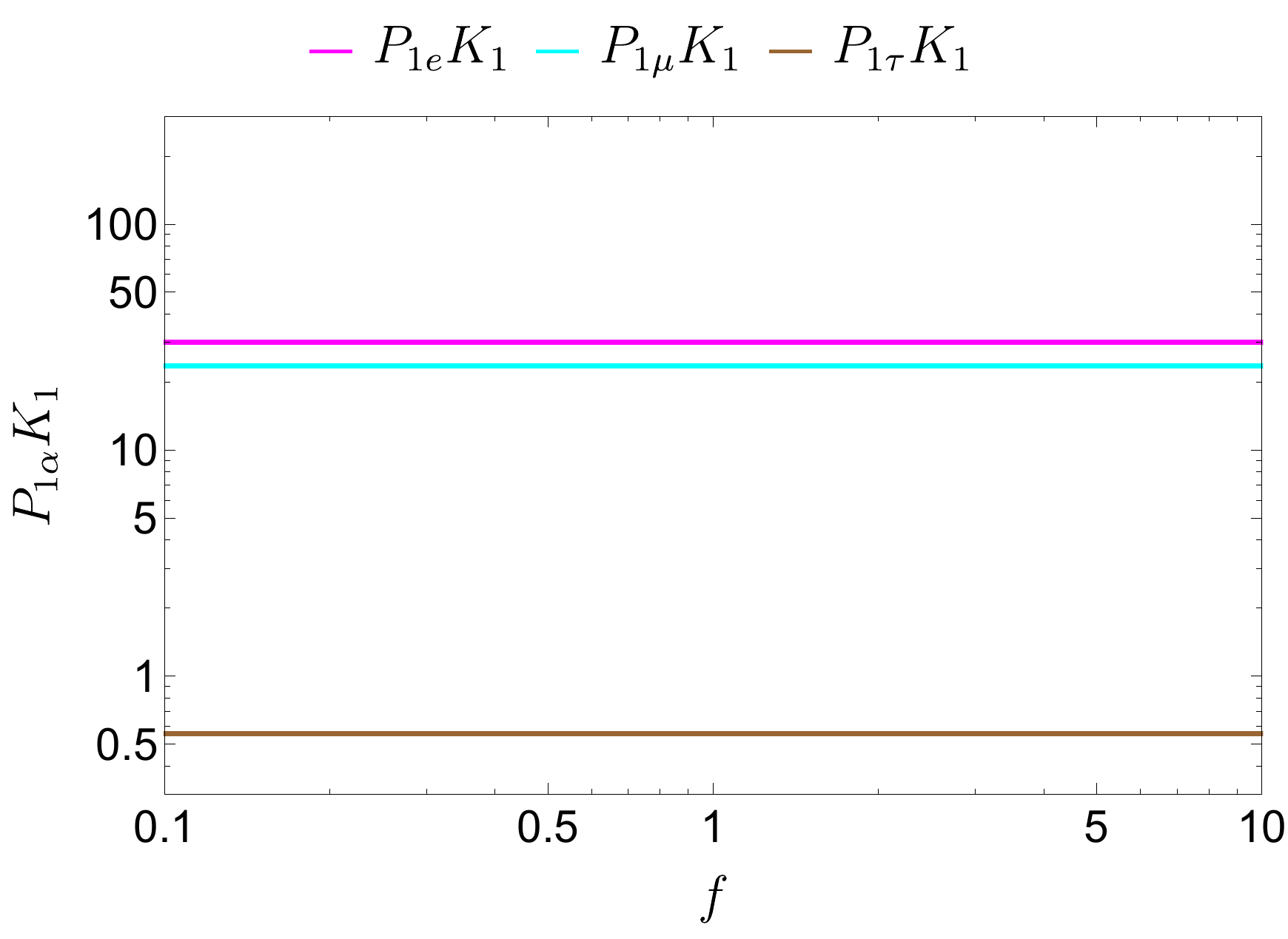}
    }
    \subfloat[\label{fig_PK_b}]{
        \includegraphics[width=0.48\textwidth]{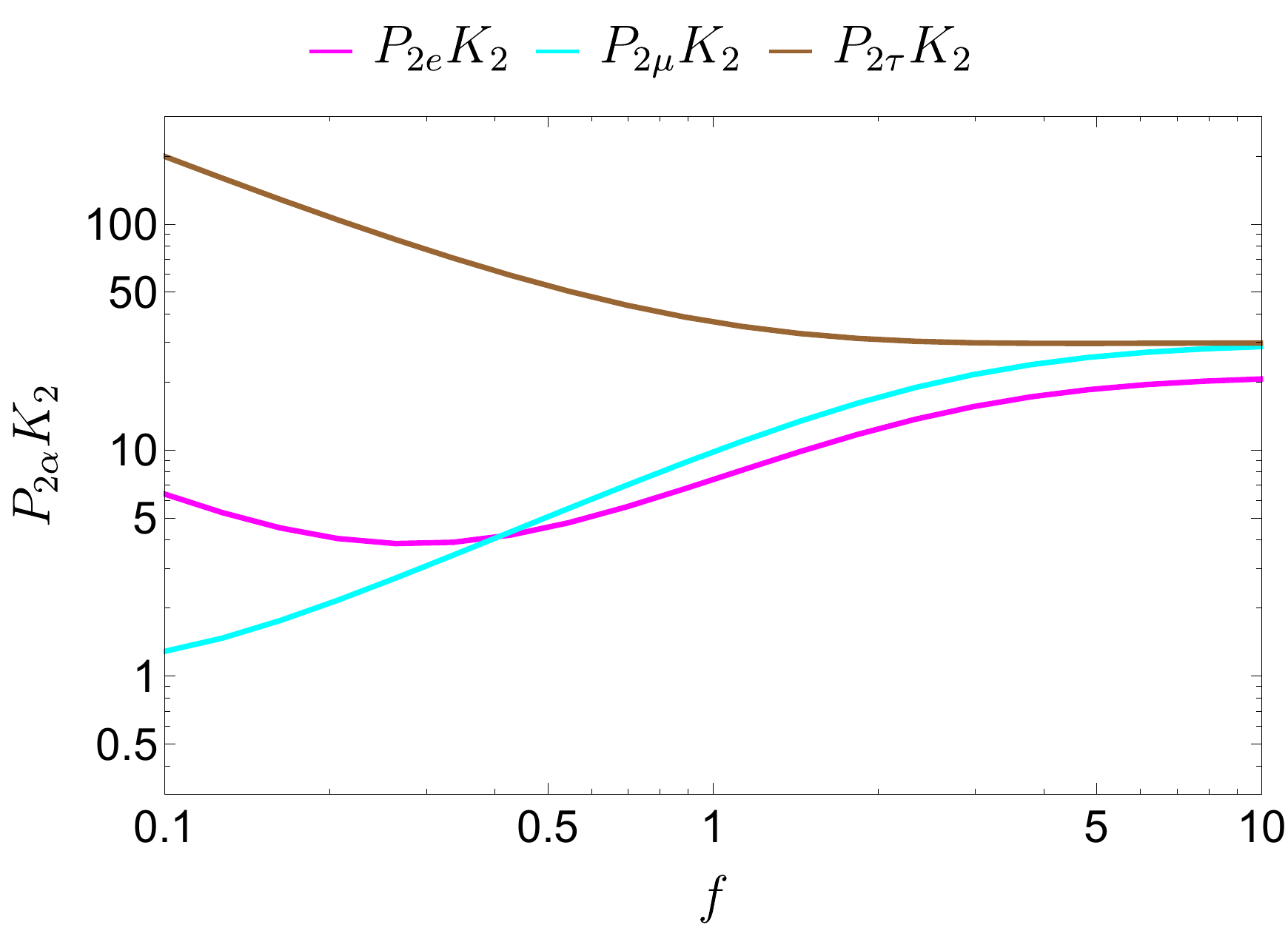} 
    }
\caption{Decay parameter times branching ratio as a function of $f$ at resonance. The asymmetry generated by $N_3$ and $N_4$ is partially washed out by $N_1$ and $N_2$, 
\textcolor{black}{and for $M_{1,2} \ll M_{3,4}$, is proportional to $e^{-P_{i\alpha}K_i}$.}
} 
\label{fig:PiKi}
\end{figure}
\blu{Let us first consider the case with $f \gg 1$. In this case, we have $M_1 \sim M_3 \simeq M_4 \gg M_2$ as shown in Eq.~\eqref{masslarge}. The washout of the asymmetry from $N_2$ at $T = M_2$ is exponential $e^{-P_{2\alpha} K_2}$ and could result in large suppression of final asymmetry if $P_{2\alpha} K_2$ is large. As can be seen in Fig.~\ref{fig:PiKi}(b), it turns out that $P_{2\alpha} K_2 \gtrsim 10$ for all flavors and hence the suppression of final asymmetry is very large.}
\begin{figure}[!ht]
    \centering
    \subfloat[$f=0.1$\label{fig_2c}]{
        \includegraphics[width=0.48\textwidth]{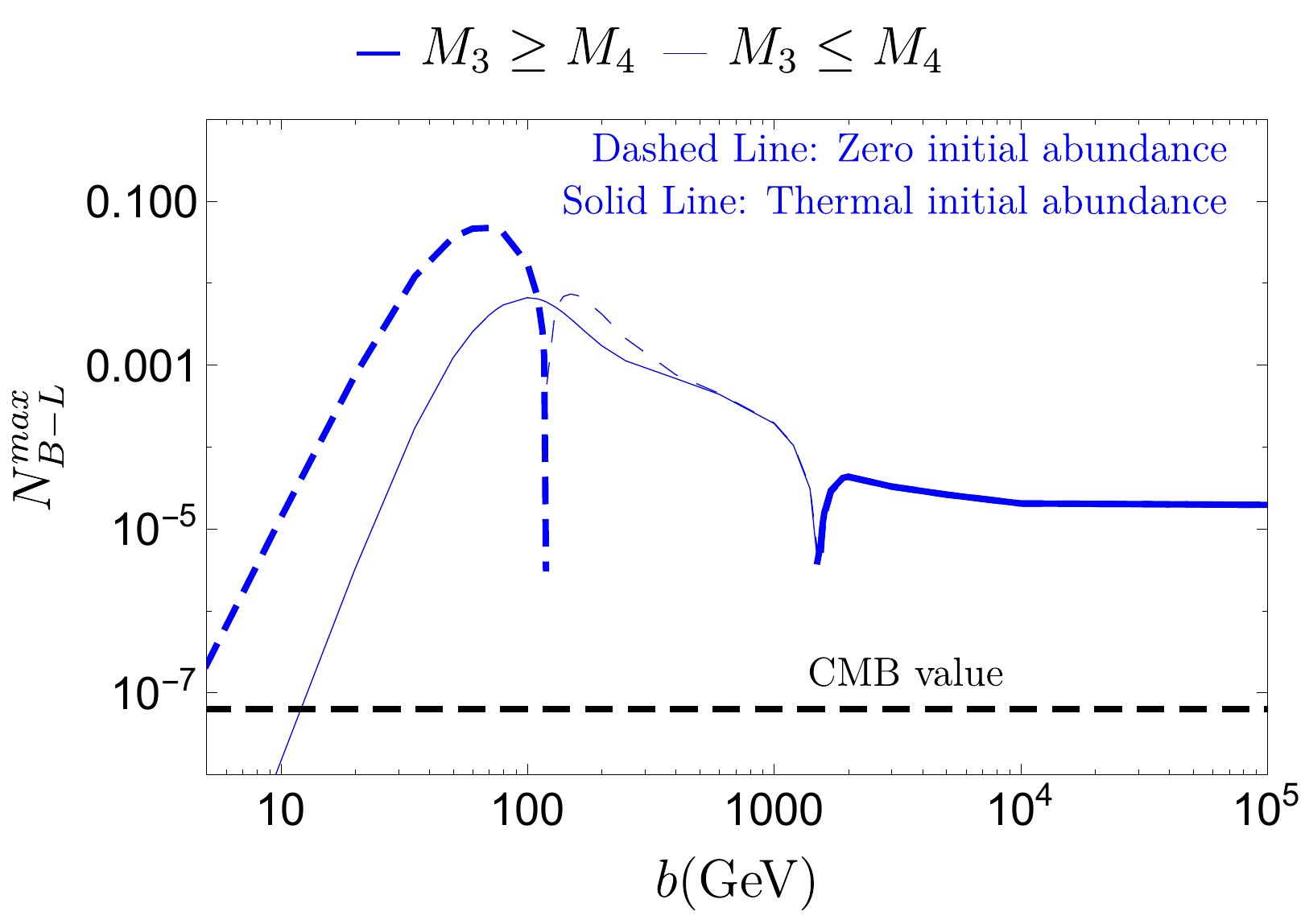}
    }
    \subfloat[$f=0.1$, without $N_1$ washout \label{fig_2e}]{
        \includegraphics[width=0.48\textwidth]{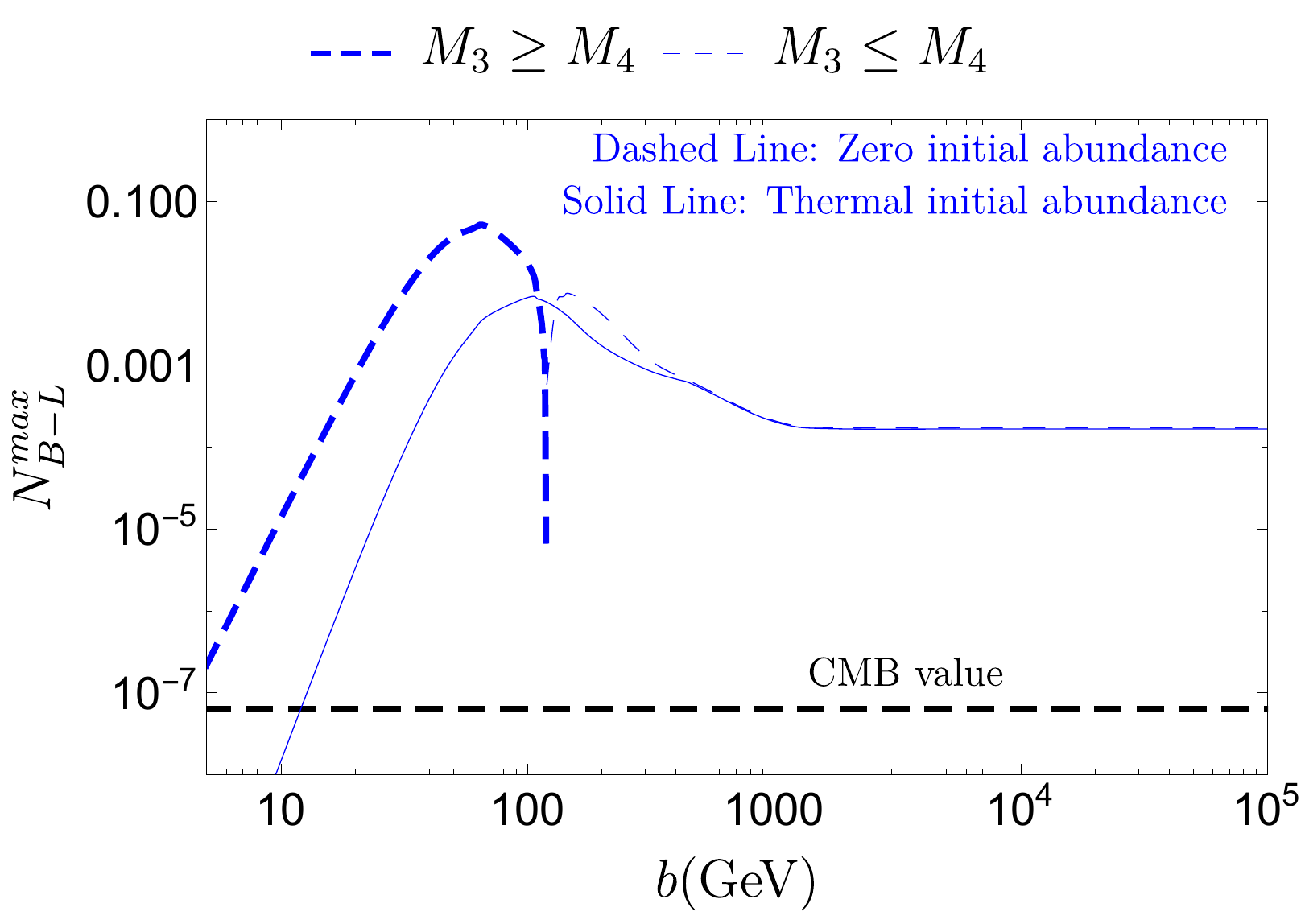}
    }\\
    \subfloat[$f=10$\label{fig_2d}]{
        \includegraphics[width=0.48\textwidth]{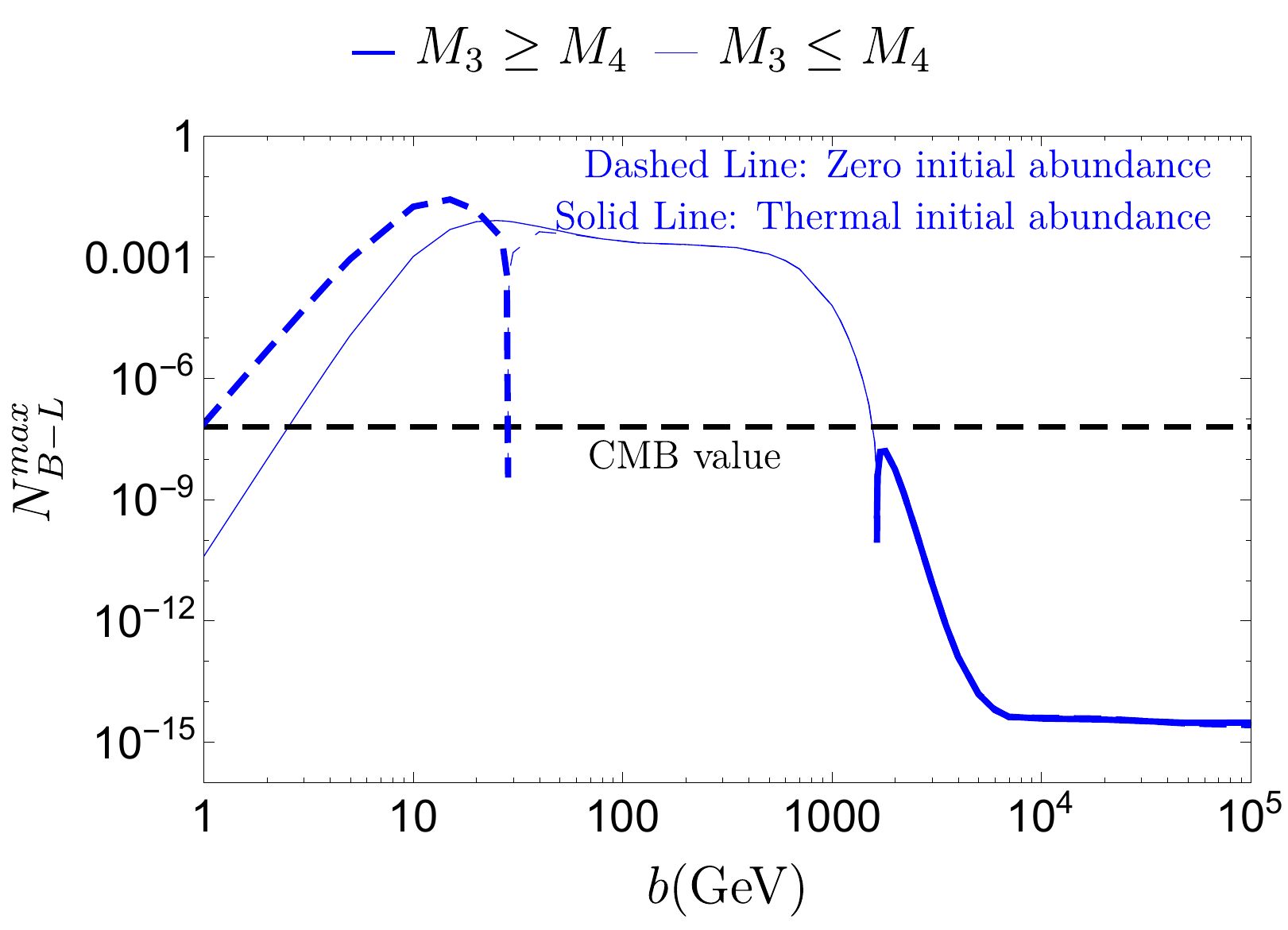} 
    }
    \subfloat[$f=10$, without $N_2$ washout  \label{fig_2g}]{
        \includegraphics[width=0.48\textwidth]{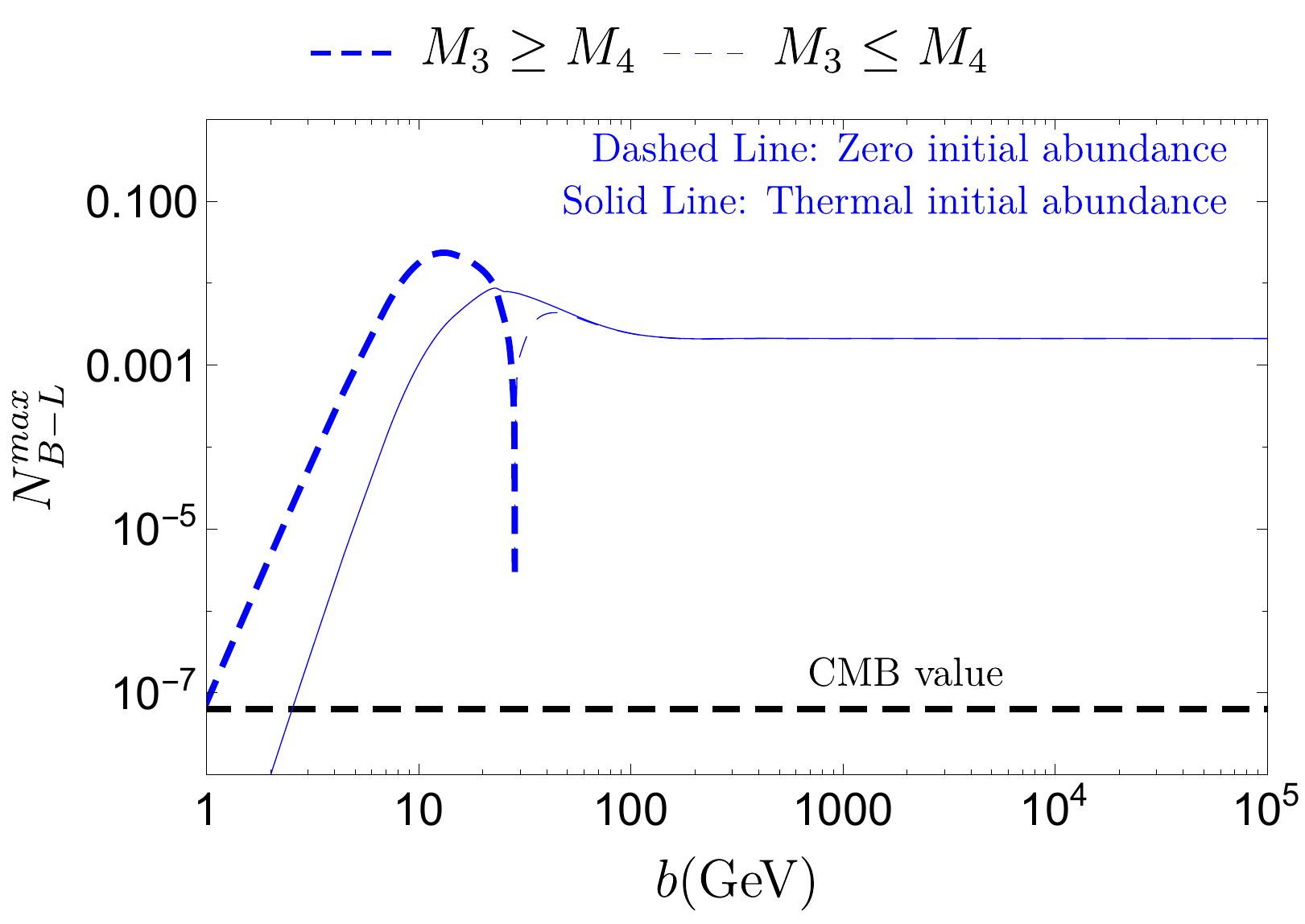}
    }
\caption{
Maximum $B-L$ asymmetry at the resonance $M_3 \simeq M_4$ for (a) $f=0.1$, (b) $f=0.1$ without considering $N_1$ washout, (c) $f=10$, and (d) $f=10$ without considering $N_2$ washout. 
Thick (thin) lines represent positive baryon asymmetry for $M_3 \gtrsim M_4$ ($M_3 \lesssim M_4$). For large $b$, the $B-L$ asymmetry saturates at a value higher than the CMB value in case (a) and (b), thus indicating that there is no upper limit on $b$. $N_1$ washout decreases the final asymmetry by only a factor of $10$, and is not very efficient. In case (c), however, the maximum $B-L$ asymmetry saturates below the CMB value for large $b$, thus setting an upper limit above which successful resonant leptogenesis is not feasible. If the $N_2$ washout is disregarded, the final asymmetry is $\mc O(10^{12})$ times large, as shown in case (d), thus implying that $N_2$ washout is efficient for $f=10$. In either case, whenever the maximum $B-L$ asymmetry is larger than the CMB value, successful resonant leptogenesis can be achieved by moving $a$ slightly away from the resonance condition given by Eq.~\eqref{rescondgen}. \reply{We have performed this analysis for the whole range $0.1<f<10$ solving the Boltzmann equations and determining the range of $b$ that can yield the observed baryon asymmetry. The case for these two particular values of $f$ have been illustrated here to explain why and when an upper bound on $b$ can arise.}}
\label{fig:whymaxb}
\end{figure}
\blu{An explicit example of this is illustrated in Fig.~\ref{fig:whymaxb}(c) and (d) for the case of $f=10$. We would need to have $M_2 \lesssim T_{sph}$ such that the washout is not effective until the baryon asymmetry is frozen. Eq.~\eqref{masslarge} then implies that $M_1\simeq M_3 \simeq M_4$ must increase with $f$. This can be seen in Fig.~\ref{fig:maxb} for $f \gg 1$.} \blu{Interestingly, there is a discontinuity on the upper bound at $f=6.2$. It is due to the specific flavor structure of the model as illustrated in Fig.~\ref{fig:gap}.
\begin{figure}[!ht] 
    \centering
    \subfloat[$f=6.1, b=1015\ \text{GeV}, M_3 \gtrsim M_4 $ \label{fig_4a}]{
        \includegraphics[width=0.48\textwidth]{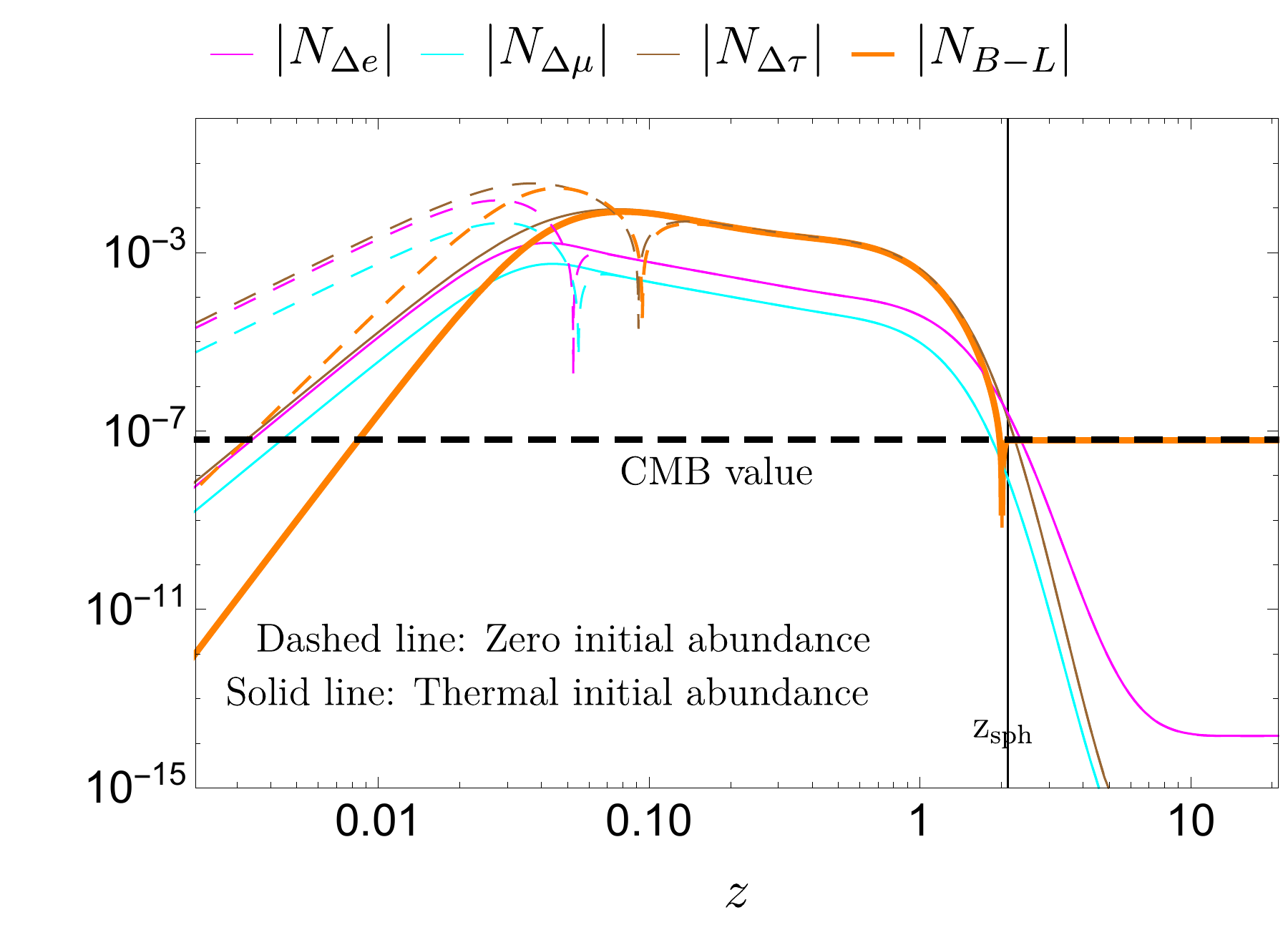}
    }
    \subfloat[$f=6.3, b=982\ \text{GeV}, M_3 \lesssim M_4 $ \label{fig_4b}]{
        \includegraphics[width=0.48\textwidth]{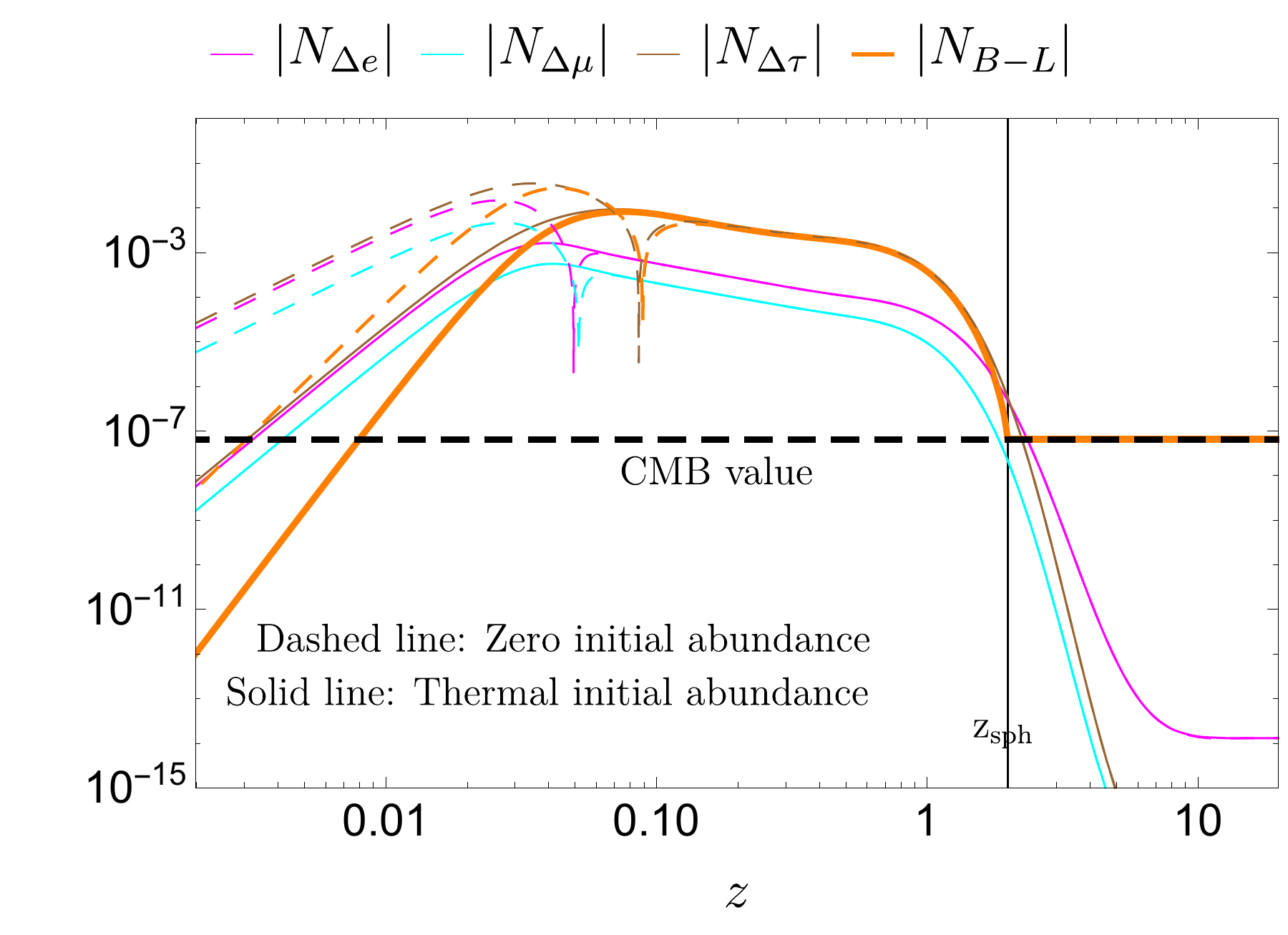}
    }
    \caption{$B-L$ asymmetry for (a) $f=6.1$ and (b) $f=6.3$. In both cases, $N_{\Delta \mu} \ll N_{\Delta e}, N_{\Delta \tau}$, and $N_{\Delta e}$ has a different sign than $N_{\Delta \tau}$. $N_{\Delta \tau}$ is greater than $N_{\Delta e}$ in (a) and smaller in (b) at $z_{sph}$, thus flipping the sign of the $B-L$ asymmetry. The positive sign of the asymmetry is achieved for $M_3 \gtrsim M_4$ in (a) and $M_3 \lesssim M_4$ in (b).}
    \label{fig:gap}
\end{figure}
In general $N_{\Delta \mu} \ll N_{\Delta e}, N_{\Delta \tau}$ at $T_{sph}$, and $N_{\Delta e}$ has a different sign than $N_{\Delta \tau}$. At $f < 6.2$, the final asymmetry is dominated by $N_{\Delta \tau}$. At $f > 6.2$, the washout of $N_{\Delta \tau}$ becomes so strong that $N_{\Delta e}$ takes over the final asymmetry and flips its sign.} \blu{Fig.~\ref{fig:maxb} also shows that as $f$ decreases, the upper bound on $M_2$ relaxes. This is because as $M_2$ is getting closer to $M_{3,4}$ (with decreasing $f$), the washout effect is no longer exponential (but goes as $1/(P_{2\alpha} K_2)$) during the asymmetry generation. In fact, the upper bound disappears at $f \lesssim 2$.
}

\blu{
To understand the absence of the upper bound for small $f$, let us focus on $f \ll 1$. In this case, we have $M_2 \sim M_3 \simeq M_4 \gg M_1$ as shown in Eq.~\eqref{masssmall}. Now the washout of the asymmetry from $N_1$ at $T = M_1$ is exponential $e^{-P_{1\alpha} K_1}$ and could result in large suppression of final asymmetry if $P_{1\alpha} K_1$ is large. From Fig.~\ref{fig:PiKi} (a), it turns out that $P_{1e} K_1, P_{1\mu} K_1 \gtrsim 10$ while $P_{1\tau} K_1 \lesssim 1$. Hence the washout of asymmetry in $N_{\Delta \tau}$ is not efficient. As shown in Fig.~\ref{fig:whymaxb}(a) and (b) for the case of $f=0.1$, the final asymmetry dominated by $N_{\Delta \tau}$ is saved from washout and is always larger than the observed value. In fact, we can see from the figure that this feature is independent of $f$. Hence there is an absence of upper bound until $f \gtrsim 2$ when the $N_2$ washout takes over.
}

\reply{It should be noted that the constraints on the parameters $a, b$ and $f$ determined in this section are valid for the low scale of resonant leptogenesis. Their values at the GUT scale can be determined by renormalization group running \cite{Antusch:2005gp}. However, these parameters have no effect in determining the mass spectrum of either the charged or the neutral fermion sector; furthermore, they are also not determined by any GUT boundary conditions in our set-up. Hence, their evolution under the renormalization group equations are not expected to impact the results obtained in this work.}


\subsection{Experimental constraints}
In this section we discuss the experimental constraints on the light sterile neutrinos as well as the possibilities to detect them in accelerator experiments. The right-handed neutrinos are typically difficult to probe in experiments due to their extremely feeble interactions. However,  experimental searches of particles of these types can be efficiently done in \textit{intensity frontier}  rather than \textit{energy frontier}. SHiP (Search  for  Hidden  Particles) \cite{Bonivento:2013jag, SHiP:2018xqw, Gorbunov:2020rjx} and DUNE (Deep Underground Neutrino Experiment) \cite{Ballett:2019bgd, Abi:2020evt} are the two most sensitive upcoming  intensity frontier experiments that are relevant to our study. 
If kinematically allowed, the sterile neutrinos can be produced in the final states from decays of heavy mesons. Subsequently, two-body (three-body) decays of the sterile neutrinos into lighter meson and a charged lepton (a pair of charged leptons and active neutrino) have the potential to be probed in SHiP as well as in DUNE.  These processes are possible due to the mixing of  sterile neutrinos with active neutrinos. Decays of the types $N\to e^-(\mu^-)\;\pi^+$ and $N\to e^-(\mu^-)\;\rho^+$  are the  most promising (corresponding decays involving kaons in the final state are less promising due to low branching fractions) for searches from D-meson decays, and among them, the $\mu^-\;\pi^+$ final state is the cleanest signature.

\begin{figure}[!t] 
    \centering
    \subfloat[$0.1<f<1$\label{mixing_a}]{\hspace{-30pt}
        \includegraphics[width=0.35\textwidth]{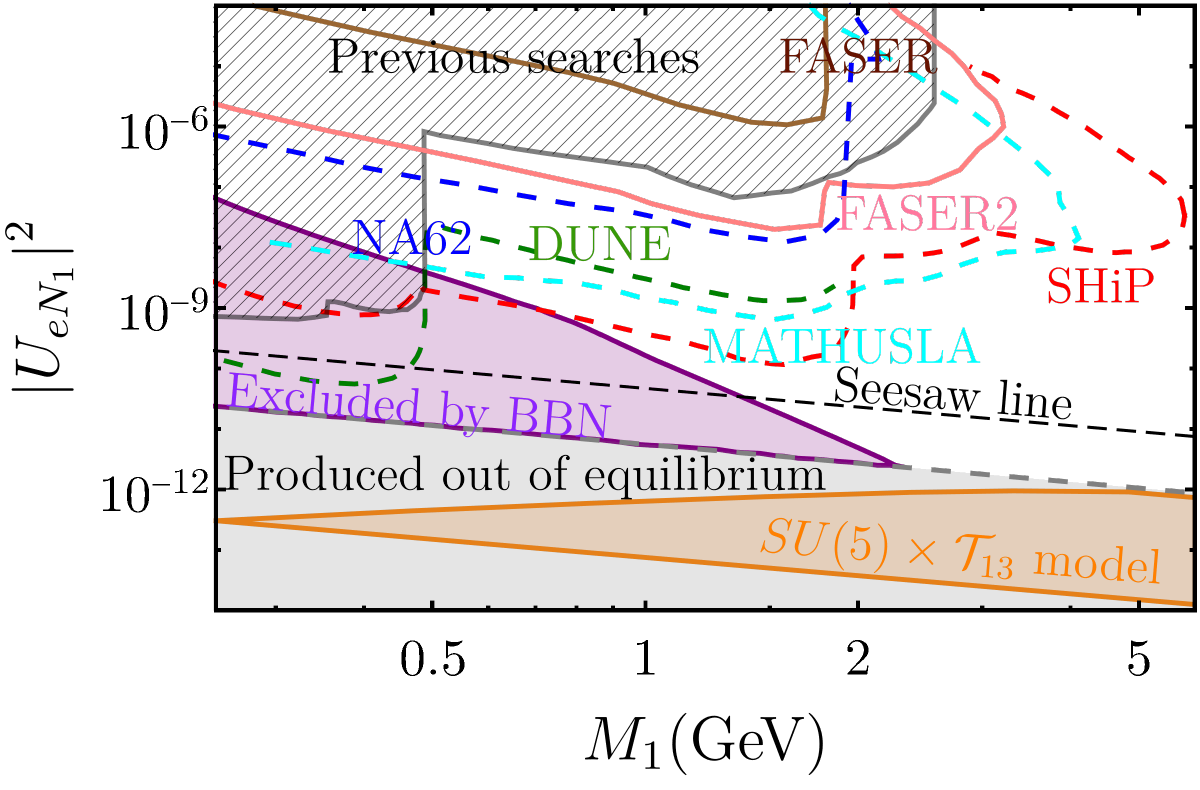}
    }
    \subfloat[$0.1<f<1$ \label{mixing_b}]{
        \includegraphics[width=0.35\textwidth]{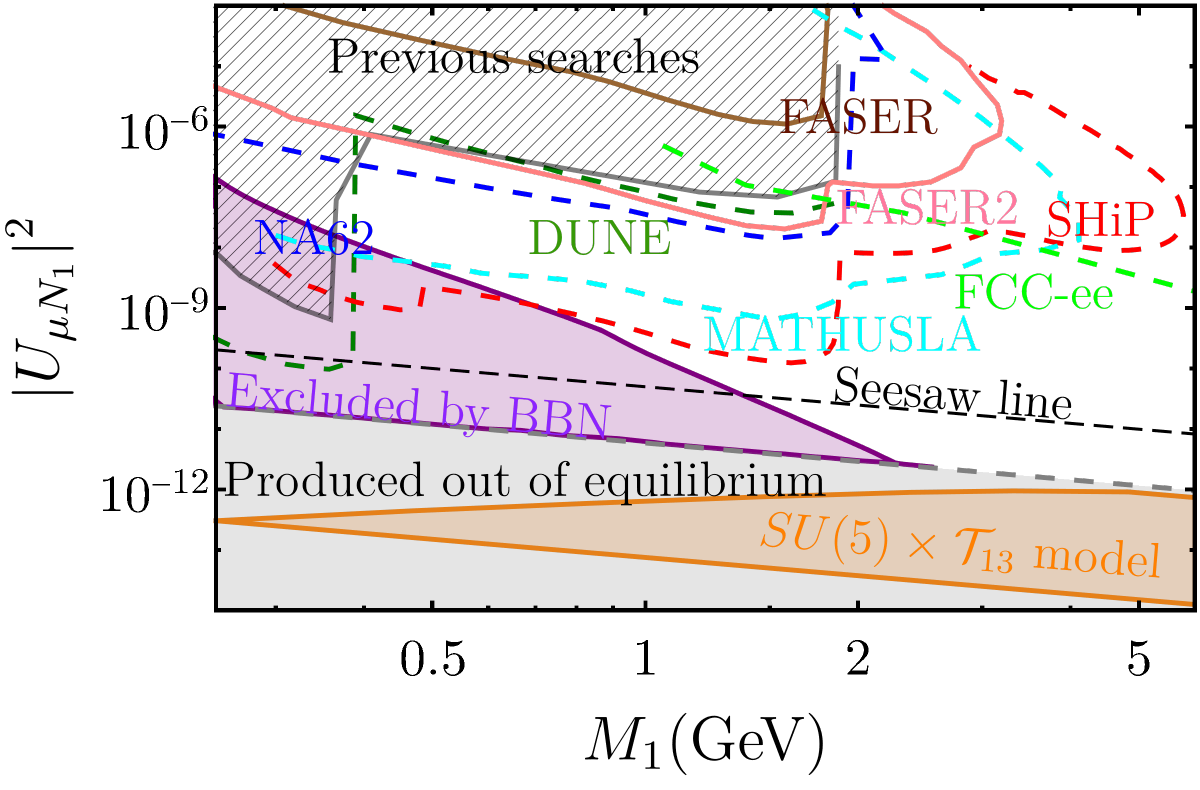}
    }
    \subfloat[$0.1<f<1$\label{mixing_c}]{
        \includegraphics[width=0.35\textwidth]{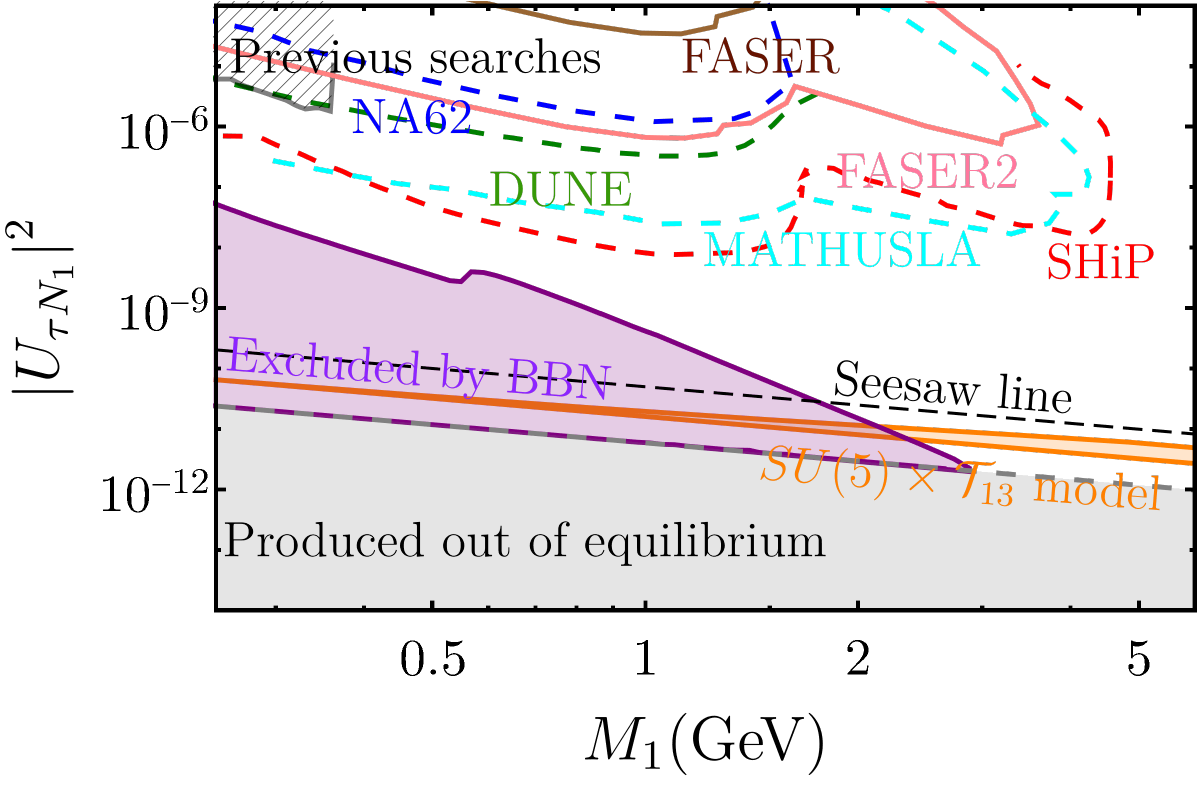}
    }\\
    \subfloat[$1<f<10$\label{mixing_d}]{\hspace{-30pt}
        \includegraphics[width=0.35\textwidth]{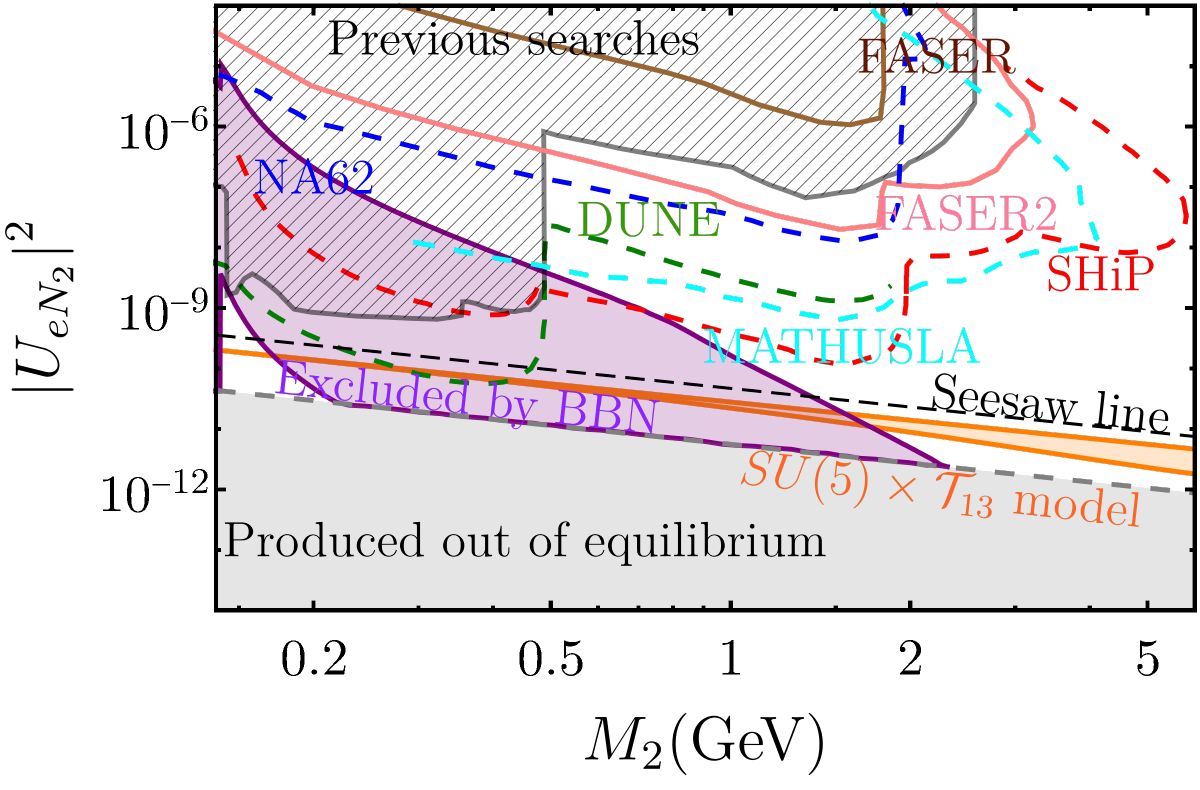}
    }
    \subfloat[$1<f<10$]{
        \includegraphics[width=0.35\textwidth]{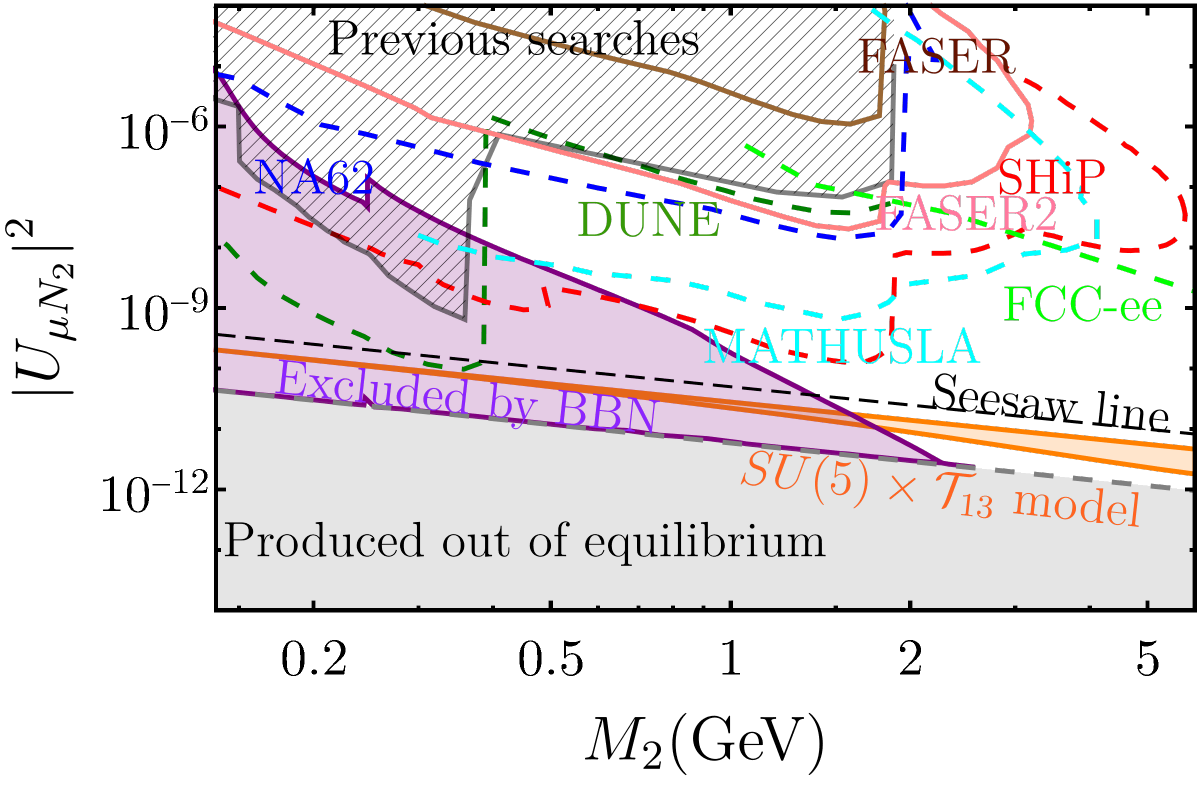}
    }
    \subfloat[$1<f<10$]{
        \includegraphics[width=0.35\textwidth]{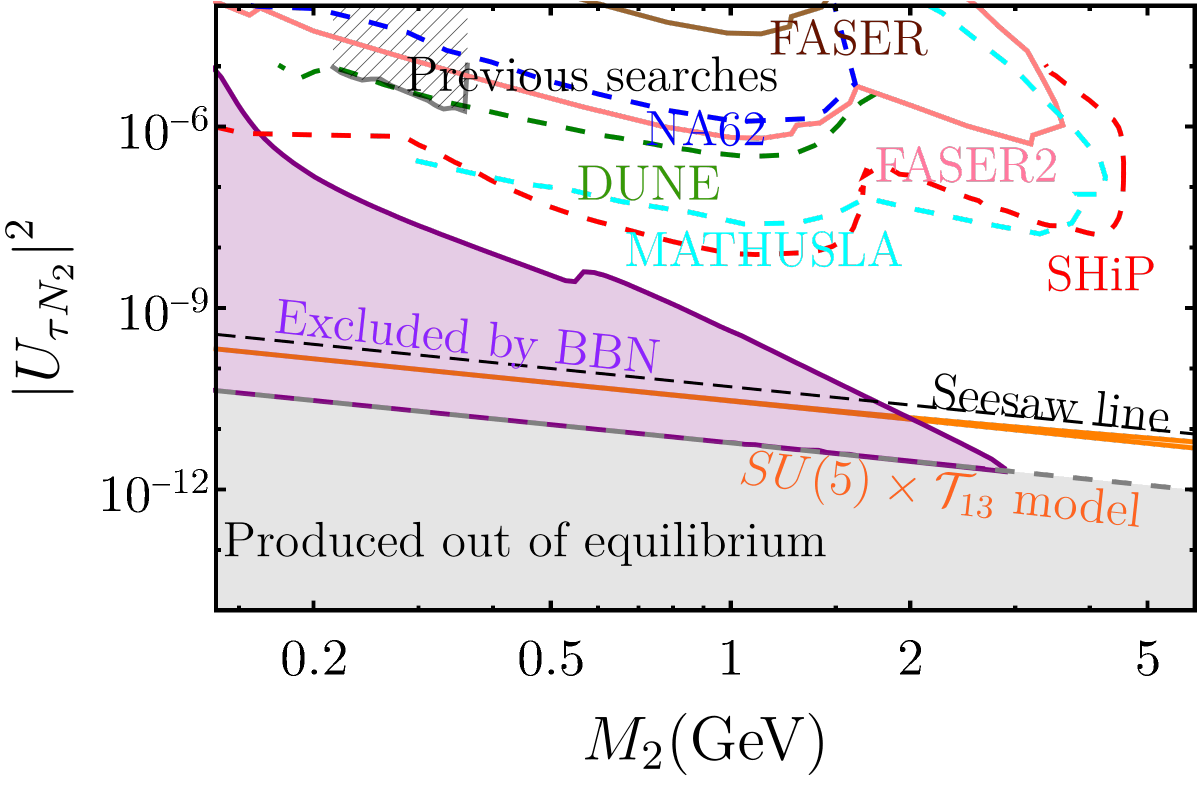}
    }
    \caption{Sterile-active neutrino mixings $|U_{\alpha N_i}|^2$ with  particular flavors at resonance for $0.1<f<1$ (upper panel) and $1<f<10$ (lower panel) are plotted (the orange region) for the model under investigation. For comparison we also show bounds from different experiments and cosmological data analysis and sensitivity of some proposed detectors and experiments. The purple region is excluded by analyzing the BBN data when the right-handed neutrinos are produced thermally and are short-lived so that their meson decay products do not alter the nuclear reactor framework  \cite{Boyarsky:2020dzc, Bondarenko:2021cpc} (also see \cite{Sabti:2020yrt} for a similar analysis without considering the meson decay). The upper hatched region represents excluded regions of the parameter space from previous searches that include accelerator experiments (for details see Ref.  \cite{Alekhin:2015byh}) such as TRIUMF \cite{Britton:1992xv, Britton:1992pg}, PS 191 \cite{Bernardi:1987ek},  CHARM \cite{Bergsma:1985is} \reply{and recent kaon decay results from NA62} \cite{NA62:2020mcv, CortinaGil:2021gga}. \reply{The green, red, blue, cyan, light green, brown and pink lines represent the sensitivity of upcoming experiments/detectors SHiP \cite{Bonivento:2013jag,SHiP:2018xqw, Gorbunov:2020rjx}, DUNE \cite{Ballett:2019bgd, Abi:2020evt}, NA62 \cite{Drewes:2018irr} , MATHUSLA  \cite{Curtin:2018mvb}, FCC-ee \cite{Abada:2019zxq}, FASER and FASER2 \cite{Ariga:2018uku, Ariga:2019ufm}, respectively.} 
    The black dashed line represents the seesaw bound for the minimal seesaw scenario assuming $m_\nu \sim 50\ \text{meV}$. The bottom gray area denotes the region where the right-handed neutrinos are produced out of thermal equilibrium and the BBN analysis of Refs.~\cite{Boyarsky:2020dzc, Bondarenko:2021cpc} are not applicable. \reply{Note that the upper and lower panels have different lower limits set by the minimum allowed mass of the respective sterile neutrino predicted by the model ($M_1 \geq 0.25\ \text{GeV}$ for $f \geq 0.1$ in the upper panel and $M_2 \geq 0.14\ \text{GeV}$ for $f \leq 10$ in the lower panel; see Fig.~\ref{fig:minb} for zero initial condition). This is why the experimental limits in the upper panel are slightly truncated from the left compared to the lower panel.}}
    \label{mixing}
\end{figure}

In the SHiP facility, $400$ $\text{GeV}$ proton beam extracted from CERN's Super Proton Synchrotron accelerator will be dumped on a high density target which aims to accumulate about $2\times 10^{20}$ protons during $5$ years of operation. Whereas D-meson decays provide stringent bounds for sterile neutrinos with masses of  $\lesssim 2$ $\text{GeV}$, SHiP has the sensitivity up to about $\sim 5$ $\text{GeV}$ associated to decays involving B-mesons. Both the two-body and three-body decays of the sterile neutrinos from B- and D-meson decays will also be probed at DUNE with high sensitivity. Considering the expected $120$ $\text{GeV}$ primary proton beams and $1.1\times 10^{21}$ protons on target per year, expected sensitivity at $90\%$ confidence level over $7$ years of data taking  \cite{Coloma:2020lgy} are shown as a function of the right-handed neutrino masses in Fig. \ref{mixing} (green dashed line).  \reply{For masses below $0.5$ GeV, kaon decays are sensitive to even smaller mixings \cite{Gorbunov:2020rjx}. These severe bounds on the masses of the sterile neutrinos and their mixings with active neutrinos arising from the projected SHiP sensitivity are presented in Fig. \ref{mixing} (red dashed line).  In this same Fig. \ref{mixing}, the upper hatched region represents excluded regions of the parameter space from previous searches that include accelerator experiments (for details see Ref.  \cite{Alekhin:2015byh}) such as TRIUMF \cite{Britton:1992xv, Britton:1992pg}, PS 191 \cite{Bernardi:1987ek},  CHARM \cite{Bergsma:1985is} and recent kaon decay results of NA62 \cite{NA62:2020mcv, CortinaGil:2021gga}. }

\reply{For comparison, we also show the sensitivities of some other proposed experiments \cite{Agrawal:2021dbo}. The blue dashed line shows the sensitivity of the NA62 experiment which is being considered for LHC run 3 assuming $10^{18}$ POT in a beam dump mode \cite{Drewes:2018irr}. 
MATHUSLA (cyan dashed line) for the HL-LHC era is proposed to be a large detector ($200$ m$\times 200$ m$\times 200$ m) on the surface above the CMS or ATLAS collecting full $3\ \text{ab}^{-1}$ of integrated luminosity \cite{Curtin:2018mvb}. The brown and pink solid lines show the sensitivities of FASER at LHC Run 3 with $150\ \text{fb}^{-1}$ and FASER2 at HL-LHC with $3\ \text{ab}^{-1}$ \cite{Ariga:2018uku, Ariga:2019ufm}. For the mixing with $\nu_\mu$, we also show the projected sensitivity of the Future Circular Collider (FCC-ee, light green dashed line) \cite{Abada:2019zxq}}.

Furthermore, due to sterile-active neutrino mixing, right-handed neutrinos are produced in the early Universe and their decays can significantly affect the BBN. If the decays into mesons are kinematically allowed, their presence in the primordial plasma can lead to over-production of light elements due to meson driven $p\leftrightarrow n$ conversion. This provides stringent bound on the lifetime ($\tau_N$) of the right-handed neutrinos, since the primordial abundances of helium and deuterium are measured with high accuracy. If the sterile neutrinos are produced thermally in the early Universe and frozen out before the onset of nuclear reactions, the corresponding strong bound on their lifetime has been derived just recently in Ref.  \cite{Boyarsky:2020dzc} (relevant earlier references can also be found therein), which gives $\tau_N \lesssim 0.02$ s. These bounds for different mixing angles as a function of sterile neutrino masses are presented in Fig. \ref{mixing} (purple shaded area). From Fig. \ref{mixing}, it can be inferred that the interesting regions of the parameter space, where DUNE has the potential to detect new physics signals, however, are in tension with the BBN constraints.

In our model, for $f < 1$ ($f > 1$), the lightest right-handed neutrino is $N_1$ ($N_2$). In all cases, their mixing elements with active neutrinos $|U_{as}|^2$ as a function their mass $M_i$ follow the seesaw expectation line $|U_{as}|^2 \sim {m_\nu}/{M_i}$ \reply{where $m_\nu \sim 50\ \text{meV}$ is some representative scale of light neutrino mass} (black dashed line in Fig.~\ref{mixing}).\footnote{\reply{It might seem surprising that the mixing parameter predictions of the model shown in Fig.~\ref{mixing} are always below the seesaw line. This figure only shows the mixing parameters for the lowest sterile neutrino, since the relevant experimental bounds are below $6$ GeV and such low masses of heavy neutrinos are allowed in the model only for the lightest one. Although its mixings with active neutrinos are below the seesaw line, the mixings of other three heavy neutrinos can be larger than the naive seesaw expectation. For example, for $f<1$, the mixing with $M_2$ is an order of magnitude larger than the seesaw expectation.}} 

\vspace{1cm}

\reply{In Fig.~\ref{mixing}, the mixing parameters of the $SU(5) \times \mc T_{13}$ model are shown for $0.1<f<1$ in the upper panel and for $1<f<10$ in the lower panel with the orange shaded region. 
Note that the lower limit of $M_1 \simeq 0.25\ \text{GeV}$ in the upper panel ($M_2 \simeq 0.14\ \text{GeV}$ in the lower panel) is determined by the minimum mass at $f=0.1$ ($f=10$).} In either case, $N_1$ or $N_2$ will always be thermalized (mixing with at least one of the active neutrino flavors lying above the regime ``Produced out of equilibrium'') and hence will be subject to the BBN bound which gives a lower bound on their mass $M \gtrsim 2$ $\text{GeV}$. 
\textcolor{black}{In order to satisfy this bound}, from Fig.~\ref{fig:minb}, we conclude that there is no gain to consider further the regimes with $f \lesssim 0.15$ and $f \gtrsim 5$.

\begin{figure}[!ht]
    \centering
    \includegraphics[width=0.48\textwidth]{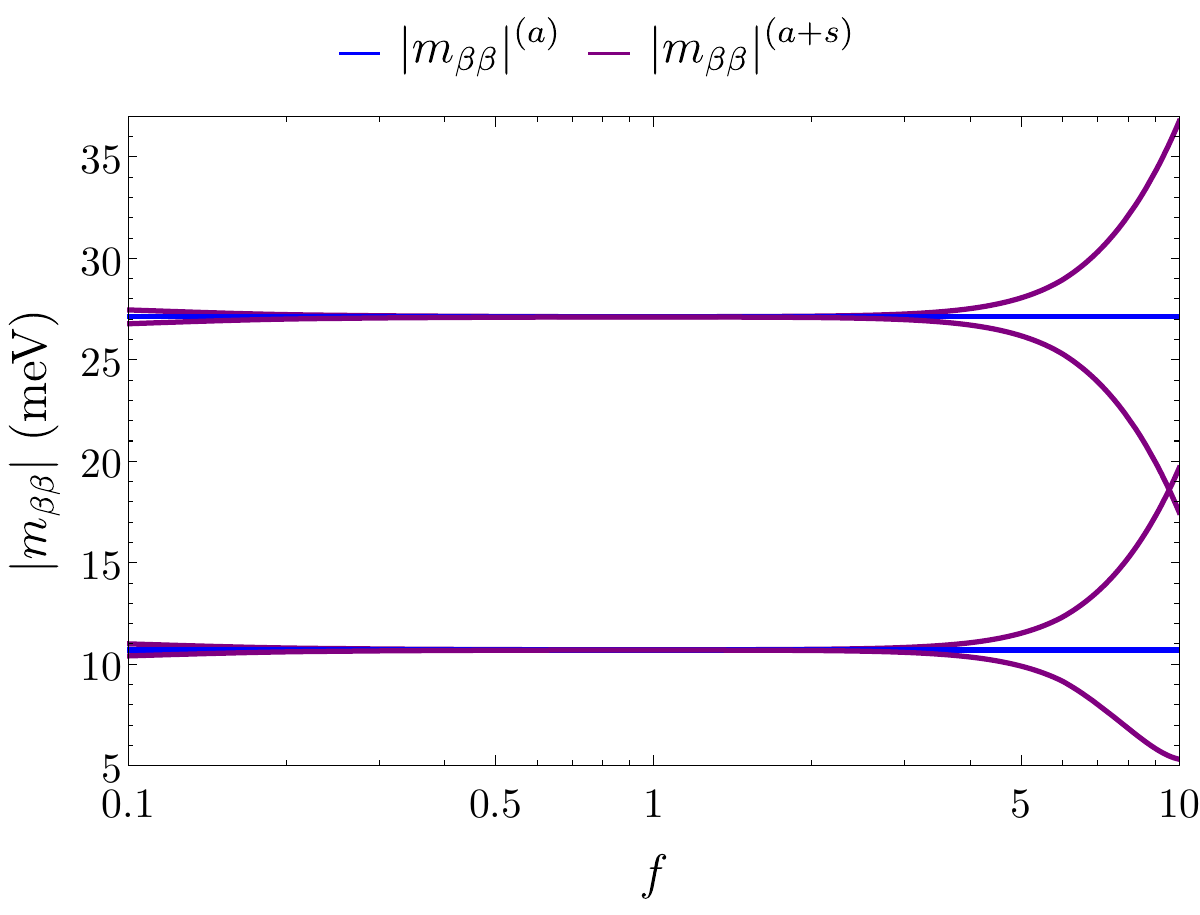}
\caption{Effective mass parameter $|m_{\beta\beta}|$ given by Eq.~\eqref{mbb}. The superscripts $(a)$ and $(a+s)$ denote the contribution coming from only active, and both active and sterile neutrinos, respectively. The former has an ambiguity \textcolor{black}{in magnitude and sign}
(see text for details), thus yielding four cases. The upper (lower) horizontal line corresponds to $|m_{\beta\beta}|^{(a)} = 27.12\ \text{meV}$ ($|m_{\beta\beta}|^{(a)} = 10.70\ \text{meV}$). For $f>4$, the contribution from the sterile neutrinos become important, and can become as large as $\mc O(10)\ \text{meV}$ near $f=10$.} 
\label{mbbfig}
\end{figure}

\reply{Before concluding this section, we briefly comment on the effects of the sterile neutrinos to the neutrinoless double beta decay parameter. The half life for the neutrinoless double beta decay for a given nucleus is \cite{deGouvea:2015euy}
\begin{align}
    \frac{1}{T_{1/2}} = \frac{G_{0\nu} M_{0\nu}^2}{m_e^2} \left|m_{\beta\beta}\right|^2,
\end{align}
where $G_{0\nu} \sim \mc O(10^{14})\ \text{yr}^{-1}$ is a phase-space factor, and $M_{0\nu}$ is a nuclear matrix element whose value as a function of neutrino masses can be obtained from Ref.~\cite{Blennow:2010th}. Sterile neutrinos mixing with active neutrinos may have nontrivial contribution to the effective mass parameter $\left| m_{\beta\beta} \right|$ in neutrinoless double beta decay. It is given by the second term in the following expression \cite{deGouvea:2015euy}:
\begin{align} \label{mbb}
    |m_{\beta \beta}| = \left| \sum_{i=1}^3 m_{\nu_i}\ {\mc U^2_\text{PMNS}}_{1i} + \sum_{I=1}^4 \frac{M_I}{1-M_I^2/p^2} U_{eN_I}^2 \right|,
\end{align}
whereas the first term describes the contribution from the standard model neutrinos. 
\textcolor{black}{The first term can be expressed by Eq.~\eqref{mbbac}, where there are four possible combinations (two possible magnitudes $27.12$ meV and $10.70$ meV with two possible overall signs) which come from four possible sign combinations of the two model parameters in Eq.~\eqref{mnu}.}
For the second term, the value of the nuclear momentum exchange $p^2$ 
can be roughly estimated to be $p^2 \sim -(0.1-0.2\ \text{GeV)}^2$ \cite{deGouvea:2015euy, Mitra:2011qr}. The second term also depends on the model parameters $b$ and $f$. 
\textcolor{black}{In general, since $|U_{eN_i}|$ increases with decreasing $M_I$, we will consider the lower bound of the sterile neutrino masses shown in Fig.~\ref{fig:minb} where the contribution is expected to be the largest.}
Using this, and putting $p^2 = -(0.1\ \text{GeV)}^2$, we show $|m_{\beta\beta}|$ as a function of $f$ in Fig.~\ref{mbbfig}.
}
\textcolor{black}{
The prefactor multiplying $U_{eN_I}^2$ in Eq.~\eqref{mbb} is maximum when $M_I^2 \approx -p^2$, and is suppressed otherwise. For $0.1<f<10$, since $M_I^2 > -p^2$, the prefactor can be approximated as $-p^2/M_I$ and the contributions become more important with decreasing $M_I$ as we move away from $f=1$ in both directions (see Fig.~\ref{fig:minb}). Since the mixing parameter $|U_{eN_2}|$ corresponding to $f>1$ is much larger than the mixing parameter $|U_{eN_1}|$ corresponding to $f<1$, cf. Figs.~\ref{mixing_a} and \ref{mixing_d}, the contribution of the sterile neutrinos to $|m_{\beta \beta}|$ is more substantial for $f>1$ than for $f<1$.}
Note that the results in Fig.~\ref{mbbfig} are still within the most stringent experimental upper limit  set by the KamLAND-Zen experiment, $|m_{\beta\beta}| < 61$-$165$ meV \cite{Dolinski:2019nrj,gando2016search}. 

\section{Conclusion and Outlook}\label{sec6}
We have considered the possibility of realizing low-scale resonant leptogenesis in a specific model based on the $SU(5)$ GUT with the $\mathcal{T}_{13}$ family symmetry. This model explains the GUT-scale mass ratios and mixing angles of both quarks and leptons with a complex TBM seesaw mixing and four right-handed neutrinos. \reply{It predicts the light neutrino masses without specifying the seesaw scale, thus prompting the investigation of both high-scale \cite{Rahat:2020mio} and low-scale leptogenesis (present work).} 
\reply{Though we have focused on the resonant regime where leptogenesis through decays is dominant over through oscillations, in the relevant regime, using density matrix equations, we have also taken into account oscillations as well as relativistic effects (helicities of right-handed neutrinos and thermal scatterings) when the baryon asymmetry freezes out while the right-handed neutrinos are still relativistic. The study where leptogenesis proceeds predominantly through oscillations can be fully explored with the same density matrix equations and will be left for future study. One expects to be able to extend the viable parameter space to a lower seesaw scale though this regime will be in tension with the BBN constraints.}

The single phase in TBM mixing, which predicts both low energy Dirac and Majorana $CP$ phases, is shown to be responsible also for $CP$ violation in resonant leptogenesis. 
We have studied resonant leptogenesis in the three flavor regime and identified a particular pair of right-handed neutrinos capable of producing resonant enhancement to the $CP$ asymmetry. 
We have found that the fourth right-handed neutrino, essential to generate viable mass spectrum for the light neutrinos, is also indispensable for low-scale resonant leptogenesis. We have determined  lower bounds on the right-handed neutrino mass spectrum for successful leptogenesis. Considering the constraints from BBN analysis,
the lowest bound on the lightest right-handed neutrino is shown to be around $2$ $\text{GeV}$. We have also found nontrivial upper bounds on the right-handed neutrino masses because of the presence of lighter neutrinos below the resonant mass which partially wash out the asymmetry generated by the resonant pair. The mixing of the sterile and active neutrinos lies within the seesaw expectation; although the regime within the sensitivity of DUNE is in tension with the BBN constraints. Future experiments designed to reach the seesaw line would be able to verify our model.


\section*{Acknowledgments}
We would like to thank Dr. Pierre Ramond, Dr. M. Jay P\'erez, Dr. Alexander J. Stuart and Bin Xu for discussion and comments on the manuscript. 
{We also thank the anonymous referee for helpful comments and constructive suggestions in improving this work.}
C.S.F. acknowledges the support by FAPESP grant 2019/11197-6 and CNPq grant 301271/2019-4. M.H.R. acknowledges partial support from  U.S. Department
of Energy under grant number DE-SC0010296.
The work of S.S. has been supported by the Swiss National Science Foundation.

\appendix
\section{Other Variants of the VEV $\vev{\varphi_{\mc B}} \equiv (b_1, b_2, b_3)$} \label{app:VEV}
The seesaw parameters relevant for leptogenesis are the right-handed neutrino masses $M_i$ and the neutrino Yukawa matrix $Y_\nu$. In this section we discuss how these parameters vary as we consider the three following VEVs: (i) $(b_1, b_2, b_2) \equiv b(1,f,1)$, (ii) $(b_1, b_2, b_2) \equiv b(f,1,1)$, and (iii) $(b_1, b_2, b_2) \equiv b(1,1,f)$.

In Sec.~\ref{sec:resonantasym} we discussed the case (i). For the Majorana matrix $\mc M$ in Eq.~\eqref{Mmat}, case (ii) and (iii) are related to case (i) in the following way:
\begin{align}
    \mc M^{(ii)} =  P_{13}\ \mc M^{(i)}\ P_{13}, \qquad  \mc M^{(iii)} = P_{23}\ \mc M^{(i)}\ P_{23},
\end{align}
where the superscript with $\mc M$ denotes the Majorana matrix for the three cases mentioned above, and $P_{jk}$ are the permutation matrices that exchanges row $j$ with row $k$:
\begin{align}
    P_{13} \equiv \left(
\begin{array}{cccc}
 0 & 0 & 1 & 0 \\
 0 & 1 & 0 & 0 \\
 1 & 0 & 0 & 0 \\
 0 & 0 & 0 & 1 \\
\end{array}
\right), \qquad P_{23} \equiv \left(
\begin{array}{cccc}
 1 & 0 & 0 & 0 \\
 0 & 0 & 1 & 0 \\
 0 & 1 & 0 & 0 \\
 0 & 0 & 0 & 1 \\
\end{array}
\right).
\end{align}
From the Takagi factorization $\mc M = \mc U_m\ \mc D_m\ \mc U_m^T$, this implies that the eigenvalues of the Majorana matrix in Eq.~\eqref{Mmass} remains same, but the unitary matrix $\mc U_m$ in Eq.~\eqref{Unitm} is transformed as
\begin{align}
    \mc U_m^{(ii)} = P_{13}\ \mc U_m^{(i)}, \qquad \mc U_m^{(iii)} = P_{23}\ \mc U_m^{(i)}.
\end{align}
The superscript with $\mc U_m$ indicate which of the three cases it represents. 

\begin{figure}[!ht] 
    \centering
    \subfloat[$(b_1, b_2, b_3) \equiv b(f,1,1)$]{
      \includegraphics[width=0.48\textwidth]{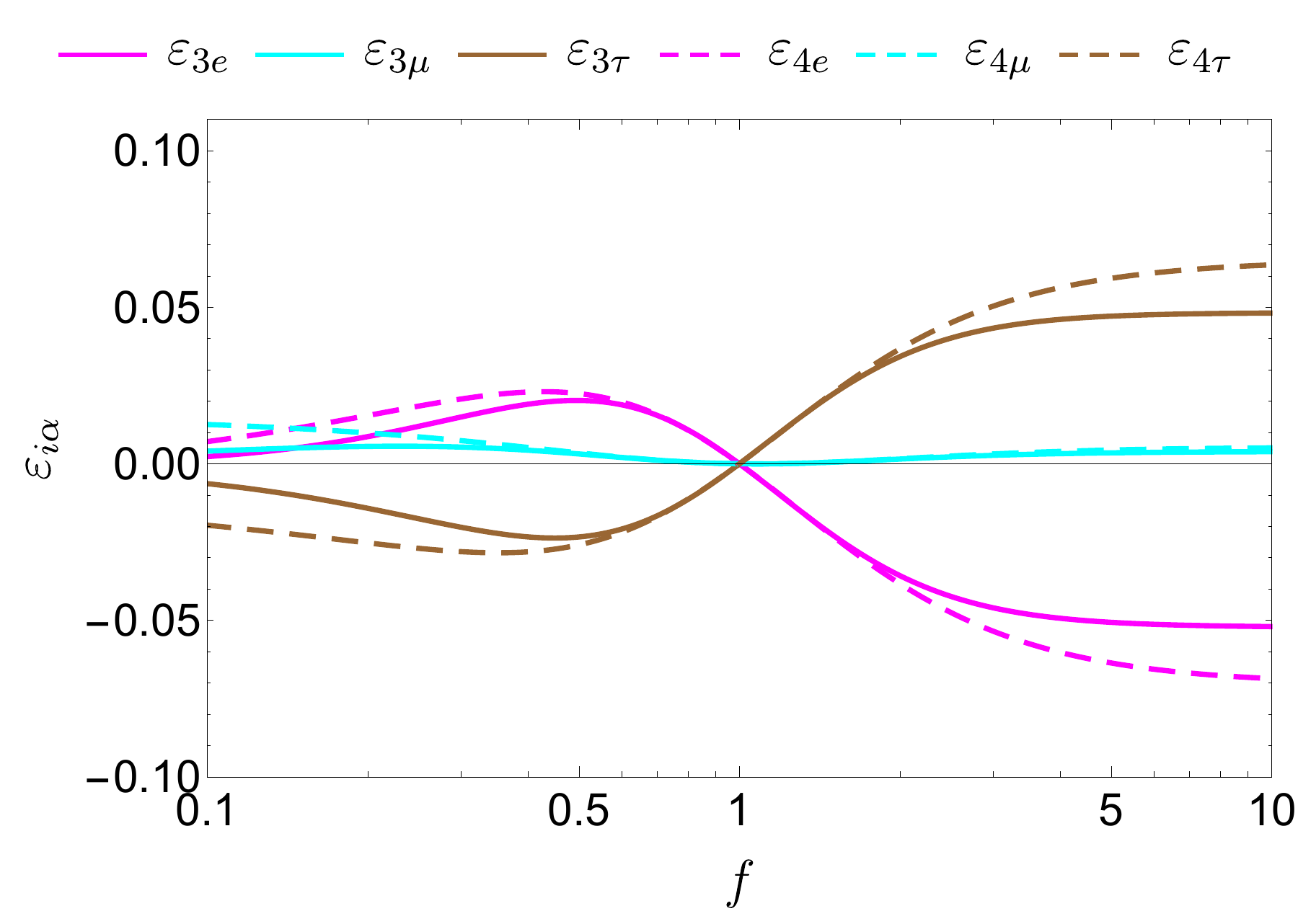}
    }
    \subfloat[$(b_1, b_2, b_3) \equiv b(1,1,f)$]{
      \includegraphics[width=0.48\textwidth]{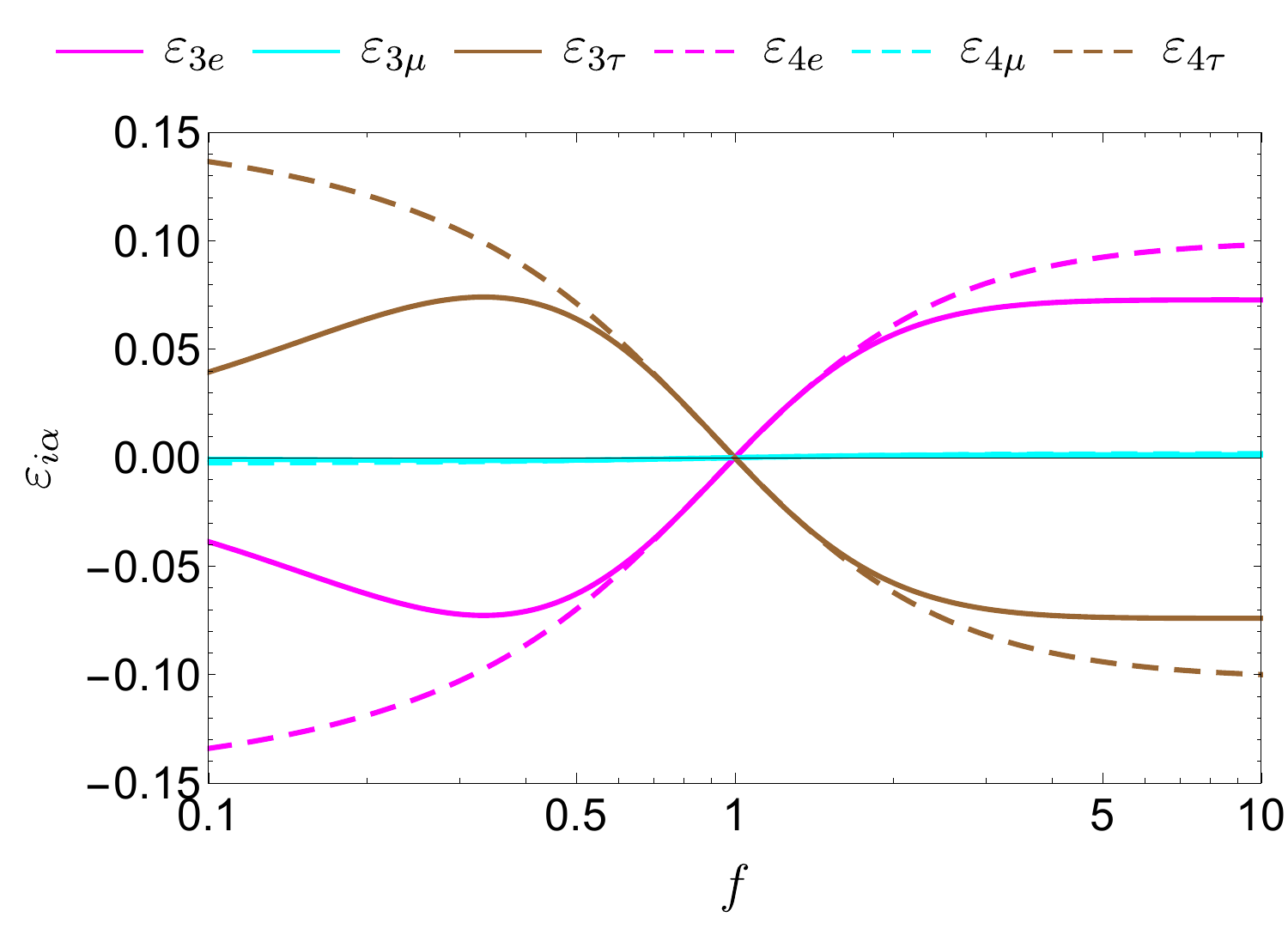}
    }
    \caption{$CP$ asymmetry parameters at the resonance $M_3 \simeq M_4$ for the plots (a) $(b_1, b_2, b_3) \equiv b(f,1,1)$ and (b) $(b_1, b_2, b_3) \equiv b(1,1,f)$. The sum of the flavored $CP$ asymmetries is zero, $\sum_\alpha \varepsilon_{i\alpha} = 0$, and hence unflavored leptogenesis is not successful in this model \cite{Rahat:2020mio}. At $f=1$, the individual flavor components vanish.
    }
    \label{fig:CPasymmetry34__11f_f11}
\end{figure}

\begin{figure}[!ht] 
    \centering
    \subfloat[$(b_1, b_2, b_3) \equiv b(f, 1, 1) $ \label{P1K1_f11}]{
        \includegraphics[width=0.48\textwidth]{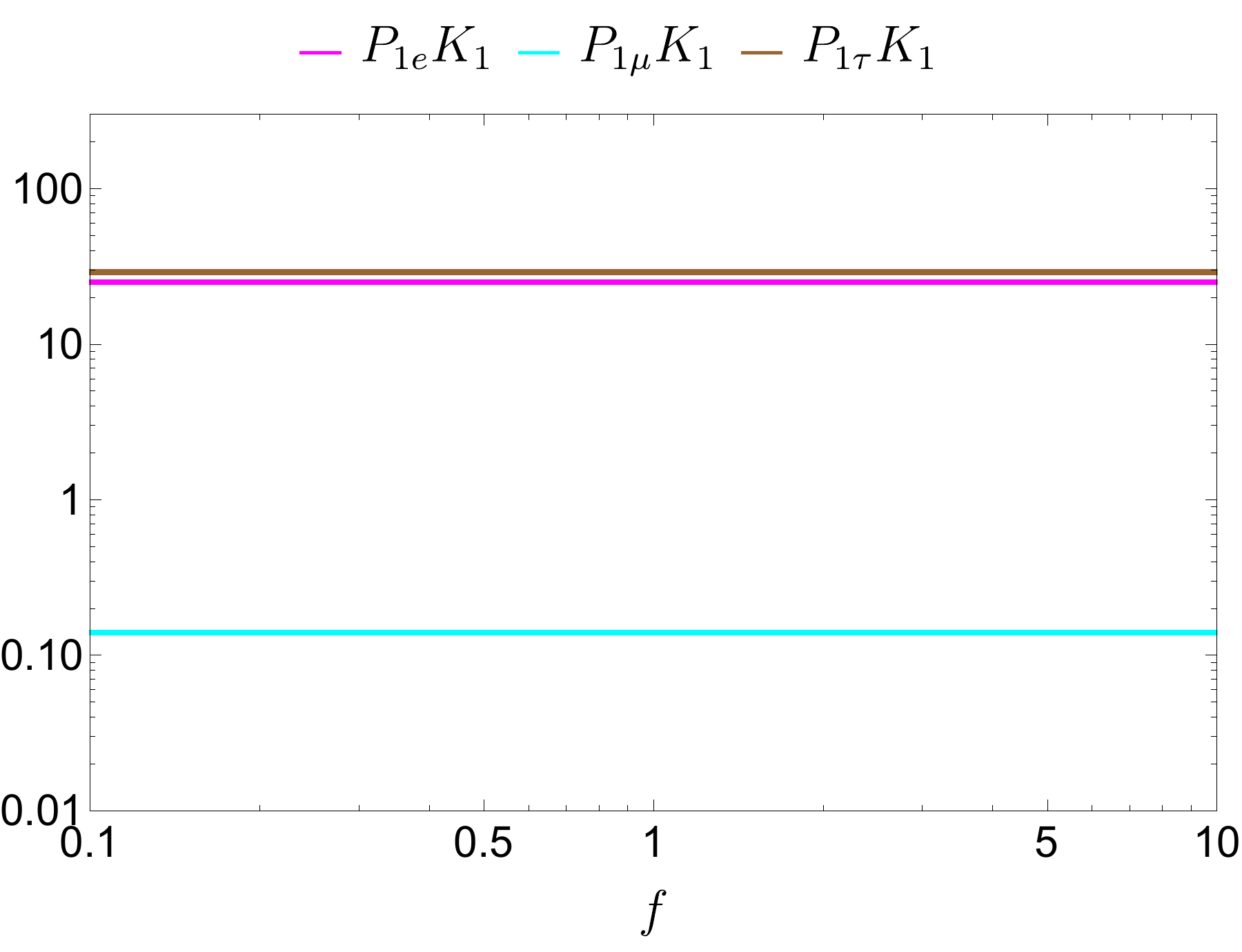}
    }
    \subfloat[$(b_1, b_2, b_3) \equiv b(f, 1, 1) $ \label{P2K2_f11}]{
        \includegraphics[width=0.48\textwidth]{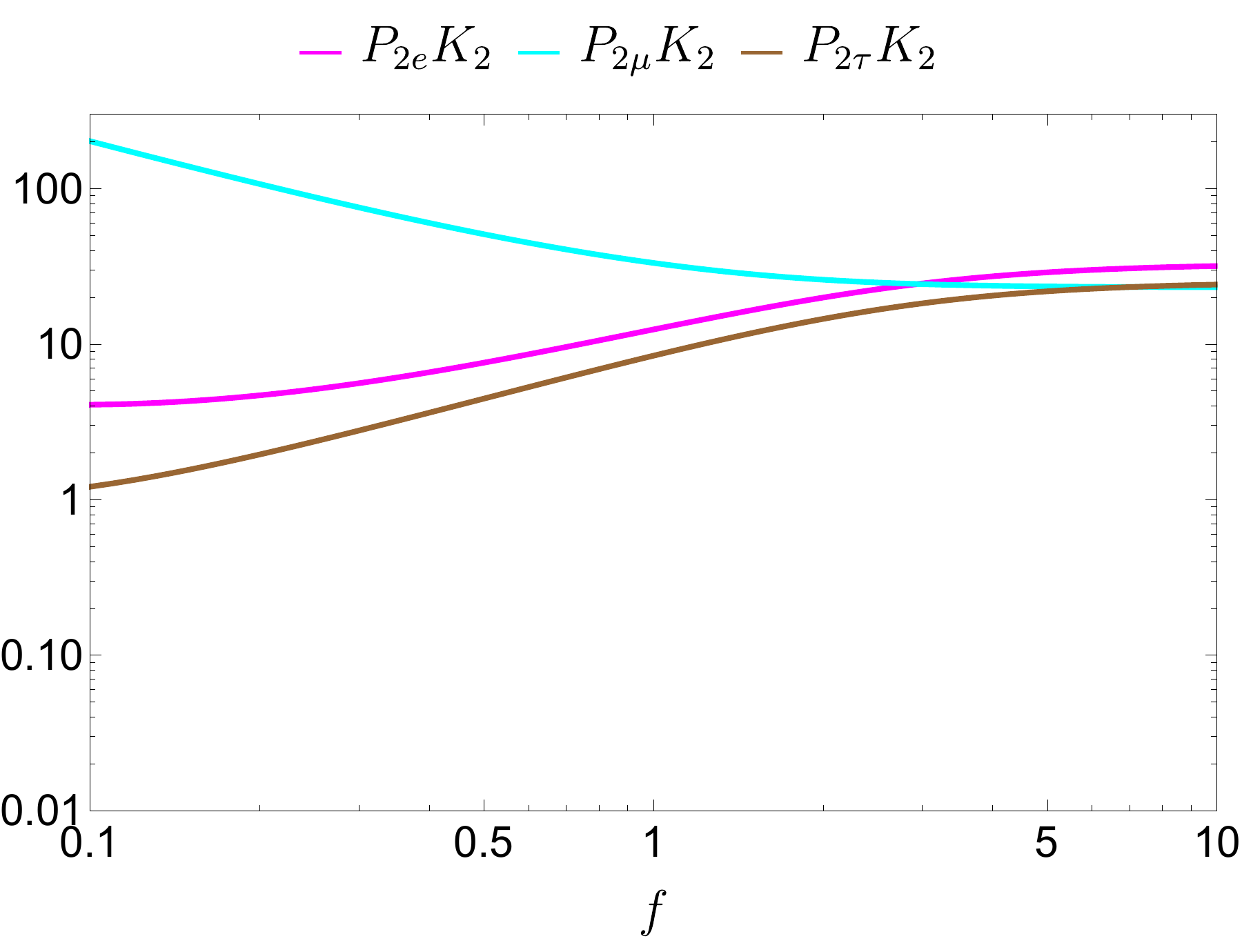}
    }\\ 
    \subfloat[$(b_1, b_2, b_3) \equiv b(1, 1, f) $ \label{P1K1_11f}]{
        \includegraphics[width=0.48\textwidth]{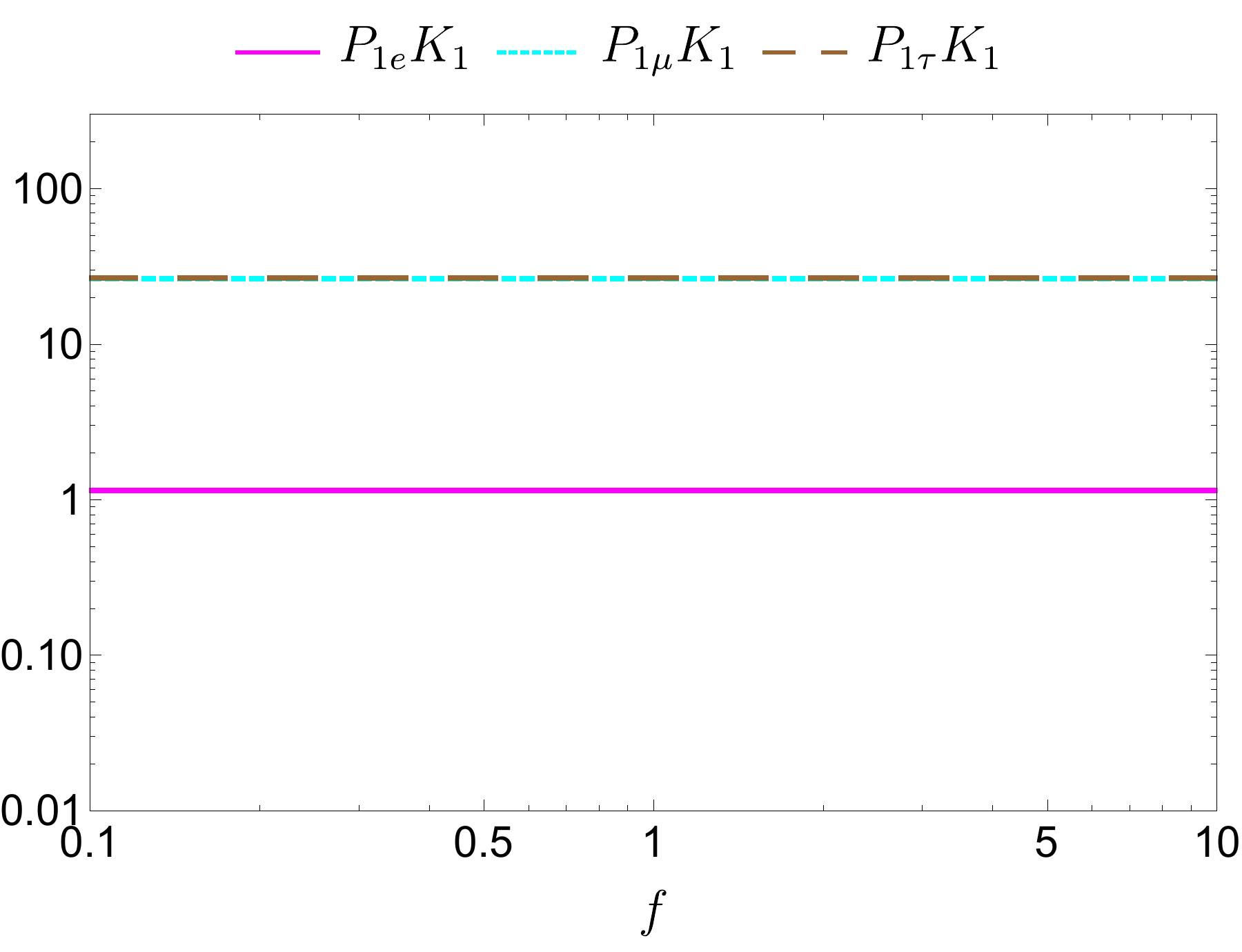}
    }
    \subfloat[$(b_1, b_2, b_3) \equiv b(1, 1, f) $ \label{P2K2_11f}]{
        \includegraphics[width=0.48\textwidth]{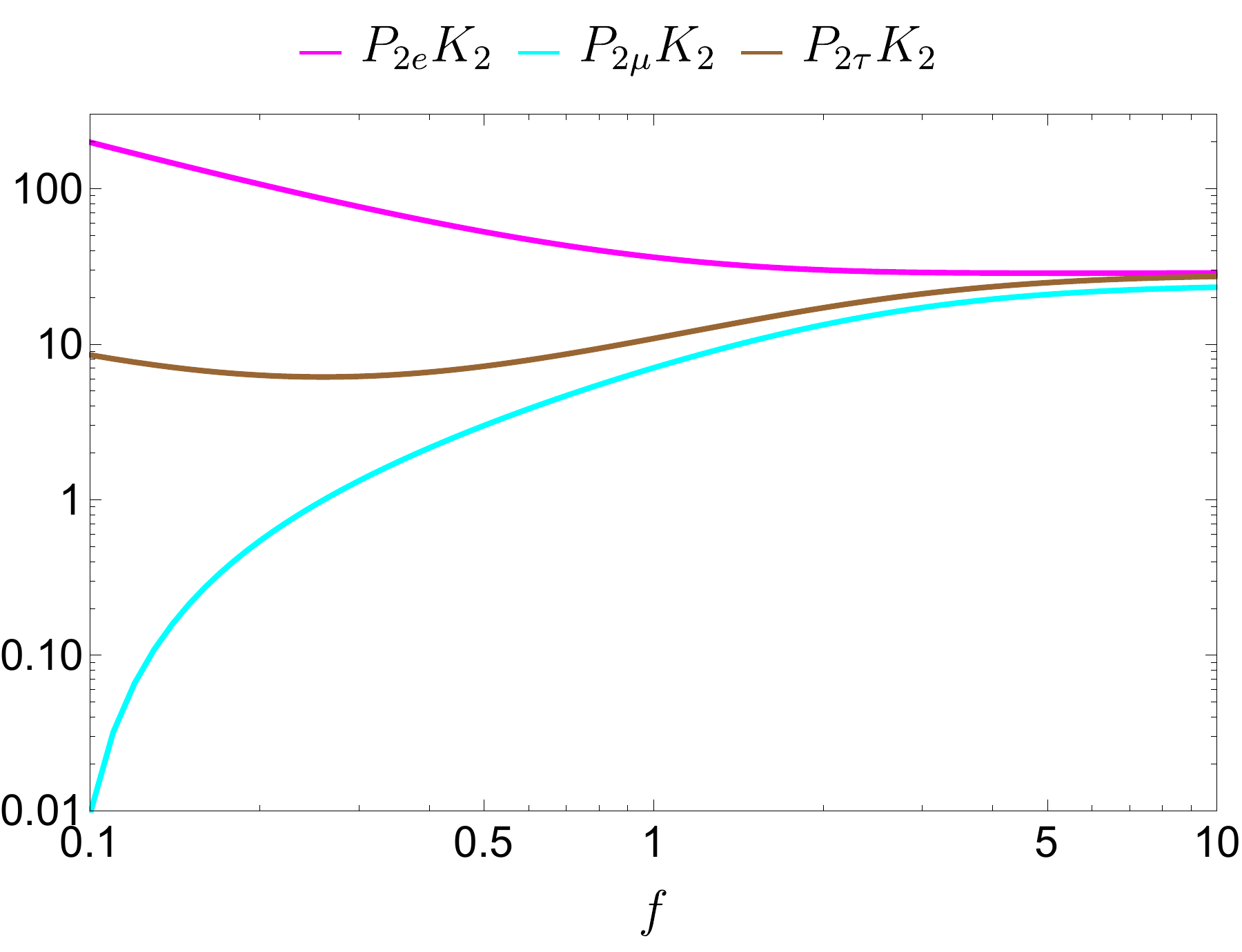}
    }
    \caption{Decay parameter times branching ratio as a function of $f$ at resonance for the plots (a), (b) $(b_1, b_2, b_3) \equiv b(f,1,1)$ and (c), (d) $(b_1, b_2, b_3) \equiv b(1,1,f)$. The asymmetry generated by $N_3$ and $N_4$ at resonance is partially washed out by $N_1$ and $N_2$, 
    \textcolor{black}{and for $M_{1,2} \ll M_{3,4}$, is proportional to $e^{-P_{i\alpha}K_i}$.}
    } 
    \label{fig:PiKi_11f_f11}
\end{figure}

The neutrino Yukawa matrix $Y_\nu$ is defined in Eq.~\eqref{first}, where $\mc U_m$ should be the appropriate unitary matrix for each case and $Y^{(0)}$ is calculated from Eq.~\eqref{Y0} with the corresponding VEV. Explicitly calculating the Hermitian matrix $Y^\dagger_\nu Y_\nu$, we find that the cases (i) and (ii) yields the same result as in Eq.~\eqref{YY}, but the case (iii) is slightly different:
\begin{align}
    &Y^{{(ii)}^\dagger}_\nu Y_\nu^{(ii)} = Y^{{(i)}^\dagger}_\nu Y_\nu^{(i)}, \\
    &Y^{{(iii)}^\dagger}_\nu Y_\nu^{(iii)} = \frac{b f m_\nu}{v^2} \nonumber \\
    &\times\!\! \left(\!\!
\begin{array}{cccc}
 1 & 0 & 0 & 0 \\
 * & \frac{1}{2}\! \left(1-\!\frac{f^3-f-\sqrt{f^2+8}}{f^2 \sqrt{f^2\!+\!8}}\!\right) & \frac{-i \sqrt{2} \left(f^2-1\right)}{f^2 \sqrt{f^2\!+\!8}} & \frac{i\beta}{f} \left(\!\!\sqrt{\frac{2 f}{\sqrt{f^2+8}}\!+\!2}+f \sqrt{1\!-\!\frac{f}{\sqrt{f^2+8}}}\!\right)  \\
 * & * & \frac{1}{2} \left(1+\frac{f^3-f+\sqrt{f^2+8}}{f^2 \sqrt{f^2+8}}\right) & \frac{\beta}{f}\left(\!\!\sqrt{2-\frac{2 f}{\sqrt{f^2\!+\!8}}}-f \sqrt{\frac{f}{\sqrt{f^2\!+\!8}}+1}\!\right) \\
 * & * & * & 6 \beta ^2 \\
\end{array}
\!\!\!\!\right), \label{YYiii}
\end{align}
where $*$ denotes the complex conjugate of corresponding transposed elements. Eq.~\eqref{YYiii} for the case (iii) is identical to Eq.~\eqref{YY} for the cases (i) and (ii), except for the off-diagonal elements in the fourth row and fourth column. However, the only real off-diagonal element is still the $(34)$ element, similar to Eq.~\eqref{YY}. Hence, the only relevant quasi-degeneracy for resonant leptogenesis remains to be $M_3 \simeq M_4$.

Due to the changes in $Y_\nu$ and $Y_\nu^\dagger Y_\nu$, leptogenesis parameters like $CP$ asymmetry, branching ratios, decay parameters etc. are quantitatively different in the cases (ii) and (iii) compared to the case (i) discussed in Sec. \ref{sec:results} and \ref{sec:resonantasym}. In Figs.~\ref{fig:CPasymmetry34__11f_f11} and \ref{fig:PiKi_11f_f11} we show the parameters $\varepsilon_{3\alpha}$, $\varepsilon_{4\alpha}$ and $P_{1\alpha}K_1$, $P_{2\alpha}K_2$ for cases (ii) and (iii).

Qualitatively, from Fig.~\ref{fig:CPasymmetry34__11f_f11}, we see that the dominant $CP$ asymmetry parameters for cases (ii) and (iii) are in the $e$ and $\tau$ flavors similar to the case (i) as shown in Fig.~\ref{fig:CPasymmetry34}. Regarding the decay parameters, for $f \gg 1$ where $M_2 \ll M_{1,3,4}$, the relevant washout effects are from $N_2$ as shown in Fig.~\ref{fig:PiKi_11f_f11} (b) and (d). From these plots, we see that washout effects are strong in all the flavors $P_{2\alpha} K_2 \gg 1$ for all $\alpha$ for cases (ii) and (iii), similar to Fig.~\ref{fig:PiKi} for case (i). Hence, one will obtain an upper bound on the right-handed neutrino mass spectrum.

For $f \ll 1$ where $M_1 \ll M_{2,3,4}$, the relevant washout effects are from $N_1$. In this case, we see there is always one flavor asymmetry $N_{\Delta \alpha}$ in which the washout is not effective. For case (ii) (Fig.~\ref{fig:PiKi_11f_f11} (a)), $N_{\Delta \mu}$ does not suffer washout while for case (iii) (Fig.~\ref{fig:PiKi_11f_f11} (c)), $N_{\Delta e}$ suffers very mild washout. Compared to case (i) (Fig.~\ref{fig:PiKi} (a)), it is $N_{\Delta \tau}$ which survives. Hence there will not be upper bound on the right-handed neutrino mass spectrum.

Regarding the active neutrino-$N_2$ mixing for cases (ii) and (iii), they are similar to those of case (i) as presented in Fig.~\ref{mixing}. As for active neutrino-$N_1$ mixing, there is an interesting correlation where the largest mixing is for those with smallest $P_{1_\alpha} K_1$. For case (i), the one with the largest mixing is with the tau flavor neutrino, for case (ii), it is with the muon flavor neutrino while for case (iii), it is with the electron neutrino.

\newpage

\section{Density Matrix Equations for Leptogenesis} \label{app:density_matrix}

At $T > M_i$, the effects from distinguishing between the two helicity states $N$ and $\bar N$ as well as scatterings could be relevant. This could be the case when we consider the mass scale of $N$ smaller than the EW sphalerons freeze-out temperature $M_i \lesssim T_{sph} \sim 131.7$ GeV. In this regime, we use the density matrix equations adapted from Ref.~\cite{Abada:2018oly} in the basis where the Majorana mass matrix $\mc D_m \equiv \text{diag}(M_1, M_2, M_3, M_4)$ is real and diagonal.
\begin{eqnarray}
zH\frac{dN_{N}}{dz} & = & -i\left[\left\langle {\cal H}\right\rangle ,N_{N}\right] -\frac{1}{2}\left\langle \gamma^{\left(0\right)}\right\rangle \left\{ Y_{\nu}^{\dagger}Y_{\nu},N_{N}-\hat{N}_{N}^{{\rm eq}}\right\} 
\nonumber \\
&& -\frac{1}{2}\left\langle \tilde{\gamma}^{\left(0\right)}\right\rangle \left\{ \mc D_mY_{\nu}^{T}Y_{\nu}^{*}\mc D_m,N_{N}-\hat{N}_{N}^{{\rm eq}}\right\} \nonumber \\
 &  & +\left\langle \gamma^{\left(1a\right)}\right\rangle N_{N}^{{\rm eq}}Y_{\nu}^{\dagger}\mu Y_{\nu}-\left\langle \tilde{\gamma}^{\left(1a\right)}\right\rangle N_{N}^{{\rm eq}}\mc D_mY_{\nu}^{T}\mu Y_{\nu}^{*}\mc D_m\nonumber \\
 &  & +\frac{1}{2}\left\langle \gamma^{\left(1b\right)}\right\rangle \left\{ Y_{\nu}^{\dagger}\mu Y_{\nu},N_{N}\right\} -\frac{1}{2}\left\langle \tilde{\gamma}^{\left(1b\right)}\right\rangle \left\{ \mc D_mY_{\nu}^{T}\mu Y_{\nu}^{*}\mc D_m,N_{N}\right\} ,\\
zH\frac{dN_{\bar{N}}}{dz} & = & -i\left[\left\langle {\cal H}^{T}\right\rangle ,N_{\bar{N}}\right]-\frac{1}{2}\left\langle \gamma^{\left(0\right)}\right\rangle \left\{ Y_{\nu}^{T}Y_{\nu}^{*},N_{\bar{N}}-\hat{N}_{N}^{{\rm eq}}\right\} 
\nonumber \\
&& -\frac{1}{2}\left\langle \tilde{\gamma}^{\left(0\right)}\right\rangle \left\{ \mc D_mY_{\nu}^{\dagger}Y_{\nu}\mc D_m,N_{\bar{N}}-\hat{N}_{N}^{{\rm eq}} \right\} \nonumber \\
 &  & -\left\langle \gamma^{\left(1a\right)}\right\rangle N_{N}^{{\rm eq}}Y_{\nu}^{T}\mu Y_{\nu}^{*}+\left\langle \tilde{\gamma}^{\left(1a\right)}\right\rangle N_{N}^{{\rm eq}}\mc D_mY_{\nu}^{\dagger}\mu Y_{\nu}\mc D_m\nonumber \\
 &  & -\frac{1}{2}\left\langle \gamma^{\left(1b\right)}\right\rangle \left\{ Y_{\nu}^{T}\mu Y_{\nu}^{*},N_{\bar{N}}\right\} +\frac{1}{2}\left\langle \tilde{\gamma}^{\left(1b\right)}\right\rangle \left\{ \mc D_mY_{\nu}^{\dagger}\mu Y_{\nu}\mc D_m,N_{\bar{N}}\right\} ,\\
zHN^{{\rm nor}}\frac{d\mu_{\Delta_{\alpha}}}{dz} & = & -\left[\left\langle \gamma^{\left(0\right)}\right\rangle \left(Y_{\nu}N_{N}Y_{\nu}^{\dagger}-Y_{\nu}^{*}N_{\bar{N}}Y_{\nu}^{T}\right) \right.
\nonumber \\
&& +\left\langle \tilde{\gamma}^{\left(0\right)}\right\rangle \left(Y_{\nu}^{*}\mc D_mN_{\bar{N}}\mc D_mY_{\nu}^{T}-Y_{\nu}\mc D_mN_{N}\mc D_mY_{\nu}^{\dagger}\right)\nonumber \\
 &  & -2\left\langle \gamma^{\left(1a\right)}\right\rangle N_{N}^{{\rm eq}}\mu Y_{\nu}Y_{\nu}^{\dagger}-2\left\langle \tilde{\gamma}^{\left(1a\right)}\right\rangle N_{N}^{{\rm eq}}\mu Y_{\nu}^{*}\mc D_m^{2}Y_{\nu}^{T}\nonumber \\
 &  & -\left\langle \gamma^{\left(1b\right)}\right\rangle \mu\left(Y_{\nu}N_{N}Y_{\nu}^{\dagger}+Y_{\nu}^{*}N_{\bar{N}}Y_{\nu}^{T}\right)
 \nonumber \\
&&  \left.-\left\langle \tilde{\gamma}^{\left(1b\right)}\right\rangle \mu\left(Y_{\nu}^{*}\mc D_mN_{\bar{N}}\mc D_mY_{\nu}^{T}+Y_{\nu}\mc D_mN_{N}\mc D_mY_{\nu}^{\dagger}\right)
\right]_{\alpha\alpha},
\end{eqnarray}
where ${\cal H}= {\cal H}_0 + V$ is the Hamiltonian including the potential from Yukawa interactions $V$, $z=m_{{\rm ref}}/T$ with $m_{{\rm ref}}$ an arbitrary mass scale, $N^{{\rm nor}}=\frac{\pi^{2}}{6\zeta\left(3\right)}$ and the Hubble rate in a radiation-dominated Universe is given by
\begin{eqnarray}
H & = & 1.66\sqrt{g_{\star}}\frac{m_{{\rm ref}}^{2}}{z^{2}M_{{\rm Pl}}},
\end{eqnarray}
with $M_{{\rm Pl}}=1.22\times10^{19}$ GeV and we will consider the
effective relativistic degrees of freedom as fixed to be $g_{\star}=106.75$ (including the four relativistic $N$, $g_\star = 113.75$ which gives only negligible effect) since we only consider $T>T_{sph}$. We have also defined $\mu \equiv -{\rm diag}(\mu_{\Delta_e},\mu_{\Delta_\mu},\mu_{\Delta_\tau})$ where the $\mu_{\Delta \alpha}$ is the chemical potential related to $B/3-L_\alpha$ charge normalized by temperature $T$ (order of one spectator effects \cite{Buchmuller:2001sr} have been neglected). Both $N_N$ and $N_{\bar N}$ are $4\times 4$ symmetric matrices of their respective number densities normalized by the photon density in the family space of $N$. 

Since the equations will be applied in the regime where $M_i \lesssim T_{sph}$, we will take $N_N^{\rm eq} = 3/8$ assuming relativistic $N$ with one degree of freedom (in the classical Boltzmann equations in Sec. \ref{sec3}, there is an additional factor of 2 since we have summed over the two spin degrees of freedom). Nevertheless, the decays of the resonant pairs are crucial for generating asymmetry (in the case of thermal initial abundance of $N_i$) and in order to describe this effect, we take
$\hat N_N^{\rm eq} = {\rm diag}(N_{N_1}^{\rm eq},N_{N_2}^{\rm eq},N_{N_3}^{\rm eq},N_{N_4}^{\rm eq})$ with
\begin{eqnarray}
N_{N_i}^{\rm eq} & = & \frac{1}{4\zeta(3)T^3} 
\int_{M_i}^{\infty} dE \sqrt{E^2 - M_i^2} E f_N^{\rm eq},
\end{eqnarray}
where $f_N^{\rm eq} = (\exp{(E/T)} + 1)^{-1}$ is the equilibrium phase space distribution of $N$ and we have approximated the $N$ mass basis to be that of the vacuum (this is reasonable for our model where neutrino Yukawa couplings are small). In our model, asymmetry from decay of the nonresonant pairs $N_1$ and $N_2$ are subdominant and this is confirmed by taking $N_{N_1}^{\rm eq} = N_{N_2}^{\rm eq} = 3/8$ without changing the results.

The thermal averaged rate is defined as
\begin{eqnarray}
\left\langle \gamma\right\rangle  & = & \frac{\int d^{3}p\, \gamma f_{N}^{{\rm eq}}}{\int d^{3}p \,f_{N}^{{\rm eq}}}.
\end{eqnarray}
In the regime where $M \lesssim T_{sph}$, $N$ is relativistic $T\gg M$ and we obtain
\begin{eqnarray}
\left\langle {\cal H}_{0}\right\rangle  & = & \frac{r}{2}{\rm diag}\left(M_{1}^{2},M_{2}^{2},M_{3}^{2},M_{4}^{2}\right),\\
\left\langle V \right\rangle  & = & \frac{rT^{2}}{8}Y^{\dagger}Y,\\
\left\langle \gamma^{(i)}\right\rangle  & = & \frac{rT^{2}}{128\pi}\gamma^{(i)},
\end{eqnarray}
where $r\equiv\frac{\pi^{2}}{18\zeta\left(3\right)T}$ and
\begin{eqnarray}
\gamma^{(i)} & = & a_{i}\left[c_{{\rm LPM}}^{(i)}+y_{t}^{2}c_{Q}^{(i)}+\left(3g^{2}+g'^{2}\right)\left(c_{V}^{(i)}-\ln\left(3g^{2}+g'^{2}\right)\right)\right], \\
\tilde \gamma^{(i)} & = & a_i c_{1\to 2}^{(i)},
\end{eqnarray}
with $a_{0}=2a_{1a}=4a_{1b}=1$. 
For the coefficients, we will ignore the mild temperature dependence and fix them as Ref. \cite{Abada:2018oly} using $c_{LPM}^{(i)}(T=10^4\ \text{GeV})$ as the reference value:
\begin{align}
\begin{tabularx}{0.8\textwidth}{X X X X}
    {$\begin{aligned}
        c_{LPM}^{(0)} &= 4.22,\\
        c_{LPM}^{(1a)} &= 3.56,\\
        c_{LPM}^{(1b)} &= 4.77,
    \end{aligned}$} &  
    {$\begin{aligned}
        c_{Q}^{(0)} &= 2.57,\\
        c_{Q}^{(1a)} &= 3.10,\\
        c_{Q}^{(1b)} &= 2.27,
    \end{aligned}$} & 
    {$\begin{aligned}
        c_{V}^{(0)} &= 3.17,\\
        c_{V}^{(1a)} &= 3.83,\\
        c_{V}^{(1b)} &= 2.89,
    \end{aligned}$} & 
    {$\begin{aligned}
        c_{1\rightarrow 2}^{(0)} &= 0.86/T^2,\\
        c_{1\rightarrow 2}^{(1a)} &= 20.4/T^2,\\
        c_{1\rightarrow 2}^{(1b)} &= 20.4/T^2.
    \end{aligned}$}
\end{tabularx}
\end{align}
The SM gauge couplings run as
\begin{align}
    g(\Lambda) &=\left(\frac{1}{g_{0}^{2}}+\frac{19}{48 \pi^{2}} \ln \frac{\Lambda}{m_{Z}}\right)^{-1 / 2}, \\
    g^{\prime}(\Lambda) &=\left(\frac{1}{\left(g_{0}^{\prime}\right)^{2}}+\frac{41}{48 \pi^{2}} \ln \frac{\Lambda}{m_{Z}}\right)^{-1 / 2},
\end{align}
where at $\Lambda = \pi T = m_Z$ the couplings are given by $g_0 = 0.652$ and $g_0' = 0.357$. 

The final $B-L$ asymmetry is given by
\begin{equation}
N_{B-L}^f = N^{\rm nor}\sum_\alpha \mu_{\Delta \alpha} (z_f = m_{\rm ref}/T_{sph}),
\end{equation}
and is related to the present day baryon asymmetry normalized by photon density as in Eq. \eqref{formulaB}:
\begin{equation}
    \eta_B \simeq 1.28 \times 10^{-2} N_{B-L}^f.
\end{equation}


%

%

%

\clearpage

\bibliography{mainresonance}
\newpage
\bibliographystyle{apsrev41}

\end{document}